\documentclass{article}
\usepackage{threeparttable}
\usepackage{longtable}
\usepackage{amssymb}
\usepackage{amsmath}
\usepackage{pifont}
\usepackage{hyperref}
\usepackage{graphicx}
\usepackage{subfigure}
\usepackage{float}
\graphicspath{{Figures/}}
\usepackage[utf8]{inputenc}    
\usepackage[numbers]{natbib}
\DeclareUnicodeCharacter{0308}{$\bullet$}
\usepackage[verbose=true,letterpaper]{geometry}
\usepackage{authblk}
\usepackage[table, svgnames, dvipsnames]{xcolor}
\usepackage{multirow}
\usepackage{hyperref}
\AtBeginDocument{
  \newgeometry{
    textheight=9in,
    textwidth=6.5in,
    top=1in,
    headheight=14pt,
    headsep=25pt,
    footskip=30pt
  }
}

\title{\textbf{Motif Discovery Algorithms in Static and Temporal Networks: A Survey}}

\author[]{Ali Jazayeri}
\author[]{Christopher C. Yang}

\affil[]{\footnotesize College of Computing \& Informatics, Drexel University, Philadelphia, PA 19104, USA}

\date{}
\begin{document}
\maketitle

\begin{abstract}
{Motifs are the fundamental components of complex systems. The topological structure of networks representing complex systems and the frequency and distribution of motifs in these networks are intertwined. The complexities associated with graph and subgraph isomorphism problems, as the core of frequent subgraph mining, have direct impacts on the performance of motif discovery algorithms. To cope with these complexities, researchers have adopted different strategies for candidate generation and enumeration, and frequency computation. In the past few years, there has been an increasing interest in the analysis and mining of temporal networks. These networks, in contrast to their static counterparts, change over time in the form of insertion, deletion, or substitution of edges or vertices or their attributes. In this paper, we provide a survey of motif discovery algorithms proposed in the literature for mining static and temporal networks and review the corresponding algorithms based on their adopted strategies for candidate generation and frequency computation. As we witness the generation of a large amount of network data in social media platforms, bioinformatics applications, and communication and transportation networks and the advance in distributed computing and big data technology, we also conduct a survey on the algorithms proposed to resolve the CPU-bound and I/O bound problems in mining static and temporal networks.}
\end{abstract}

\textbf{\textit{Keywords:}} Subgraph mining, Motif discovery, Network mining, Temporal networks, Static networks

\section{Introduction}
\sloppy
Complex systems characterized by adaptation, self-organization, and emergence are observed in different domains \cite{ottino2004engineering}. One of the common building blocks of these systems are motifs defined as recurring patterns appear more frequently in the networks representing these systems than in some reference networks \cite{Alon2007motifs, MiloS2002motifs, Kashtan2004generalizations, Mangan2003Forward, Milo2004Superfamilies}. Also, it is shown that the topological organization of complex networks and their frequent patterns are mutually related \citep{Topological2004Complex}. The conventional approach to detect these patterns is to iteratively generate candidate patterns and compute their frequencies, identify their significant and frequent ones, and use them to create new candidates. The frequency computation is performed either by counting or enumeration of patterns. In either case, it requires coping with the complexities associated with graph isomorphism (not known to be in the P or NP-complete \citep{Fortin96thegraph, Read1977isomorphism}) and subgraph isomorphism (known to be in NP-complete \citep{GareyJohnson1980}) problems.

It is shown that static or dynamic network modeling of a complex system changing over time might result in different insights \citep{Tangetal2010TemporalDistance, Kempeetal2002Connectivity, Pan2011temporal, Holme2012Temporal, Holme2015colloquium}. One approach is to represent the system as an attributed static network. In this representation, the attributes of vertices or edges summarize the temporal aspect or appearance of these components. However, it is shown that this approach cannot represent the temporality of the system completely \citep{Kovanenetal2011Temporalmotifs, Nicosia2013Temporal}. Moreover, the concepts with known definitions in static networks need to be re-defined to be applicable in temporal networks, such as identification of central components of the network and controllability \citep{Tangetal2010centralitymetrics, Kostakos2009Temporal, Lietal2017temporalnetworks, Ravandi2019driver} or the definition of concepts used for motif discovery \cite{Redmond2013Isomorphism, Huang2014Traversals}.  

The problem of motif discovery in networks representing complex systems have been an avenue for extensive research in the past few decades, for example in transportation networks \cite{kaluza2010complex}, food webs \cite{paulau2015motif}, economic and financial networks \cite{Ohnishi2010}, brain networks \cite{Sporns2004brain, sporns2007identification, martens2017brain}, metabolic pathways \cite{Youetal2006MetabolicPathways, lacroix2006motif}, networks similarity calculation \citep{Przulj2007graphletdistribution}, conserved structures in biological networks \citep{Albert2004motifs, Kelleyetal2003Pathways, Sharanetal2005Conserved, wuchty2003evolutionary, Shen2002transcriptional}, communication and human mobility networks \citep{Zhaoetal2010Communicationmotifs, li2014statistically, Schneider2013human}, cattle movements \cite{Bajardi2011Movements}, and social insects interactions \cite{Waters2012Insect, quevillon2015social}. 

The input data to the problem of motif discovery generally is one or a few giant networks. And the output is the patterns identified as frequent or significant based on their exact or approximated frequencies. Another similar problem not covered in this survey is mining frequent subgraphs in a set or sequence composed of a large number of networks. The latter problem is called network-transaction setting. Each network in the set or sequence would be a transaction. The objective is to find the subgraphs in common in a pre-defined number of networks in the set. In other words, in the motif discovery, the algorithm mines the input data to discover the frequent patterns in a single network, with an exact or approximated value of the number of appearances of each pattern in the network. In the network-transaction problem, the algorithm mines a set of networks to find the patterns that appear in at least a pre-defined number of transactions of the set. In general, the number of appearances of patterns in each member of the set is not of interest. Therefore, the main difference between these two problems is in the frequency computation approach. The algorithms proposed for motif discovery can be modified and applied to the network-transaction setting. However, the algorithms developed for network-transactions settings cannot necessarily be used for motif discovery \cite{KuramochiKarypis2006MiningGraphData}.

Both problems have attracted increasing popularity in the past years in both static and temporal networks, and numerous algorithms have been proposed to solve them. These algorithms have been reviewed in multiple studies. Some of the review papers focus on both problems and some just on either a single large network or a set of networks. Table \ref{tab:review_papers} summarizes these review papers with the \textit{Problem setting} column showing the problem covered in each of them. Some papers have reviewed the motif discovery algorithms and their applications in specific domains (mostly biological). For example, in \citep{Bruno2010Trends} (also refer to \citep{parthasarathy2010survey}), the mining approaches proposed for motif discovery in biological networks are reviewed and categorized in three categories based on the source of the network: gene regulatory networks, protein-protein interaction networks, and metabolic networks. This survey paper categorizes motifs into two groups, structural motifs, in which the motifs are mined, taking only the topological structure of the motif into account, and node-colored motifs, in which the functional characteristics of the nodes are considered as well. In this case, the nodes or edges are not anonymous (unlabeled) elements anymore. Another survey paper related to motif discovery in protein-protein interaction networks in the single network setting is written by Ciriello and Guerra \cite{Ciriello2008review} (also Chapters seven \citep{Willy2009Discovering} and eight \citep{Wan2009Motifs} of \cite{Li2009Biological}). They categorized the algorithms into two categories composed of exact frequency algorithms, in which the frequency of subgraphs are exactly computed, and approximate frequency algorithms, in which the frequency of subgraphs are approximated. In the latter case, the subgraphs meeting the frequency or significant conditions are collected. A similar categorization is provided in \citep{Masoudi2012review}, and it is concluded that the main differences between these two categories are the higher speeds achievable by approximate frequency algorithms for mining frequent subgraphs (at the expense of accuracy) and the larger size of mined subgraphs. A more comprehensive review of frequent subgraph mining and motif discovery algorithms in both settings for bioinformatics applications is provided in \citep{Mrzic2018bioinformatics}.

\begin{table}[h!]
  \begin{center}
    \caption{Summary of motif discovery review papers}
    \label{tab:review_papers}
    \resizebox{\textwidth}{!}{\begin{tabular}{llll} 
    \hline
    \hline
      \textbf{Authors} & \textbf{Title} & \textbf{Year} & \textbf{Problem setting}\\
      \hline
      		Washio and Motoda \cite{Washio2003graph}  &   State of the art of graph-based data mining &  2003 &  Network-transaction/Single network\\
			W\"{o}rlein et al. \cite{Worlein2005Quantitative}  &   A quantitative comparison of the subgraph miners MoFa, gSpan, FFSM, and Gaston &  2005 &  Network-transaction\\
			Yan and Han \cite{yan2006discovery}  &   Discovery of Frequent Substructures &  2006 &  Network-transaction/Single network\\
			Han et al. \cite{Han2007status}  &   Frequent pattern mining: current status and future directions &  2007 &  Network-transaction\\
			Ciriello and Guerra \cite{Ciriello2008review}  &   A review on models and algorithms for motif discovery in protein-protein interaction networks &  2008 &  Single network\\
			Ribeiro et al. \cite{Ribeiro2009Strategies}  &   Strategies for network motifs discovery &  2009 &  Single network\\
			Wan and Mamitsuka \cite{Wan2009Motifs}  &   Discovering Network Motifs in Protein Interaction Networks &  2009 &  Single network\\
			Cheng et al. \cite{Cheng2010survey}  &   Mining Graph Patterns  &  2010 &  Network-transaction/Single network\\
			Parthasarathy et al. \cite{parthasarathy2010survey}  &   A survey of graph mining techniques for biological datasets  &  2010 &  Network-transaction/Single network\\
			Bruno et al. \cite{Bruno2010Trends}  &   New Trends in Graph Mining: Structural and Node-Colored Network Motifs  &  2010 &  Single network\\
			
			Krishna et al. \cite{Krishna2011comparative}  &  A comparative survey of algorithms for frequent subgraph discovery &  2011 &  Network-transaction/Single network\\
			Masoudi-Nejad et al. \cite{Masoudi2012review}  &   Building blocks of biological networks: a review on major network motif discovery algorithms &  2012 &  Single network\\
			Jiang et al. \cite{Jiang2013survey}  &   A survey of frequent subgraph mining algorithms &  2012 &  Network-transaction/Single network\\
			Rehman et al. \cite{Rehman2014Performance}  &   Performance Evaluation of Frequent Subgraph Discovery Techniques &  2014 &  Network-transaction\\
			Ramraj and Prabhakar \cite{Ramraj2015Survey}  &   Frequent Subgraph Mining Algorithms – A Survey &  2015 &  Network-transaction\\
     		G\"{u}venoglu and Bostanoglu \cite{Guvenoglu2018qualitative} & A qualitative survey on frequent subgraph mining &  2018 &  Network-transaction/Single network\\
     		Mrzic et al. \cite{Mrzic2018bioinformatics} & Grasping frequent subgraph mining for bioinformatics applications &  2018 &  Single network\\
     		
      \hline
      \hline
    \end{tabular}}
  \end{center}
\end{table}

One of the approaches followed by a few review papers for comparison of algorithms proposed for motif discovery is the re-implementation of multiple algorithms on the same platform. This approach makes it possible to compare the performance of algorithms based on the adopted strategies for candidate generation, frequency computation, and the size and accuracy of motifs detected without being biased with the framework adopted for implementation and the experience of developers. For example, this approach is adopted in \cite{Ribeiro2009Strategies} in which a few motif discovery algorithms proposed for the single network setting are re-implemented. The authors suggest that in some applications, it might be beneficial to use a combination of algorithms to have the advantages offered by each algorithm. However, in most of the review papers in Table \ref{tab:review_papers}, the algorithms are compared qualitatively and based on the theoretical investigation of approaches toward the motif discovery problem.

In nearly all the review papers in Table \ref{tab:review_papers}, the algorithms reviewed and evaluated operate on static networks (either a single large network or a set of relatively small networks). Besides, most of the algorithms considered are designed for problems in which the network data can be held in the main memory without the support of implementing parallel or distributed computing. G\"{u}venoglu and Bostanoglu \cite{Guvenoglu2018qualitative} have considered a few algorithms proposed for temporal networks or implemented for parallel computation; however, in both cases, the number of algorithms reviewed is minimal. In this paper, we provide a classification of algorithms proposed for motif discovery in static and temporal networks. Then, we review algorithms developed for static and temporal networks. Furthermore, a section is considered for algorithms implemented by having the big network data in mind, both from memory and computational requirement perspectives. In the following, the terminology adopted in this paper is defined, and classification for motif discovery algorithms proposed in the literature is provided.

\section{Terminology and Network Classification}\label{terminology}
The primary concepts used in this paper are described in Appendix \ref{app_prel_concepts}. These concepts are the fundamental concepts of network or graph theory \citep{Bollobas1998, BondyMurty1976, Diestel2005, GrossYellenZhang2013}. In the following subsection, the input data of motif discovery problems reviewed in this paper are described. Then, in subsection \ref{secondary_conc}, the concepts more specifically related to the literature of motif discovery problems are explained. 

\subsection{Networks Classification}
The motif discovery algorithms can be categorized based on their different strategies adopted for candidate generation and frequency computation. However, at the highest level, we differentiate the algorithms based on whether the input data is related to a static network or a temporal network. Here, these two types of input data are briefly discussed. Then, after the introduction of secondary concepts, classification of algorithms proposed in the literature of motif discovery is provided. 

\subsubsection{Static single network setting} The algorithms in this category operate on a single static network as the input data. The static network is defined as an ordered pair, $N=(V, E)$. The first term, $V$, is called the set of \textit{vertices} or \textit{nodes} of the network. The second term, $E$, is called the set of \textit{edges} of the network. The edges represent the interactions between pairs of vertices. In cases where there are labels or attributes associated with vertices and edges, the network is represented as $N=(V, E, L_V, L_E)$. Here, $L_V$ and $L_E$ are two functions that map the vertices and edges of the network to their corresponding labels, respectively \cite{KuramochiKarypis2006MiningGraphData}.

\begin{figure}[ht] 
\centering 
\subfigure{ 
    \includegraphics[scale=0.3]{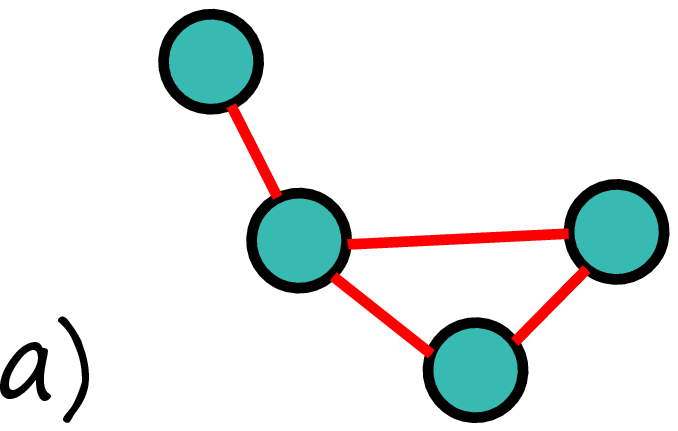} 
    \label{fig:motif_SN_a} 
} 
\subfigure{ 
   \includegraphics[scale=0.3]{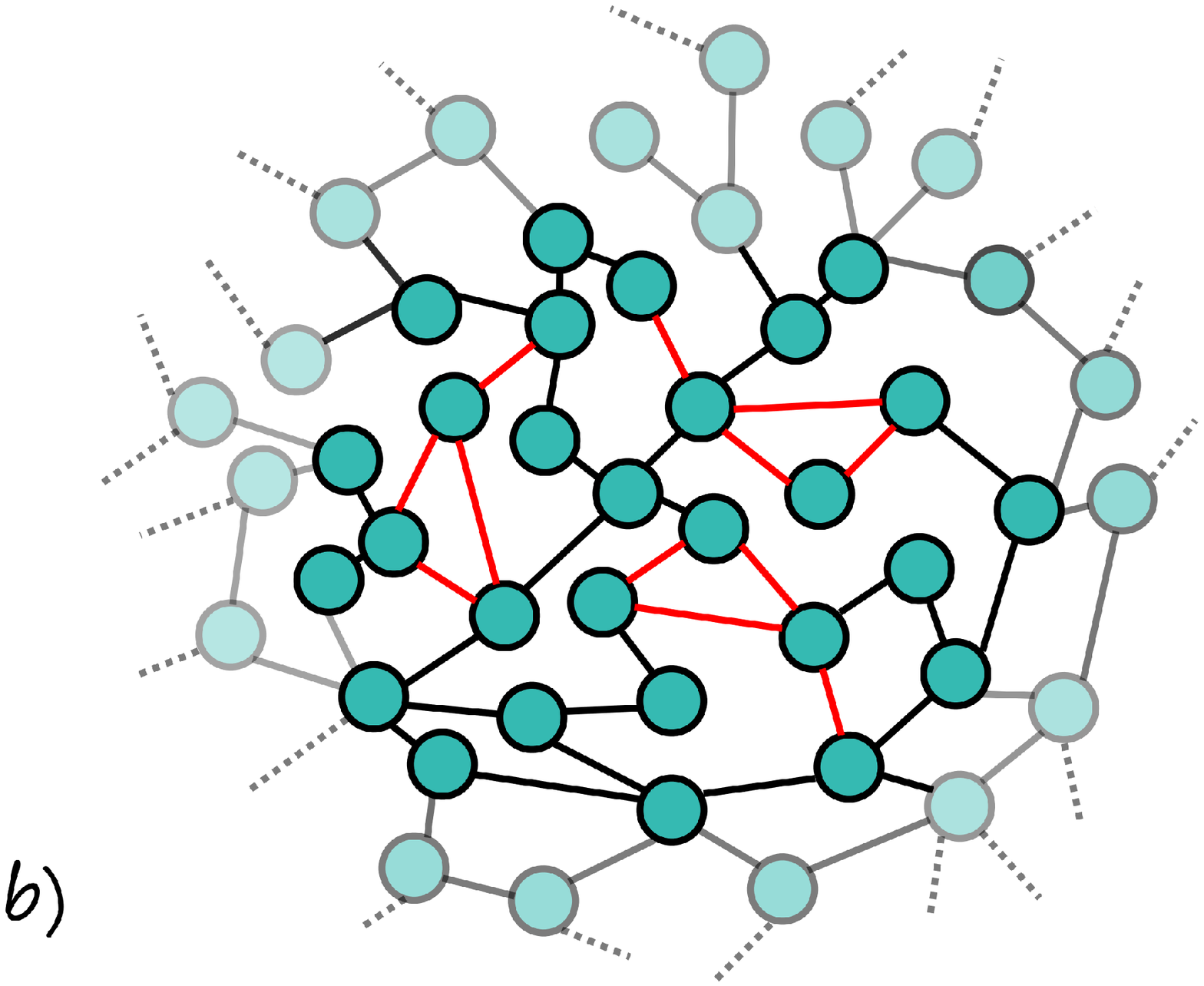} 
    \label{fig:motif_SN_b} 
} 
\caption{Motifs in a single giant static network. \subref{fig:motif_SN_a}: 4-vertex, 4-edge motif. \subref{fig:motif_SN_b}: Small portion of a signle giant network including three non-overlapping motifs of \subref{fig:motif_SN_a}} 
\label{fig:motif_SN} 
\end{figure}

As the behavior of a large number of entities being observed, recorded, and stored in many different domains and there is a tremendous increase of computational resources, there is an increasing interest in analyzing single large networks and their characteristics in the last few decades \citep{BarabasiAlbert1999, AlbertJeongBarabasi1999, WattsStrogatz1998}. However, frequent subgraph discovery in these networks is mostly developed after the subclass related to a set of independent static networks introduced earlier. Single giant networks generally represent a system of large interacting distinct entities, such as people in a communication network, or users of the internet. Although these entities are unique and have unique labels (names or IP addresses) because the patterns of interaction are important, they are generally assumed unlabeled (or for practical simplicity and usage, a single label considered for all the vertices and edges). The frequent subgraphs in these networks are traditionally called \textit{motifs} defined as inter-connected patterns or subgraphs occurring significantly more frequently in the network of interest than the randomized network counterparts \citep{MiloS2002motifs}. The trivial subgraphs are single nodes and edges, and most of the proposed algorithms start from these simple subgraphs and extend them to find the motifs. The motif discovery in single networks has been around for a while. However, due to the computational complexities inherent in the subgraph and graph isomorphism problems, the patterns which can be detected by the proposed algorithms are very limited in size (size is defined as the number of vertices or edges in the discovered motifs) and barely exceeds 10-15 vertices. The details of technical difficulties, algorithms, and tools proposed for motif discovery will be explained in the next sections. Figure \ref{fig:motif_SN} shows a small schematic of a single giant network and a motif. As can be seen, the motif shown in Figure \ref{fig:motif_SN_a} is composed of four vertices and four edges. There are three embeddings of this motif shown in Figure \ref{fig:motif_SN_b}. As you may notice, the free edge of the motif (the one not included in the triangle), can be selected in several ways for each embedding of the motif. However, all other embeddings, which can be chosen in Figure \ref{fig:motif_SN_b} has at least one edge in common with embeddings already identified and shown in this figure. The strategy adopted for counting these \textit{overlapping} embeddings is one of the differentiating factors among algorithms. It will be discussed in subsection \ref{secondary_conc} in more detail.

\subsubsection{Temporal single network setting}
\begin{figure}[ht] 
\centering 
\subfigure{ 
    \includegraphics[scale=0.3]{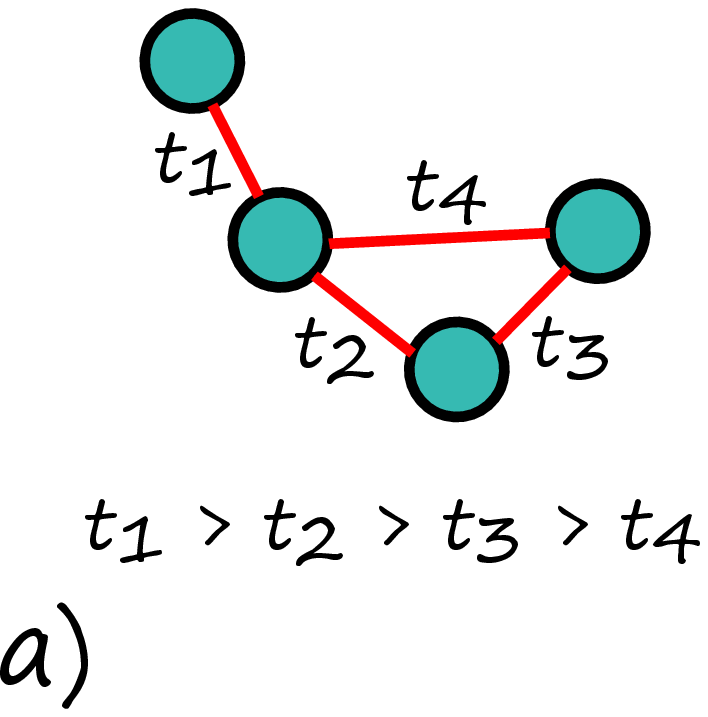} 
    \label{fig:motif_DN_a} 
} 
\subfigure{ 
   \includegraphics[scale=0.3]{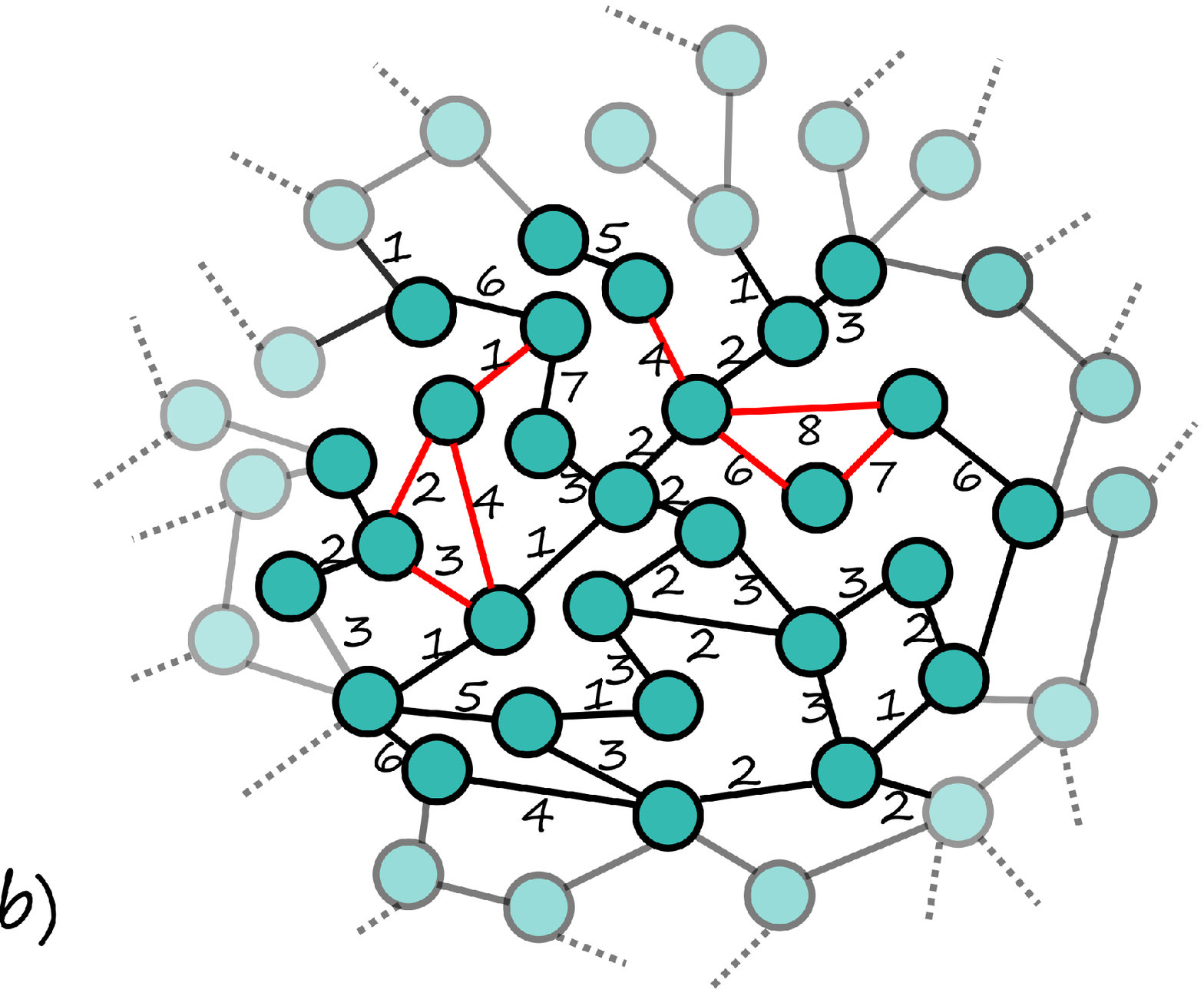} 
    \label{fig:motif_DN_b} 
} 
\caption{Motifs in a single giant temporal network. \subref{fig:motif_DN_a}: 4-vertex, 4-edge motif. \subref{fig:motif_DN_b}: Small portion of a signle giant network including two non-overlapping motifs of \subref{fig:motif_DN_a}} 
\label{fig:motif_DN} 
\end{figure}

The representation of a dynamic system as a static network is an assumption that might not be true in all circumstances, specially when the consideration of vertices and edges as a fixed part of the network influences the measures of interest \cite{Pan2011temporal, Nicosia2013Temporal, Holme2012Temporal, Holme2015colloquium}. Therefore, there has been a large effort proposing different network modeling methodologies to come up with better strategies to preserve the temporality and dynamic aspects of the system. The input data in this class is composed of one single temporal network. The motif discovery algorithms in this category assume different dynamic properties for the network. The network might be considered temporal based on the temporal changes in the vertex set, edge set, attribute values, or any combination of these components. Therefore, there is no single definition of temporal networks that can be used for all the algorithms. In general, the temporal network is defined as an ordered tuple, $N=(V, E, L_V^t, L_E^t)$. The $V$ and $E$ are similar to static networks. However, in cases where the set of vertices or edges of the network are changing over time, two time mappings, $t_V$ and $t_E$, are defined which map vertices and edges to their active timestamps \cite{lietal2018temporalHIN}. The $L_V^t$ and $L_E^t$ are temporal labeling functions that map the vertices or edges to their attributes over time. This representation can be simplified based on the properties considered dynamic by each algorithm in this category. Besides, based on the definition considered for the temporal network, the graph and subgraph isomorphism tests verify the temporality and sequence of appearance, disappearance, and changes of edges, vertices, and their attributes as well (Figure \ref{fig:motif_DN}). More details about the dynamic properties of temporal networks will be provided in section \ref{tsns}.

\subsection{Secondary Concepts}\label{secondary_conc}
After introducing the general concepts used throughout this survey (Appendix \ref{app_prel_concepts}) and the two types of network data considered, the concepts more specifically related to the literature of motif discovery are introduced in the following.

\textbf{Measures of frequency and significance:} In the problem of frequent subgraph mining in a set of networks, the frequency of a subgraph is considered as the number of members of the set containing the subgraph. On the contrary, the frequency of a subgraph in a single network is defined as the number of embeddings or instances of the subgraph in the network. Besides, the common practice is to compute and report the significance of subgraphs in addition to their frequencies. 

There are multiple strategies or ``concepts'' for computation of frequency and, consequently, the significance of subgraphs. These strategies are based on the level of overlapping permitted among different embeddings of the subgraph being counted \cite{schreiber2004towards, Ciriello2008review, Schreiber2005Frequency}. The algorithms adopting the first concept of frequency, called $F_1$, enumerate or count all the embeddings of a subgraph toward the subgraph frequency calculation. These embeddings should be at least different by a vertex or an edge. However, they can have multiple vertices or edges in common. The algorithms adopting the second concept of frequency, called $F_2$, consider valid embeddings as the instances of the subgraph that do not share any edges (edge-disjoint). In this case, the embeddings might still have some vertices in common. In the third concept of frequency, called $F_3$, not only the embeddings do not share any edges, but also they should not have any vertices in common (node-disjoint). Therefore, in the motif discovery problem, the frequency of a subgraph is defined based on the number of embeddings of that subgraph using one of these frequency concepts. Then, the subgraph frequencies are used to compute subgraph significance measures. These measures are computed for each subgraph separately and are generally a function of the frequency of subgraph in the original network and the corresponding frequencies of the same subgraph in some randomized versions of the original network. For the detection of frequent/significant subgraphs or motifs, the user should provide a threshold to which the frequency/significance measures of each subgraph is compared.  The subgraphs with measures more than the user-defined threshold are considered motifs or frequent subgraphs. For example, $z$-score and \textit{abundance} $(\Delta)$ are used in \citep{Ciriello2008review} as measures of significance of subgraphs:


\begin{align}\label{eq:supp}
z(s_k) &= \frac{f_{orig}(s_k)-\bar{f}_{rand}(s_k)}{\sigma_{f_{rand}(s_k)}}\\
\Delta(s_k) &= \frac{f_{orig}(s_k)-\bar{f}_{rand}(s_k)}{f_{orig}(s_k)+\bar{f}_{rand}(s_k)+\epsilon}
\end{align}

In which the $f_{orig}(s_k)$ represents the frequency of subgraph $s_k$ in the original network and $\bar{f}_{rand}(s_k)$ and $\sigma_{f_{rand}(s_k)}$ represent the mean and standard deviation of the frequency of the same subgraph in the randomized versions of the original network. The $\epsilon$ is considered to avoid over-growth of $\Delta(s_k)$ in cases where the frequency of subgraph is very low in the original and randomized versions of the network. In \citep{Ribeiro2009Strategies}, three measures are reviewed and discussed for subgraphs to verify if they can be considered as motifs: over-representation, minimum frequency, and minimum deviation. The first measure is the evaluation by $z$-score, and the minimum frequency is used to verify that the subgraph is frequent enough (has a frequency more than the user-defined threshold). The minimum deviation is defined as follows for a subgraph $s_k$. This measure should be more than a user-defined threshold as well.

\begin{align}\label{eq:supp}
D(s_k) &= \frac{f_{orig}(s_k)-\bar{f}_{rand}(s_k)}{\bar{f}_{rand}(s_k)}
\end{align}

The randomized versions of original networks (\textit{null models}) can be generated in different ways. The most common approach in motif discovery literature is to create random networks with the same degree distribution as the original network. However, other strategies are available, which can preserve other characteristics of the original network. The generation of a random network from an original network has a rich literature and interested readers may refer to \citep{Milo2003random} for static networks and to \citep{Holme2012Temporal, Holme2015colloquium} for temporal networks.

Although most of the algorithms proposed in the literature use the above frequency and significance measures, the measures available in the literature are not limited only to these cases. For example, in \citep{Onnela2005Intensity}, the concepts of intensity and coherence are introduced applicable to weighted networks. The intensity and coherence are defined as the geometric mean, and the ratio of geometric to the arithmetic mean of weights of the subgraphs, respectively. Then, the $z$-score can be modified to \textit{motif intensity} and \textit{motif coherence} scores by replacing the subgraph frequency measures with their intensities. The intensity and coherence measures are used for mining motifs, considered as subgraphs with high values of intensity (in relation to a reference network). The concept of frequency as the number of subgraphs then would be a special case of intensity. In \citep{Chen2006NeMoFinder}, the \textit{unique} concept is used as a measure of significance of subgraphs. This measure is defined as the ratio of the frequency of each subgraph in the original network to the frequency of the same subgraph in the randomized versions of the original network (also, please refer to \cite{parthasarathy2010survey}). 

In many of the algorithms proposed in the literature, the edge-disjoint or node-disjoint subgraph embeddings are used for computation of frequency and significance. In \citep{Bringmann2008Frequent}, a new measure, called ``minimum image-based support'', is proposed for frequency computation allowing overlapping subgraph embeddings. This measure is computed as follows for a candidate subgraph $s$. For each vertex $v_i$ in subgraph $s$, the number of unique vertices in the original network to which the $v_i$ can be mapped is counted as the frequency of $v_i$, $f(v_i)$. The frequency of the subgraph, $f(s)$, cannot be more than the minimum of frequencies obtained for $f(v_i)$ for $v_i \in V(s)$. Therefore, considering $f(s)=min|f(v_i)|$, the frequency $f(s)$ of subgraph $s$ would be an upper bound for actual frequency of subgraph $s$ in the original network (and an upper bound for two other common measures computed based on \textit{simple overlap} and \textit{harmful overlap} definitions \citep{fiedler2007support, Kuramochi2005Sparse, Cheng2010survey} and therefore guarantee a superset of subgraphs). Therefore, it can improve the computational complexity of frequency computation in the single network setting. In mfinder \citep{Kashtanetal2004}, the concentration of a subgraph is considered as the frequency of subgraph to the frequency of all the subgraphs in the network with the same number of vertices.

\begin{align}
	C(s_k) &= \frac{f(s_k)}{\sum{f(s_k)}}
\end{align}

In \citep{Picardetal2008exceptionality}, it is discussed that the $z$-score is based on the assumption that the frequencies of motifs are following normal distribution and generating $p$-values need a large number of randomized versions of the input network. These assumptions might not be correct or feasible in many situations. Instead, they show that geometric Poisson distribution provides more accurate approximations of motif counts and $p$-values in giant networks in comparison with the assumption of normal distribution.

\begin{figure}[ht] 
\centering 
\subfigure{ 
    \includegraphics[scale=0.3]{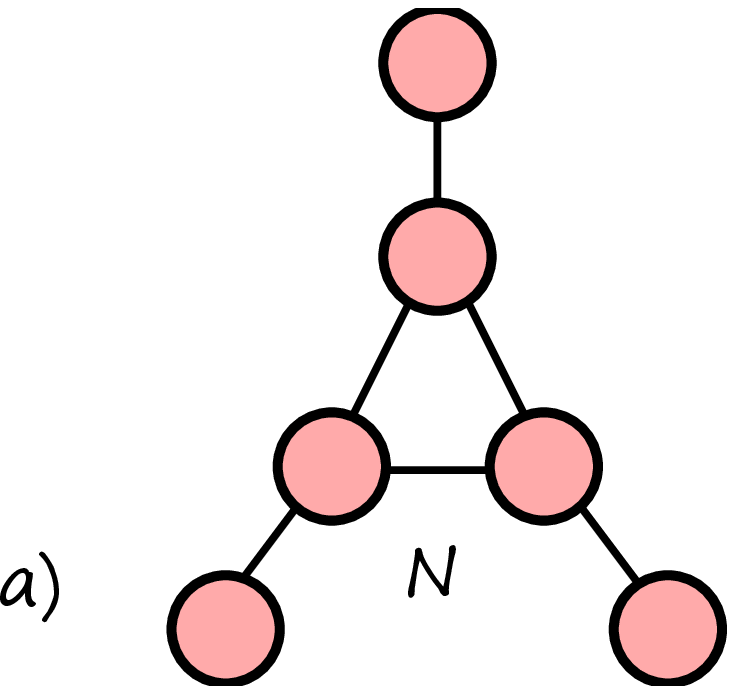} 
    \label{fig:downward_g} 
} 
\subfigure{ 
   \includegraphics[scale=0.3]{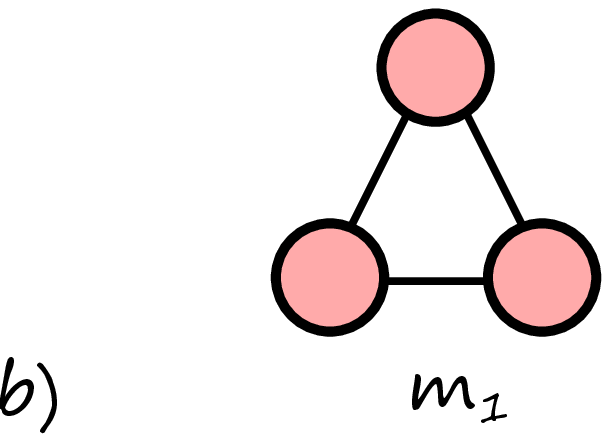} 
    \label{fig:downward_s1} 
} 
\subfigure{ 
   \includegraphics[scale=0.3]{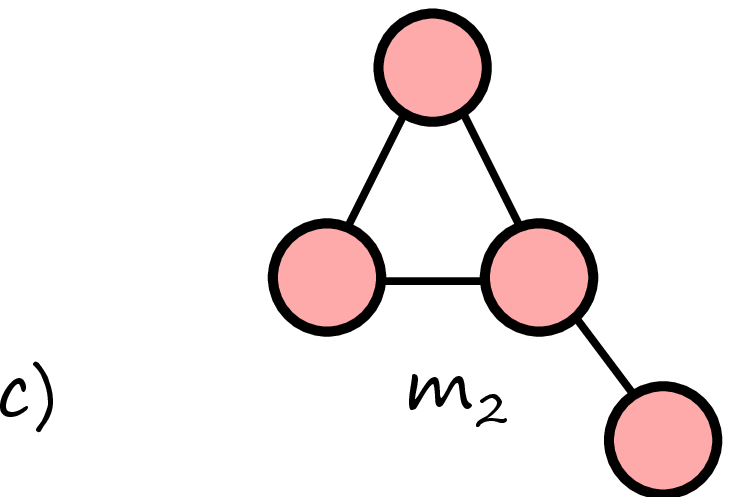} 
    \label{fig:downward_s2} 
} 
\caption{Network and its two subgraphs $(m_1 \subseteq m_2 \subseteq N)$. \subref{fig:downward_g}: network composed of 6 vertices and 6 edges. \subref{fig:downward_s1}: Subgraph $m_1$ with $f(m_1)=1$ in \subref{fig:downward_g}. \subref{fig:downward_s2}: Subgraph $m_1$ with $f(m_2)=3$ in \subref{fig:downward_g}.} 
\label{fig:downward} 
\end{figure}

\textbf{Pruning strategies:} The general approach to the discovery of motifs in single networks is generating candidates (starting from single vertices and edges), selecting the frequent and significant candidates by computing their frequencies or some measures of significance, and using these frequent candidates as the core for the generation of new candidates. This set of tasks is iteratively performed until some stopping conditions are met. The graph isomorphism problem arises when candidates representing identical networks should be identified to avoid redundancy. And the subgraph isomorphism problem arises when the frequency of distinct candidates should be computed in the original network. The proposed algorithms for motif discovery adopt different strategies or pruning mechanism to reduce the computational complexity or memory requirements of the discovery process. One of the most important strategies adopted is downward closure or anti-monotonicity property. In the context of motif discovery problem, this property implies that if a motif, $m$, has the frequency of $f(m)$, then none of the candidates growing from this motif should have a frequency of more than $f(m)$. Therefore, for two subgraphs $m_1$ and $m_2$, if $m_2\subseteq m_1$, then $f(m_2)\geqslant f(m_1)$. This property is extensively used in association rule mining problems \citep{agrawal1994fast} in which the ordering of different candidates can be performed with lower complexities than candidate subgraphs. The usefulness of this property in the motif discovery problem depends on the adopted concept of frequency discussed earlier. Figure \ref{fig:downward} shows an example of how this property might not be applicable in motif discovery problems. This strategy is not useful when the adopted strategy for frequency computation is $F_1$, in which overlapping embeddings without any limitations is permitted. It implies that for problems that discovery of overlapping instances of motifs are of interest, this property does not hold. For example, in some biological networks in which vertices and edges might be involved in multiple functional processes \citep{Chen2006NeMoFinder}. Also refer to \citep{VanetikShimonyGudes2006, VanetikGudesShimony2002} for relevant discussions. Other pruning strategies are size-based \citep{BorgeltBerthold2002}, in which the candidate generation is stopped once the size of candidates exceeds a pre-defined value, and structural pruning  \citep{BorgeltBerthold2002, YanHan2002gSpan} in which candidates for which their isomorphic representations are already assessed are pruned before further evaluation. These strategies are general and known strategies to reduce the effects of the complexities of the motif discovery problem. There are other strategies designed for different steps of the motif discovery problem in different algorithms. We will discuss these strategies when the corresponding algorithms are reviewed.

The number of motifs discovered in single networks is generally very large. One approach is to limit the search to special subclasses of motifs (such as induced, closed, and maximal), which either based on their structural or frequency properties meet some conditions.

\textbf{Closed subgraphs:} A motif $m_1$ is considered as a closed motif if there is no other motif such as $m_2$ which $m_1 \subseteq m_2$ and both with the same frequency. The closed motifs can be used to generate all the non-closed motifs  \citep{yan2006discovery}.

\textbf{Maximal subgraphs:} A motif $m_1$ is considered as a maximal motif if there is no other subgraph such as $m_2$ which $m_1 \subseteq m_2$ and $m_2$ is frequent \citep{yan2006discovery}.

Besides, the algorithms can be categorized based on their approach to the structural and frequency exactness. Here we refer to the first category as exact frequency algorithms and the second category as exact isomorphic algorithms. These two categories are defined as follows:

\textbf{Exact isomorphic algorithms:} To compute the frequency of a candidate, the algorithms in this category just consider the exact isomorphic embeddings of the candidate in the network. On the contrary, we consider a second class in which some structural variations are allowed among the candidate and its embeddings in the network being mined. Therefore, in latter algorithms, some similarity functions are used to compute the similarity of the candidate with the embeddings it has in the network.

\textbf{Exact frequency algorithms:} Another approach toward the classification of algorithms is based on the approach adopted toward the computation of frequency. Some of the algorithms produce or count the complete list of embeddings of the candidates in the network to find the motifs. We call these subclass as exact frequency algorithms. In another group of algorithms, instead of mining all the embeddings of a candidate, the frequencies of the candidates are approximated by mining some sample areas of the network. This group is called approximate frequency algorithms as they do not produce the exact frequency of motifs.

Based on the concepts introduced so far, the algorithms proposed for motif discovery can be categorized based on their input data and consideration of temporality. Furthermore, in the review of these algorithms, we noticed that the main differentiating factors in each of these two categories (static and temporal networks) are different. In static networks, the approaches adopted for frequency computation can differentiate the algorithms. Based on this difference, the algorithms mine the single network either to find the complete list of embeddings of subgraphs or just to approximate the frequencies. Besides, some of the algorithms proposed for static networks mine all the subgraphs with a maximum of 4-5 vertices (graphlets). The algorithms proposed for temporal networks, on the other hand, can be categorized based on the features assumed to be dynamic by the algorithm, either the vertices or edges or their associated labels and attributes. Also, some algorithms are proposed for mining temporal networks in the form of data streams.  Based on these findings, the following classification is considered (Table \ref{tab:classification}). In the following sections, the algorithms in each category are reviewed.

\begin{table}[h]
\centering
\caption{Classification of motif discovery algorithms}
        \begin{tabular}{lll}
            \hline
            \hline  
            \textbf{Network Data} & \textbf{Problem Setting} & \textbf{Mining Category}\\\hline
           
\multirow{7}{*}{Single network} & \multirow{4}{*}{Static single network setting} & Exact frequency algorithms\\
                                 &                      & Approximate frequency algorithms\\
                                 &                      & Other approaches\\  
                                 &                      & Graphlet mining \\\cline{2-3}              
                                 & \multirow{3}{*}{Temporal single network setting } & Dynamic attributes\\
                                 &                      & Dynamic topology \\
                                 &                      & Network data streams\\
\hline  
\hline      
        \end{tabular}
    \label{tab:classification}
\end{table}

\section{Motif Discovery}
In the motif discovery problem, when vertices are all distinctly labeled, we have the minimum amount of complexity for graph and subgraph isomorphism.  On the other end of the spectrum, having all the vertices/edges unlabeled or identically labeled, just the structure of the network is of importance. In some cases, it is assumed that vertices can be grouped based on their functionalities, roles, and properties in the network. In this case, these roles might be considered as labels of vertices, which consequently make the problem less computationally expensive. For the classification of motif discovery algorithms based on this approach, please refer to \citep{Bruno2010Trends}. In temporal motif discovery, in addition to structure, consideration of the temporal changes of network components is critical as well \citep{Gurukaretal2015COMMIT}. 

The most common approach to identify frequent or significant motifs in a single network setting is based on the frequency of the motifs in the input network in comparison with the randomized versions of the original network. In the following, the algorithms proposed for the motif discovery are classified based on the temporality of the network data. Then, each class will be broken down into different subclasses based on the approaches adopted for motif discovery or the dynamics of the network.

\subsection{Static single network setting}
One common classification scheme for the algorithms proposed for mining frequent subgraphs in a set of networks is based on the search and candidate generation and enumeration strategies. The same scheme can be used for the algorithms proposed for motif discovery in a single giant static network. For example, the SIGRAM \citep{KuramochiKarypis2006MiningGraphData} (SIngle GRAph Miner) is proposed for mining frequent motifs for a single giant network. This algorithm is composed of two sister algorithms: HSIGRAN and VSIGRAM. They are exact isomorphic algorithms and can find all the motifs with (exact or estimated) frequencies of edge-disjoint instances more than a user-defined threshold. The HSIGRAN utilizes a breadth-first search strategy similar to the FSG algorithm \citep{KuramochiKarypis2001FSG}, and VSIGRAN uses a depth-first search strategy. 

In network-transaction settings, the frequency is generally computed as the number of transactions containing candidate subgraphs, no matter the frequency of appearance in each transaction. In contrast, in single network settings, the main challenge is counting the embeddings of subgraphs in the input network. Therefore, it is very reasonable to categorize proposed algorithms based on the adopted strategies to approach this problem. In the following, the algorithms proposed for a static single network setting are classified into two groups based on their approach toward counting embeddings of motifs; exact frequency algorithms and approximate frequency algorithms. For a summary of the algorithms reviewed in this section, please refer to Table \ref{tab:statmotif}. Being an exact frequency algorithm does not imply that the algorithm can detect all the frequent subgraphs. It means that for the subgraphs mined, the exact frequencies are computed.

\subsubsection{Exact frequency algorithms}
In each iteration of algorithms in this category, one or multiple candidates are generated. For each of these candidates, these algorithms can find all the embeddings in the network. Therefore, for the candidates generated, the exact frequencies are computed. However, being able to create all the candidates depends on the strategy they adopt to traverse the search space. Therefore, not all of them necessary can create all the candidates ending in the frequent subgraphs. The SUBDUE \citep{CookHolder1994SUBDUE, HolderCookDjoko1994SUBDUE} is one of the first algorithms developed for motif discovery in single large networks. This algorithm is developed using the minimum description length (MDL) principle \citep{Rissanen1989} to iteratively find motifs. After each iteration, all the embeddings of each discovered motif are replaced by one vertex with an updated label corresponding to the motif which the new vertex represents. The discovered subgraphs are evaluated based on their ability to compress the input network (considering the bits required to encode the subgraph and the input network after replacing all the embeddings of the subgraph in the input network by the new vertex). The iterative nature of SUBDUE, along with the replacement of discovered subgraphs, makes it possible to finally represent the data in a hierarchical form, which increases the interpretability of the network representation. The SUBDUE can be set up to discover embeddings with a predefined level of structural variations. The allowed variations are controlled by defining costs for structure modification operations such as insertion, deletion, and substitution of the vertices and edges. In addition to the MDL principle, the discovery process can be guided by domain-dependent or independent background knowledge, which positively or negatively bias the discovery process toward specific types of motifs. The SUBDUE does not use a graph isomorphism technique, and it is shown that in very large networks (or transactions), or networks with high degrees of randomness, SUBDUE is not very successful \citep{CookHolderKetkar2006MiningGraphData}. Besides, it might miss some of the motifs due to the greedy search strategy adopted by this algorithm \citep{InokuchiWashioMotoda2000, InokuchiWashioMotoda2003Complete}. 

The B-GBI (Beam-wise Graph-Based Induction) algorithm \citep{matsuda2002BGBI} utilizes a beam-wise search strategy for mining frequent motifs. This algorithm is an improved version of the GBI \citep{yoshida1994graph, matsuda2000GBI1}, and CLIP \citep{YoshidaMotoda1995CLIP}, two other previously developed subgraph mining approaches. The B-GBI can be implemented with different evaluation functions besides subgraph frequency. Therefore, it can find discriminatory or contrast subgraphs. The idea is to find the most ``typical" patterns iteratively. The typicality can be defined based on some measures of interest, such as the frequency of subgraphs. In each iteration, multiple patterns are selected, ranked based on their measure of typicality, and replaced with a new vertex. The B-GBI is greedy, and finding the most typical patterns or frequent subgraphs is not guaranteed. To prevent miscounting identical subgraphs created in different iterations or from different typical patterns, B-GBI uses canonical labeling. The canonical labeling or form is a code that represents the isomorphism class of a network. In other words, all the isomorphic networks have identical canonical labeling. In  B-GBI, the labels of vertices and their degrees are used to produce canonical labeling of networks and to narrow down the combinatorial search space. 

The B-GBI is modified to CL-GBI (ChunkingLess Graph-Based Induction) \citep{Oharaetal2006MiningGraphData} in which the frequent subgraphs are not collapsed. Instead, they are used to create  \textit{pseudo-nodes}. This modification helps to detect overlapping embeddings of induced and general frequent subgraphs, which can be considered a significant feature. The CL-GBI can be applied to both single network and network-transaction settings.

The HSIGRAM and VSIGRAM algorithms for mining a complete set of frequent subgraphs in a single, not necessarily connected, network are proposed in \citep{Kuramochi2004Sparse, Kuramochi2005Sparse}. It is assumed that the network is undirected, sparse, and labeled. They use \textit{overlap networks} for frequency computation. In the overlap network, the vertices represent non-identical embeddings of a subgraph. Each pair of vertices in the overlap network are connected if the corresponding embeddings of the two vertices have at least one edge in common. The maximum independent set (MIS) of this overlap network is used to find the set of edge-disjoint embeddings for frequency calculation (the approximate versions of MIS also are proposed to improve the computational time). Both algorithms use canonical labeling \citep{KuramochiKarypis2001FSG} for checking the graph isomorphism. The main difference between the two algorithms is the search strategy. The HSIGRAM employs a breadth-first search strategy to generate size-$k+1$ candidates from size-$k$ subgraphs. To join the pairs of size-$k$ subgraphs, they have to have the same $k-1$ subgraph. For each size-$k$ subgraph, there might be up to $k$ size-$k-1$ subgraph. Therefore, to modify the search space, a set composed of two size-$k-1$ subgraphs of each size-$k$ subgraph is generated, including the two size-$k-1$ subgraphs with the smallest canonical labeling (or one size-$k-1$ subgraph if the two smallest canonical labelings represent the same network). If the sets for the two size-$k$ subgraphs intersect, they are joined to form a size-$k+1$ candidate subgraph.

On the other hand, VSIGRAM, which is faster, employs a depth-first search strategy, in which, to prevent generating any duplicate candidates for each subgraph of size $k+1$ a unique parent of size $k$ is specified. The child subgraph of size $k+1$ can be generated just from this parent. The idea behind this algorithm is then used in FPF, \textit{frequent pattern miner} algorithm \citep{schreiber2004towards, Schreiber2005Frequency} to mine all the frequent patterns with specific size under different concepts of frequency introduced earlier (e.g., $F_2$ and $F_3$). This algorithm is implemented in MAVisto \citep{Schreiber2005MAVisto} proposed for mining a complete set of motifs in a single network under three different frequency concepts for a given number of nodes or edges. MAVisto can compute the $p$-value and $z$-score of motifs detected in comparison with the frequency of the same motifs in randomized versions of the network. The gApprox \citep{Chenetal2007gApprox} is an exact frequency algorithm in this category, allowing for isomorphism with structural variations. The gApprox uses two vertex penalty and edge penalty to estimate the (dis-)similarity of two subgraphs. The user provides the maximum approximation similarity. The gApprox also uses the edge-disjoint mechanism to make the downward-closure property possible.

GREW \citep{KuramochiKarypis2004GREW} is proposed to find exact isomorphic and vertex-disjoint motifs with a frequency of more than a user-defined threshold. This algorithm mines iteratively frequent motifs and generate new candidates by combining previously identified frequent motifs connected by at least one edge. It also keeps track of frequent motifs and collapses the frequent ones into a new vertex with a new label representing the collapsed motif. By following this candidate generation approach, the size of subgraphs being mined in each iteration doubles, which makes the algorithm faster than algorithms generating candidate by adding one edge or vertex in each iteration at the expense of potentially losing some frequent subgraphs. In comparison with the SUBDUE, the developers of GREW show that it can find larger motifs (up to size 16, the largest they found with SUBDUE is 10) in shorter times.

In \citep{Parida2007topologicalmotifs}, the maximal motifs are defined as motifs, which are both edge-maximal and vertex-maximal, meaning that no edges and no vertices can be added to the motifs without changing their occurrence list. For storing the location of vertices and edges, a compact list is created. This list is produced once and prevent any duplicate generation of locations. In \citep{Kashani2009Kavosh}, the Kavosh algorithm is proposed for finding the frequent motif of size $k$. This algorithm can mine overlapping embeddings. The developers show that it can mine motifs of size more than eight. For the graph isomorphism problem, Kavosh relies on the nauty algorithm \citep{mckay1981, McKayPiperno2014}. The nauty is known as one of the most efficient algorithms for graph isomorphism \citep{Fortin96thegraph}. For enumeration, a new efficient algorithm is proposed in Kavosh in which subgraphs of size $k$ are enumerated. First, for each vertex $v$, all the subgraphs that include $v$ are mined. Then, the $v$ is removed from the network. This vertex-based discovery is iteratively repeated for all the vertices remained in the network. The significance of the motifs is computed in comparison with randomized versions of the original network with the same degree distribution. Another algorithm in this category, NeMoFinder \citep{Chen2006NeMoFinder}, can mine frequent motifs up to size 12. Starting from size-2 trees, it first mines all the size-$k$ trees. Using frequent size-$k$ trees, NeMoFinder partition the original network into multiple networks. The general subgraphs are created from identified trees as new candidates. The frequencies of candidates are checked in the partitions using a modified version of the SPIN algorithm \citep{Huanetal2004SPIN, Huanetal2004SPINtechreport} based on different frequency concepts. The frequency and uniqueness of discovered subgraphs are computed and compared with randomized versions of the original network.

\begin{figure}[ht] 
\centering 
\includegraphics[scale=0.5]{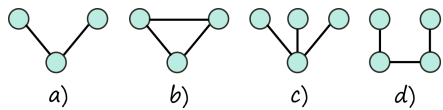}
\caption{The four basic building patterns used for motif discovery in \citep{Elhesha2016disjoint}}.
\label{fig:building_patterns} 
\end{figure}

It is discussed in \citep{zhu2011mining} that the number of frequent subgraphs increases by the size of the subgraph. It implies that if larger frequent subgraphs are composed of smaller frequent ones, and these smaller ones are already detected, then we might be able to generate the larger frequent subgraphs with a complexity less than a standard pattern growth approach. In \citep{Elhesha2016disjoint}, it is proved the all the connected motifs with more than three edges can be constructed from four basic building patterns (Fig. \ref{fig:building_patterns}). Based on this observation, the algorithm proposed in \citep{Elhesha2016disjoint} first identifies the maximum independent set of embeddings of these four patterns. All other candidates are generated with a join operation of a frequent subgraph and any of these four basic patterns if they share at least one edge. Then, the edge-disjoint embeddings for candidates are counted. The algorithm is designed to mine all the frequent subgraphs with a specific size given a support threshold.

\subsubsection{Approximate frequency algorithms}
The algorithms proposed for motif discovery in a single giant network is generally very limited on the size or the number of motifs they can discover. It is due to the exponential increase in the computational complexity of the problem with the size of the input data or the size of the motifs. Therefore, to identify frequent motifs, many approaches have been proposed which approximate the frequency of subgraphs instead of exhaustive counting or enumeration of subgraphs. The SEuS \citep{ghazizadeh2002seus} is one of the algorithms in this category developed for detecting frequent motifs in a single network (also, it can be applied to network-transaction setting). Although the SEuS allows the instances of motifs to overlap, it meets the downward closure property by stopping the expansion of subgraphs as soon as the subgraph is not frequent any more. The SEuS is an interactive mining algorithm in the sense that it allows the user to stop the mining process for interim exploratory data analysis on the detected motifs. It is an approximate frequency algorithm due to its utilization of a data summary for counting the frequency of motifs as an upper bound for the motifs' actual frequencies. The candidate generation method is level-wise and performed by adding single edges, which may or may not add a new vertex to the motifs. Later, the exact frequency of motifs that have an estimated frequency more than the threshold can be evaluated and confirmed. For the graph isomorphism problem, the SEuS uses the nauty package \citep{mckay1981, McKayPiperno2014}. The developers of SEuS compare their results with SUBDUE and show that in larger databases, SEuS outperforms SUBDUE. The efficiency of SEuS and SUBDUE in small databases is comparable.

Another algorithm in this category is mfinder developed based on a ``sampling method for subgraph counting'' \citep{Kashtanetal2004}. This algorithm samples the subgraphs employing a systematic method to prevent potential biases and estimate subgraphs' frequency in the original network. Increasing the number of sampling iterations improves the accuracy of estimated frequencies. However, they show that even with small sampling iterations, good precision can be obtained. Another similar tool is FANMOD \citep{WernickeRasche2006FANMOD, FANMOD_Wernicke2006} developed based on the RAND-ESU, as an ``unbiased subgraph sampling algorithm'' \citep{wernicke2005faster}, which is able to find motifs with the number of vertices equal or less than eight. This algorithm can enumerate or sample from a giant network, adopts nauty package \citep{mckay1981, McKayPiperno2014} for checking graph isomorphism and preventing over-counting.

Grochow and Kellis \cite{GrochowKellis2007symmetrybreaking} introduces a \textit{motif-centric} approach based on a \textit{symmetry-breaking technique} which produces larger motifs (up to 15 vertices). It searches exhaustively (or by sampling) for query subgraphs in a giant network. Therefore, a subgraph set is produced, and then the algorithm is applied to the set. The algorithm introduces a technique to avoid over-counting subgraphs due to their structural symmetries. Using network invariants such as vertices' degrees and vertices' neighbors' degrees, it mitigates the complexities of the isomorphism problem. The candidate subgraphs are evaluated by comparing their frequencies in the input network and the randomized version using \textit{z}-score. The proposed algorithm also can be used for anti-motif discovery (subgraphs significantly less frequent than their counterparts in the randomized networks). However, they applied their algorithm on relative small networks (1379/2493 and 685/1052 vertices/edges). Another algorithm for mining motifs and anti-motifs is proposed in \citep{baskerville2006subgraph}. This algorithm introduces a novel heuristic approach for graph isomorphism, which makes it possible to mine motifs and anti-motifs of size 8. This heuristic is based on a set of invariants (invariant vertex labels and network labels). These invariants can label and differentiate all the subgraphs uniquely up to size eight. MODA \citep{OmidiFalkMasoudi2009MODA} adopting a pattern growth approach has been able to find larger motifs (more than eight vertices). It also uses the symmetry-breaking technique and starts from trees of size $k$, finds the frequent subgraphs, and iteratively expands and evaluates them. Also, for improving efficiency, MODA can use a sampling approach, which in this case, the frequency would be approximated. The developers express that this approach does not work acceptably for motifs with more than ten vertices. 

Many of the algorithms in this category proposed in the literature adopt one of the Markov chain Monte Carlo or color-coding strategies for estimating frequencies. The general approach for the first group is to create a network composed of (distinct) induced candidate subgraphs as vertices. In the generated network, two vertices are connected if the two subgraphs that each pair of vertices represent are different by one vertex. Then a random walk is started. It is assumed that random walk stops uniformly on the vertices after some walks, (\textit{mixing time}). And, repetition of the random walk provides the number of times each subgraph has been visited. The number of visits of each vertex is an estimation of the frequency of the induced subgraph that vertex represents. The second approach is color-coding. These two methods are compared in \citep{Bressanetal2017CountingGraphlets}. For color coding, different approaches can be used. In \citep{Bressanetal2017CountingGraphlets}, a modified version of the color-coding method by \citep{Alonetal1995Colorcoding} is adopted in which it starts with sampling and counting with non-induced tree graphlets (treelets) and is generalized to induced graphlets (For the discussions on graphlets and treelets, please refer to subsection \ref{Graphlet}). They show that, in general cases, the frequencies estimated by the Monte Carlo are not always reliable. On the other hand, the color-coding approach provides a more accurate estimation, even for larger networks and graphlet sizes. However, it is not cost-free. Based on the experiments reported in \citep{Bressanetal2017CountingGraphlets}, for induced subgraphs composed of three to seven vertices, the running time is comparable. However, the color-coding approach needs more space.

In \citep{Liuetal2015stochasticmotifs}, an approach is proposed for the detection of stochastic network motifs. The stochastic motifs are defined as frequent subgraphs with stochastic edges, i.e., edges of each motif have a probability of presence, contrary to other approaches that assume that edges are either present or not. Their approach can be implemented using regular subgraph sampling and network isomorphism tests. In \citep{Liuetal2015stochasticmotifs}, the approaches proposed by FANMOD \citep{WernickeRasche2006FANMOD, FANMOD_Wernicke2006} are used. Then, a finite mixture model is adopted for the identification of motifs, assuming that edges appearing in a motif are independent of each other. Using the synthetic and real-world data, they show that following a stochastic approach is more robust in the detection of network motifs in comparison with baseline deterministic approaches. Two random walk-based approaches, mix subgraph sampling (MSS) and pair-wise subgraph random walk (PSRW), are proposed in \citep{Wang2014Statistics}. The sampling is conducted based on the subgraph random walk approach as a modified version of the regular random walk over $k$-vertices connected induced subgraph (CIS) relationship network, $G^{(k)}$. The relationship network is composed of the set of all the $k$-vertices CISs as vertices. Then two vertices are connected if the two corresponding $k$-vertices CISs have $k-1$ vertices in common. The main difference between the PSRW and MSS is in the sampling approach adopted. In PSRW, the $k$-vertices CISs are sampled by applying subgraph random walk over $G^{(k-1)}$. In MSS, the subgraph random walk is applied to $G^{(k)}$ to estimate the concentration of subgraph classes of size $k-1$, $k$, and $k+1$ at the same time. They show that adopting these random walk-based sampling approaches, it would be possible to produce a more accurate and unbiased estimation of motif frequencies with a significantly lower number of samples. In \citep{Han2016Waddling}, the Waddling Random Walk (WRW) is introduced as a sampling-based approach based on random-walk. Another contribution of this algorithm is that it allows directing the mining process toward specific motifs. They show that their approach outperforms other state-of-the-art sampling-based motif discovery approaches.

A domain-specific algorithm in this subcategory is RiboFSM \citep{Gawronski2014RiboFSM} developed for mining frequent subgraphs in \textit{directed dual graphs} in the context of RNA structures. In directed dual graphs, vertices represent complementary regions, and edges represent unpaired nucleotides. Complementary regions are regions of two sequences with complement nucleotides. The RiboFSM is a sampling-based algorithm, and significant patterns are identified in comparison with the randomized version of the original network.

To identify the significant motifs in the algorithms proposed for motif discovery based on sampling, it is quite common to compare the frequency of subgraphs with the frequency of the same subgraphs in the randomized versions of the original network. It requires both generating randomized versions of the original network and implementation of motif discovery algorithms on each of the generated randomized networks. Both steps are generally computationally expensive. The analytical formulation of the distribution of motifs in random networks could eliminate both the generation and motif discovery in randomized networks. For this purpose, multiple approaches have been developed to minimize this cost. In \citep{FANMOD_Wernicke2006}, a randomized version of the input network with the same degree distribution without explicit generation of the random network is used to identify significant subgraphs. Another approach is proposed in \citep{matias2006network} in which probabilistic models are developed to estimate the mean and variance of the frequency of subgraphs of size 3 and 4 in random networks. Therefore, it would be possible to determine the significance of motifs based on the mean and variance of these subgraphs without generating randomized networks.

\subsubsection{Other approaches}\label{sgn_otherapproaches}
The algorithms proposed for motif discovery in single giant networks can be categorized in one of the categories discussed above. However, there are some algorithms that can be differentiated from other algorithms in the literature by their unique features. For example, MotifMiner \citep{Parthasarathy2002, Coatney2005MotifMiner} is developed for mining frequent motifs in large biological networks of vertices with spatial coordinates, assuming that the function of biochemical compounds relies on their spatial structure. The proposed toolkit for this algorithm can handle mining both intra-molecule and inter-molecule motif discovery. The interaction forces between pairs of atoms are inversely related to the distance between them. Using this fact, MotifMiner just includes atom-pairs into frequent motifs in which the Euclidean distance between one vertex and at least one another vertex in the motif is less than a user-specified threshold. This process is called \textit{range pruning}. Two networks are assumed to be isomorphic if employing a set of spatial translations on one network can produce the second network. The MotifMiner uses two approaches to handle potential noises: discretization and equi-width binning of Euclidean distances for handling minor fluctuations and recursive fuzzy hashing for relaxing exact matching. Also, it uses three approaches for improving performance: depth-first pruning, dynamic duplicate screening (that discard duplicates candidate during the running time), and analyzing polymer backbone (when the global structure of compounds are of interest).

Another approach called SpiderMine is proposed in \citep{zhu2011mining} in which the objective is to mine the top $k$ largest frequent subgraphs in a single network ($k$ is specified by user). This approach identifies the probabilistically promising growth paths in the network that result in the largest frequent subgraphs. To accomplish this mining objective, SpiderMine mines \textit{$r$-spiders}, frequent subgraphs with a head node in which path distances of all the vertices in the subgraph from the head is less than $r$ (i.e., subgraph is \textit{$r-bounded$} from the head). It is proved that SpiderMine can find the top-$k$ largest patterns with a probability of $1-\epsilon$ in which the $\epsilon$ is also a user-specified error threshold.

GRAMI \citep{Elseidy2014GraMi} is another algorithm in this group which instead of enumerating or counting all the embeddings of a subgraph, finds a minimum set of embeddings which makes a candidate frequent. GRAMI solves a constraint satisfaction problem at each iteration of the algorithm to find the frequent subgraphs. They adopt the minimum image-based support discussed earlier and proposed in \citep{Bringmann2008Frequent} for computing the frequency. The GRAMI is customized for special tasks as well. The first extension is pattern mining instead of subgraph mining, in which patterns are considered as substructures which incorporate indirect relationships between vertices. Therefore, it can replace edges with paths. Two other extensions of GRAMI are CGRAMI and AGRAMI. The CGRAMI is developed to include user-defined \textit{structural} constraints, such as the number of vertices and edges or maximum degree of vertices in mined subgraphs, or \textit{semantics} constraints, such as subgraphs containing a specific set of vertex or edge labels or with a particular number of different labels. The AGRAMI, developed for scalability purposes, is an approximate frequency version of GRAMI, which, although might miss some of the frequent subgraphs, all the returned subgraphs are frequent.

\subsubsection{Graphlet Mining}\label{Graphlet}
\begin{figure}[ht] 
\centering 
\includegraphics[scale=1]{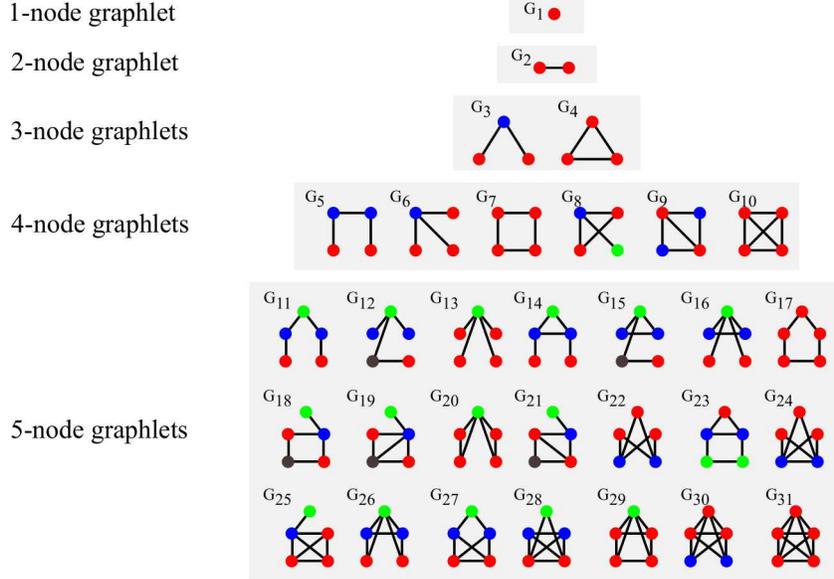}
\caption{The graphlets composed of up to 5 vertices (also refer to \citep{Bhuiyan2012GUISE, Przulj2007graphletdistribution}). The vertices with the same color in each graphlet create an orbit. These vertices can map onto each other in an isomorphism of a graphlet to itself.}.
\label{fig:graphlets} 
\end{figure}

A subcategory of motif discovery is graphlet mining. Graphlets are defined as connected and induced subgraphs. In the literature, the size of graphlets rarely exceeds 4 or 5 vertices. The graphlet of size $k$ is called $k$-graphlet. The 1-graphlet and 2-graphlet are the vertices and edges of the network, respectively, and, therefore, are considered as trivial graphlets. A graphlet is called a \textit{tree graphlet} (or treelet) if it is a tree, otherwise it is called a \textit{cyclic graphlet} \citep{Bhuiyan2012GUISE, Bressanetal2017CountingGraphlets}. Figure \ref{fig:graphlets} shows all the connected graphlets up to size 5 vertices. The frequencies of graphlets are used to produce \textit{graphlet frequency distribution} (GFD). The GFD is a vector of the frequency of different graphlets. This distribution is suggested to be a fingerprint of the network. Please note that in creating the GFD, we are interested in all the graphlets, either frequent or infrequent (or it can be assumed that the frequency threshold is very low, and therefore all the subgraphs composed of less than 4 or 5 vertices are captured). Graphlet mining is generally proposed for single network settings. Having the distribution of graphlet frequencies, then different networks can be compared and classified. For some of the applications of graphlet mining, refer to \citep{Ahmedetal2015Graphlet, Chen2016graphlet}.

Multiple exact frequency and approximate frequency algorithms have been proposed for graphlet mining. For example, ESCAPE \citep{Pinaretal2016ESCAPE} is proposed for \textit{counting without enumeration} of subgraphs composed of 5 or less vertices. ESCAPE, utilizing a set of axioms, uses the counts of some of the subgraphs to compute the count of other subgraphs. In other words, there are 6 and 21 connected (and 5 and 11 disconnected) subgraphs of size 4 and 5, respectively. The authors express that using a precisely selected subset of these subgraphs, the frequency of other subgraphs can be calculated. Also, the counting of the subgraphs in the selected subset can be performed pretty fast. The main idea is that for counting the graphlets, we do not need to enumerate all the embeddings of the subgraphs exhaustively. Instead, we can use the counts of other already counted graphlets to compute the frequency of other graphlets. They show that this algorithm is significantly faster than approaches including the enumeration of all the embeddings. They also use a modified approach developed previously for counting triangles to be used for graphlets of sizes 4 and 5, which makes the algorithm even faster than other similar approaches without enumeration step (such as PGD \citep{Ahmedetal2015Graphlet} discussed in section \ref{bounded_SSNS}). Another counting without enumeration approach is GUISE \citep{Bhuiyan2012GUISE}; an approximate algorithm developed based on a Markov chain Monte Carlo uniform sampling strategy and random walk. It approximates the frequency of graphlets without enumeration in samples drawn from the input network. GRAFT \citep{Rahman2014Graft} is another approximate frequency algorithm for the creation of graphlet frequency distribution for subgraph up to 5 vertices. This algorithm samples edges based on a stratified sampling strategy. Then, for each edge, the partial count of graphlets having this edge is counted. The frequency of different graphlets is estimated using these partial counts. 

Focusing on relative frequency or \textit{concentration} of $k$-graphlets (instead of exact frequencies), an approximate frequency random walk-based algorithm proposed in \citep{Chen2016graphlet}. This algorithm can find graphlets without any size limitation. For estimating relative frequency or percentage of different graphlets of size $k$, this algorithm adopts a random-walk based sampling approach. Based on the \textit{strong law of large numbers} for Markov chains, the authors show that this algorithm provides an unbiased estimator of actual relative frequencies with an analytical bound on the sample size.

A exact frequency algorithm (implemented as a tool called RAGE) is proposed in \citep{Marcus2012RAGE} for counting graphlets of size four or less. The algorithm can accurately count all the (both induced or non-induced) graphlets in the network. Based on the positions that vertices can take in different graphlets, the proposed algorithm counts the graphlets with respect to these positions. A similar approach is adopted in \citep{Hocevar2014graphlet} in which orbits are considered as automorphism groups which vertices of graphlets can participate in, meaning that the vertices are in the same orbit if we can map the vertices onto each other in an isomorphism of a graphlet to itself. There are 30 graphlets composed of 2 to 5 nodes, forming 73 orbits. These orbits are represented as colored vertices in Figure \ref{fig:graphlets}. The vertices with the same color are considered in the same orbit, implying we can map them to each other in an isomorphism of the graphlet to itself. Hočevar and Demšar \cite{Hocevar2014graphlet} propose Orca, an algorithm which, instead of direct enumeration of graphlet embeddings, count the frequency of orbits (or the number of times a specific vertex appears in different graphlets) by the combinatorial relationships and creation of systems of equations among them. They show that employing this approach outperforms some of the state-of-the-art algorithms for graphlet counting. In this case, the orbits frequency distribution can be used as a set of features for evaluation and comparison of different networks.

\begin{table}[H]
  \begin{center}
  \begin{threeparttable}
    \caption{Algorithms for motif discovery in static single network setting}
    \label{tab:statmotif}
    \begin{tabular}{l|c|c|c} 
    \hline
    \hline
    \rowcolor{lightgray}
      \textbf{Algorithm} & \textbf{Exact isomorphic\tnote{a}} & \textbf{Complete\tnote{b}} & \textbf{General\tnote{c}}\\
      \hline
      \hline
      \rowcolor{Gainsboro!50}
      \multicolumn{4}{l}{Exact frequency algorithm} \tabularnewline      
      (H/V)SIGRAM \citep{Kuramochi2004Sparse, Kuramochi2005Sparse}& \checkmark & both & \checkmark \\
      SUBDUE \citep{HolderCookDjoko1994SUBDUE,CookHolder1994SUBDUE, CookHolderKetkar2006MiningGraphData}& both & - & \checkmark  \\
      B-GBI \citep{matsuda2002BGBI}& \checkmark & - & \checkmark \\ 
      CL-GBI \citep{Oharaetal2006MiningGraphData}  & \checkmark & \checkmark & \checkmark \\
      FPF (MAVisto) \citep{schreiber2004towards, Schreiber2005MAVisto}  & \checkmark & \checkmark & \checkmark \\
      gApprox \citep{Chenetal2007gApprox} & - & \checkmark & \checkmark \\
      GREW \citep{KuramochiKarypis2004GREW}& \checkmark & - & \checkmark \\ 
		Parida \cite{Parida2007topologicalmotifs} & \checkmark & \checkmark & - \\
		NeMoFinder \citep{Chen2006NeMoFinder} & \checkmark & \checkmark & \checkmark\\
		Kavosh \citep{Kashani2009Kavosh} & \checkmark & \checkmark & \checkmark \\           
	  Elhesha and Kahveci \cite{Elhesha2016disjoint} & \checkmark & \checkmark & - \\
      \rowcolor{Gainsboro!60}
      \multicolumn{4}{l}{Approximate frequency algorithm} \tabularnewline      
      SEuS \citep{ghazizadeh2002seus}& \checkmark & - & \checkmark  \\
      mfinder \citep{Kashtanetal2004}& \checkmark & - & \checkmark \\ 
		FANMOD \citep{WernickeRasche2006FANMOD}& \checkmark & both & \checkmark \\
		RAND-ESU \citep{wernicke2005faster} & \checkmark & - & \checkmark \\
		Grochow and Kellis \cite{GrochowKellis2007symmetrybreaking}& \checkmark & both & \checkmark \\
		MODA \citep{OmidiFalkMasoudi2009MODA} & \checkmark & both & \checkmark \\
		Baskerville and Paczuski \cite{baskerville2006subgraph} & \checkmark & - & \checkmark \\
		Liu et al. \cite{Liuetal2015stochasticmotifs}& - & - & \checkmark \\
		RiboFSM \citep{Gawronski2014RiboFSM} & \checkmark & - & - \\
	 \rowcolor{Gainsboro!50}
      \multicolumn{4}{l}{Other approaches} \tabularnewline      
      MotifMiner \citep{Parthasarathy2002, Coatney2005MotifMiner} & - & - & \checkmark \\
      SpiderMine \citep{zhu2011mining} & - & \checkmark & \checkmark \\  
      GRAMI \citep{Elseidy2014GraMi}   & \checkmark & \checkmark & \checkmark \\
      AGRAMI \citep{Elseidy2014GraMi}   & \checkmark & - & \checkmark \\
      CGRAMI \citep{Elseidy2014GraMi}   & \checkmark & - & - \\
      \rowcolor{Gainsboro!50}
      \multicolumn{4}{l}{Graphlet mining} \tabularnewline                        
		ESCAPE \citep{Pinaretal2016ESCAPE} & \checkmark & \checkmark & - \\      
		RAGE \citep{Marcus2012RAGE} & \checkmark & \checkmark & - \\
		GUISE \citep{Bhuiyan2012GUISE}  & \checkmark & - & - \\
		Chen et al. \cite{Chen2016graphlet} & \checkmark & - & - \\
		Orca \citep{Hocevar2014graphlet} & \checkmark & \checkmark & - \\
		GRAFT \citep{Rahman2014Graft} & \checkmark & - & - \\
		Bressan et al. \cite{Bressanetal2017CountingGraphlets} & \checkmark & - & - \\
		WRW \citep{Han2016Waddling} & \checkmark & - & \checkmark \\
		PSRW \citep{Wang2014Statistics} & \checkmark & \checkmark & - \\
		MSS \citep{Wang2014Statistics} & \checkmark & \checkmark & - \\				
      \hline
      \hline
    \end{tabular}
    
      \begin{tablenotes}
      \small
      \item[a] If the algorithm mines exact isomorphic subgraphs.
      \item[b] If the algorithm mines all the frequent subgraphs.
      \item[c] If the algorithm mines different types of subgraphs (in contrast to special types such as induced, closed, or maximal subgraphs).
    \end{tablenotes}
      
  \end{threeparttable}
  \end{center}
\end{table}

\subsection{Temporal single network setting}\label{tsns}
In cases where attributes of vertices and edges are dynamics or the set of vertices or edges of the networks are changing over time, overlooking these changes might impact the findings negatively. Therefore, algorithms proposed in this subclass are concerned with the dynamic or temporal changes of the network through insertion or deletion of vertices and edges, or through changes of weights or labels of vertices or edges over time. The algorithms in this category can be categorized based on the changes they focus on, the changes in the attributes of vertices and edges, or the topological changes in the network in the form of vertex and edge insertion and deletion. For a summary of the algorithms reviewed in this section, please refer to Table \ref{tab:temporalmotif}.

\subsubsection{Dynamic attributes}
One of the proposed algorithms in this subclass is Trend Motif \citep{JinMcCallenAlmaas2007TrendMotif, JinMcCallenAlmaas2007TrendMotiftechreport}. In this algorithm, motifs are defined as subgraphs that show consistent changes (trends) over sub-intervals of time (either decreasing or increasing). The trend can be evaluated at both vertex-level (when the weight of the vertex is increasing or decreasing over sub-intervals of time) and subgraph level. They consider the maximal trends as well (trends which are not a subset of other trends). Two subgraphs are considered isomorphic if their corresponding labeled induced forms are isomorphic, and vertex-level trends are identical in both subgraphs. And the motifs are considered frequent if there is more than a pre-defined ratio of isomorphic edge-disjoint (with at most one vertex in common among instances) subsets of vertices in comparison with the randomized version of the network. Both frequency and $z$-score significance can be computed for the identification of motifs. The Trend Motif adopts a depth-first search strategy, finds the maximal trends at the vertex-level first, and remove all the vertices that do not have a significant trend to narrow down the search space. In this algorithm, each vertex might have two types of attributes, a dynamic attribute and a fixed label, at the same time. For example, in \citep{JinMcCallenAlmaas2007TrendMotif}, the dataset is composed of a time series of shares of different countries to the global economy. The vertices are countries with known labels, while each can have positive or negative trends of the share of the country over time. Then, the objective is to find motifs (composed of countries) showing consistent dynamics over all instances of the motif (motif might have vertices with positive or negative trends). The considered datasets in \citep{JinMcCallenAlmaas2007TrendMotiftechreport} have 196/375/52, 116/887/250, 116/607/250, and 6105/8815/18 vertices/edges/snapshot and are relatively small.

\subsubsection{Dynamic topology}
In \citep{BorgwardtKriegelWackersreuther2006}, a sequence of $n$ networks (or time series of networks) representing one single network is defined as $N_{ts}=\lbrace N_1,\ldots ,N_n\rbrace =(V,E,m)$ in which all the networks in the sequence have the same set of vertices, $V$. However, the edges might be deleted or inserted over time. In this definition, $E$ is the set of all the edges in the $n$ network and $m$ represents a mapping from $E$ to a binary, $m:E\longrightarrow \lbrace0|1\rbrace^n$, showing the (lack of) presence of edges in the network via a string. Then, $N_{2}=(V_{2},E_{2},m_2)$ is considered a \textit{topological subgraph} of $N_{1}=(V_{1},E_{1},m_1)$ if $V_2 \subseteq V_1$, $E_2 \subseteq E_1$, and $m_1(e_1)=m_2(e_2)$ for all $e_1 \in E_1$ and $e_2 \in E_2$. Also, $N_{2}=(V_{2},E_{2},m_2)$ is considered a \textit{dynamic subgraph} of $N_{1}=(V_{1},E_{1},m_1)$ if $V_2 = V_1$, $E_2 = E_1$, and $m_2(e_2)$ is a substring of $m_1(e_1)$ for all $e_1 \in E_1$ and $e_2 \in E_2$. In other words, in the topological subgraph we have a subset of vertices and edges of the supergraph but both on the same $n$ sequence. The detected subgraphs have all the vertices and edges of the supergraph, but just in a sub-interval of the $n$ sequence. A dynamic subgraph is frequent if it appears in a more than a pre-defined number of times in the sequence. Considering that instances of a subgraph can start and end from and in the same or different networks in the sequence, two sub-groups of frequent dynamic subgraph mining can be defined; synchronous (when they start from and end in the same network in the sequence), and asynchronous. Also, it is expressed that subgraph mining on these sequences of networks can be performed in two ways. Mining frequent topological subgraphs and then identification of dynamic subgraphs in the mined frequent topological subgraphs. Or, integrating the two steps of mining frequent topological subgraph  and mining frequent dynamic subgraphs. For the mining frequent topological subgraph, the algorithms proposed in the literature of subgraph mining in static networks can be adopted. In \citep{BorgwardtKriegelWackersreuther2006}, the GREW algorithm \citep{KuramochiKarypis2004GREW} has been modified to find frequent dynamic subgraphs in each iteration.

In \citep{Kovanenetal2011Temporalmotifs}, the $\Delta t$-adjacent edges and $\Delta t$-connected edges are defined. Two edges that have one vertex in common and the duration between the start time of the second edge and end time of the first edge is not longer than $\Delta t$ is called $\Delta t$-adjacent. A sequence of edges in which all the consecutive pairs are $\Delta t$-adjacent is called $\Delta t$-connected edges. Then, a \textit{valid temporal subgraph} is defined as a subgraph in which all the $\Delta t$-connected edges of all the vertices in the subgraph are consecutive. For every edge in the network, there is a maximal subgraph in which all the edges are $\Delta t$-connected, and no other edges can be added to the subgraph and keep the subgraph still $\Delta t$-connected. The proposed algorithm finds all the maximal subgraphs for each edge in the network; among them, all the valid subgraphs are identified, and isomorphic subgraphs and temporal motifs are mined. They explain that comparing motifs concentration in the original network and randomized version of the network might be an option. However, due to biases that might arise by using null-models (randomized network), they propose comparing the motifs concentration in different regions of the same network or at different times in the network's temporal range.

Another algorithm in this category is COMMIT \citep{Gurukaretal2015COMMIT} proposed for mining communication motifs in dynamic interaction networks. In this algorithm, the temporality of edges is shown as multi-edges in cases where there is more than one communication between two vertices at different times. This algorithm converts the temporally connected components of the giant network into interaction sequences (a component to be temporally connected should have every pair of vertices to be $\Delta t$-connected). Then, frequent subsequences in interaction sequences are mined to form candidate subgraphs. These candidates are then converted back to networks, and motifs are mined. The mapping of connected components to the sequence interactions is performed using a total ordering on the network edges. For doing that, vertices are labeled with their degrees, and edges are labeled with the degrees of their two endpoints. They also define an edge containment constraint to guarantee the presence of subgraph relations in the sequence space, (i.e. if subgraph $s_2$ is a subgraph of $s_1$, $s_2 \subseteq s_1$, the corresponding sequences generated using the map function, $\mathcal{M}$, show containment relations too, $\mathcal{M}(s_1)$ contains $\mathcal{M}(s_2)$). After mapping the connected components of the network to sequence interactions, the subgraph mining problem is reduced to mining frequent subsequences, which can be performed with lower computational complexity. However, to make sure that no two different subgraphs are mapped to the same sequence, the mined frequent subsequences are converted beck to networks to finalize the true list of motifs.

Two concepts of communication motifs and maximum-flow motifs are introduced in \citep{Zhaoetal2010Communicationmotifs}. For communication motifs, a time window $W$ is considered, which represents the duration of information validity. For each edge in the motifs, there is at least one $W$-adjacent edge in the motif. Also, to make it possible to identify motifs representing information propagation, they introduce the $L$-support defined as an upper-bound for the number of embeddings that contribute to the total frequency of the motif at different time points. This strategy consequently forces the algorithm to detect the motifs not only frequent but distributed over time. Representing edges in the form of $e_i=\langle u_i,v_i,t_i,\delta_i \rangle$, in which the $u_i$ and $v_i$ are the two endpoints of the communication, $t_i$ is the start time of communication, and $\delta_i$ is the duration of communication, they quantify the probability of information propagation in each motif looking at pairs of edges. The two connected edges, $e_i=\langle u_i,v_i,t_i,\delta_i \rangle$ and $e_j=\langle u_j,v_j,t_j,\delta_j \rangle$ maximize the flow if and only if for all other edges in the motif such as $e_k=\langle u_k,v_k,t_k,\delta_k \rangle$, we have $\delta_j/(\delta_i\times (t_j-t_i))>\delta_k/(\delta_i\times (t_k-t_i))$. To evaluate the significance of the motifs, they use randomized versions of the input network.

An algorithm is proposed in \citep{Paranjapeetal2017Motifs} for counting temporal $k$-node and $l$-edge motifs (they focus more specifically on star and triangle patterns). The considered network data is a single network with edges labeled with time points in which the edge is active and represented as $(u_i,v_i,t_i)$ where $u_i,v_i \in V$. The times in which edges are active, $t_i$, are unique, which makes it possible to order all the edges according to $t_i$ values strictly. They define $\delta$-temporal motifs as motifs in which all the edges are active during $\delta$ units of time, and their induced static motif is connected. The motifs considered are very small, having a maximum of three nodes and three edges. This algorithm is based on counting (without enumeration). Another algorithm is proposed in \citep{lietal2018temporalHIN} for mining temporal motifs in heterogeneous information networks (HINs), and more specifically, \textit{fusiform} motifs in which two node types are connected through multiple intermediate node types. The proposed algorithm uses dynamic programming. 

\subsubsection{Network data streams}
In \citep{Rayetal2014StreamingGraphs}, an algorithm, StreamFSM, is proposed for mining frequent subgraph in a single giant network when the graph data is updated with new streams of the network data in the form of updates of vertices and edges (including the addition of new vertices and edges). The cases of deletions of vertices and edges are not considered because it requires updating and recalculation of the frequency for the subgraphs with now-deleted vertices or edges. The frequency of subgraphs, therefore, is changing with time.  After the new data is received, it is added to the giant network. The neighborhood region for each new edge in the giant network (1-hop  neighborhood around two endpoints of the added edge) is extracted. Each extracted neighborhood is then considered as a transaction. A network database is created from these transactions in each iteration. Then, subgraph mining algorithms in network-transaction setting are used to mine frequent subgraphs in the network database (In \citep{Rayetal2014StreamingGraphs}, gSpan \citep{YanHan2002gSpan} is mentioned). The frequency threshold for this step is considered very low to make sure the frequent subgraphs are retrieved (however, this approach does not guarantee to discover all the frequent subgraphs). Then, the mined frequent subgraphs are used to update the repository of candidate subgraphs for the giant network. By each update, some of the subgraphs which have not been considered frequent might change to be frequent after the update. Also, after each update, frequent subgraphs can be reported from this repository based on another pre-defined frequency threshold. 

An algorithm is proposed in \citep{Mukherjeeetal2018Countingmotifs} for mining motifs in a topologically evolving network. It can count both overlapping and edge-disjoint embeddings of subgraphs. To count the edge-disjoint embeddings, they use an \textit{overlap network} for each subgraph composed of all the (overlapping) embeddings of the subgraph as vertices. The vertices in the overlap network are connected if their corresponding embeddings share at least one edge. The creation of a set of edge-disjoint embeddings is performed by iteratively finding the vertex (embedding) with the smallest degree, recording this embedding, and removal of all other nodes connected to this vertex. The final list of recorded embeddings gives the list of edge-disjoint embeddings for each subgraph. After having these lists of all the edge-disjoint embeddings for the network at iteration $0$, in the next iteration of network evolution, these lists are updated. If an edge is deleted, all the embeddings which include this edge will be removed as well, and the list of all the embeddings is updated. If a new edge is added, the neighborhood of this new edge is examined for the embeddings containing the new edge, and the list is updated. To speed up these operations, an \textit{edge compressed bitmap} is created containing the list of all the embeddings for each edge in the network. To update the list of edge-disjoint embeddings, if a new edge is added to the network, the process is similar to the case of a list of all embeddings, unless the new edge creates new embeddings with edges already included in the list of edge-disjoint embeddings, which in this case the new edge has no impact on this list. The deletion of an edge from the network, however, needs more investigation. The elimination of an edge might result in the removal of an embedding from the list of edge-disjoint embeddings. It should be checked if there is another embedding in the network which has other edges (but not the deleted edge) in common with the deleted embedding (and it was removed in the first place due to the presence of the now-deleted embedding), if yes, this embedding should be added to the list. For finding this embedding, the neighborhood of the deleted edge is examined. They evaluated their algorithms on relatively small motifs, composed of three (with two or three edges) or four vertices with three edges.

\begin{table}[h!]
\begin{center}
\begin{threeparttable}
  
    \caption{Algorithms for motif discovery in temporal single network setting}
    \label{tab:temporalmotif}
    \begin{tabular}{l|c|c|c} 
    \hline
    \hline
    \rowcolor{lightgray}
      \textbf{Algorithm} & \textbf{Exact isomorphic\tnote{a}} & \textbf{Complete\tnote{b}} & \textbf{General\tnote{c}}\\
      \hline
      \hline
	 \rowcolor{Gainsboro!50}
     \multicolumn{4}{l}{Dynamic attributes} \tabularnewline 
		Trend Motif \citep{JinMcCallenAlmaas2007TrendMotif, JinMcCallenAlmaas2007TrendMotiftechreport} & \checkmark & \checkmark & \checkmark \\ 
       		    
	 \rowcolor{Gainsboro!50}
     \multicolumn{4}{l}{Dynamic topology} \tabularnewline
     Dynamic GREW \citep{BorgwardtKriegelWackersreuther2006} & \checkmark & - & \checkmark\\
	 Kovanen et al. \cite{Kovanenetal2011Temporalmotifs}& \checkmark & \checkmark & \checkmark\\
	 COMMIT \citep{Gurukaretal2015COMMIT}& \checkmark & - & \checkmark  \\
	Paranjape et al. \cite{Paranjapeetal2017Motifs} & \checkmark & \checkmark & - \\ 
	Li et al. \cite{lietal2018temporalHIN}& \checkmark & \checkmark & -   \\
	\rowcolor{Gainsboro!50}
     \multicolumn{4}{l}{Network data streams} \tabularnewline                
     StreamFSM \citep{Rayetal2014StreamingGraphs}& \checkmark & - & \checkmark \\
     Mukherjee et al. \cite{Mukherjeeetal2018Countingmotifs}& \checkmark & - &\checkmark\\    
      \hline
      \hline

    \end{tabular}

      \begin{tablenotes}
      \small
      \item[a] If the algorithm mines exact isomorphic subgraphs.
      \item[b] If the algorithm mines all the frequent subgraphs.
      \item[c] If the algorithm mines different types of subgraphs (in contrast to special types such as induced, closed, or maximal subgraphs).
    \end{tablenotes}
      
  \end{threeparttable}
  \end{center}
\end{table}

\section{Algorithms for CPU-bound and I/O bound problems}\label{bounded_SSNS}
One of the main assumptions made in the algorithms proposed for the motif discovery problem is that the data of the single network can fit into the main memory, or the computational resources of the local machine suffice the processing steps required for candidate generation and frequency computation. With the increasing amount of network data being produced in different domains, this assumption might not be valid in some applications. Having these (I/O or CPU-bound) limitations in mind, some algorithms are proposed for motif discovery when the single network cannot be held in the main memory, or the processing should be performed in parallel or distributed mode. In the following, some of these algorithms are reviewed, and their characteristics are summarized in Table \ref{tab:ParallelAlg}.

The SUBDUE algorithm introduced earlier as a compression-based subgraph mining algorithm is developed for applications where the network data can be stored in the main memory. This algorithm is modified in \citep{Chakravarthyetal2004DBSubdue} as DB-SUBDUE for mining subgraphs using relational database management systems based on SQL. This algorithm creates two basic tables, \textit{Vertices}, composed of rows representing each vertex with a unique id and not a necessarily unique label, and \textit{Edges}, composed of rows that include the identifiers of the corresponding vertices and not necessarily unique label. The algorithm extends subgraphs one edge at a time. The development of equivalent to minimum description length principle adopted in the main-memory version of SUBDUE is considered as a future objective. The Enhanced DB-SUBDUE (EDB-SUBDUE) is proposed on top of SUBDUE and DB-SUBDUE, which can handle cycles and overlaps in the input network. In \citep{Cooketal2001ParallelSUBDUE}, three SUBDUE-based approaches are proposed for parallel and distributed computing; FP-SUBDUE (functional parallel approach), SP-SUBDUE (static partitioning), and DP-SUBDUE (dynamic partitioning). The SP- and DP-SUBDUE use distributed memory architectures. In the FP-SUBDUE, there are one master and multiple slave processors. The slave processors search for subgraphs that can potentially compress the network reasonably and report that to the master processor. If the slave processors do not have any subgraph to search for in the network, masters ask another slave processor (if it has more than a pre-defined number of subgraphs to search for) to send the candidate to the first slave processor. Keeping a global queue of reported subgraphs by slave processors, the master processor can decide which subgraphs should be kept and which ones should be discarded. In DP-SUBDUE, similarly, there is one master processor and multiple slave processors. However, the whole input network is used by each processor to find subgraphs. The DP-SUBDUE is designed to make sure each subgraph can be discovered just by one slave processor. The role of the master processor is quality control of the slave processors' performances and collecting the frequent subgraphs. It also receives information from slave processors of examined potentially unfruitful subgraphs and informs other slave processors about them. Therefore, they can discard duplicate candidates. They discuss that this approach has generated the poorest performance among the three methods. The SP-SUBDUE has the best performance among these three with the highest scalability, which similarly works with one master and multiple slave processors. In SP-SUBDUE, the input network is partitioned among slave processors. These processors mine their partitions individually and report their best subgraphs to other processors. In this way, other processors would be able to examine the best subgraphs identified by other processors on their own partition. Finally, the master processor collects the final list of subgraphs and determines the best subgraphs among them. 

In \citep{Ahmedetal2015Graphlet}, PGD is proposed for parallel counting of induced (connected/ disconnected) 2-, 3- and 4-vertex motifs. This method can be implemented in both parallel and distributed (and hybrid) forms. The main contribution of this method is using intuitive facts for counting 3-, 4-vertex motifs. In addition, some \textit{combinatorial arguments} are expressed, which help to derive the count of 4-vertex motifs from the counts of 3-vertex motifs and other already counted 4-vertex motifs. These combinatorial arguments help to compute the exact count of some motifs based on the counts of other motifs without referring directly to the input network. The experiments performed in \citep{Ahmedetal2015Graphlet} show that this algorithm is significantly faster than FANMOD \citep{WernickeRasche2006FANMOD}, RAGE \citep{Marcus2012RAGE}, and Orca \citep{Hocevar2014graphlet}. 

FASCIA \citep{SlotaMadduri2013Counting} is proposed for approximate counting and enumeration of (tree-structured) subgraphs in a large network. This approach can be used iteratively for enumeration or counting of children of subgraphs already identified as frequent. This method is a shared memory parallel implementation (using OpenML parallelization) of a modified version of the color-coding technique proposed in \citep{Alonetal1995Colorcoding}. Using parallelization, it is shown that FASCIA is able to mine frequent subgraphs with up to 12 vertices.

The p-MotifMiner \citep{Wang2004pmotifminer} is developed to improve the efficiency of MotifMiner (discussed in subsection \ref{sgn_otherapproaches}). The improvement is accomplished by parallelization of subgraph isomorphism and noise handling tasks using the privatize-and-reduce principles for parallelization. In this algorithm, the candidate generation and recursive fuzzy hashing pruning are independently parallelized, and then these two tasks are combined to complete the algorithm. Similar to MotifMiner, the parallel version is applicable to both intra- and inter-structure motif discovery in which for the inter-molecular motif discovery, the intra-molecular motif discovery is iteratively run for a queue of networks. In \citep{Ribeiro2012Parallel}, two parallel strategies are proposed for the parallelization of FANMOD for approximate frequency (sampling-based) and exact frequency motif discovery. The motif discovery process is composed of three phases, \textit{pre-processing}, \textit{work}, and \textit{aggregation}. The strategies proposed for parallelization use the same first and third phases of FANMOD. However, they employ different approaches in the second phase in which the subgraphs are analyzed, and their frequencies are computed. The first strategy is a master-worker, and the second one is distributed. In the former, one of the workers is responsible for load balancing, and other workers are responsible for the frequent motif discovery. In the distributed strategy, all workers contribute to load balancing. They examine their working queue frequently and request work from other workers if they have an empty queue.

Another parallel algorithm in this category called Subenum is proposed in \citep{Shahrivari2015Multicore}. It is designed for enumerating all the frequent subgraphs in a single network designed on multicore and multiprocessor machines. The main contribution of this algorithm is that it employs an edge-based enumeration mechanism. In this algorithm, a shared queue of edges is created from which threads pick edges for subgraph generation and enumeration. This algorithm uses external storage to keep the list of all non-isomorphic canonical labelings. For subgraph isomorphism, a two-stage check is performed. First, a heuristic, \textit{ordered labeling}, is used to remove some of the isomorphic subgraphs. In the second step, the nauty algorithm \citep{mckay1981, McKayPiperno2014} is employed to remove all the duplicates not identified in the first step.

\begin{table}[ht]
  \begin{center}
  \begin{threeparttable}
    \caption{Algorithms proposed for CPU-bound and I/O bound problems}
    \label{tab:ParallelAlg}
    \begin{tabular}{l|c|c|c} 
    \hline
    \hline
    \rowcolor{lightgray}
      \textbf{Algorithm} & \textbf{Exact isomorphic\tnote{a}} & \textbf{Complete\tnote{b}} & \textbf{General\tnote{c}}\\
      \hline
      \hline

	 DB-SUBDUE \citep{Chakravarthyetal2004DBSubdue} & \checkmark & \checkmark & \checkmark \\
     FP-/DP-/SP-SUBDUE \citep{Cooketal2001ParallelSUBDUE}& - & - & \checkmark\\
     PGD \citep{Ahmedetal2015Graphlet}& \checkmark & \checkmark & - \\
     FASCIA \citep{SlotaMadduri2013Counting}& \checkmark & - & \checkmark\\
	 p-MotifMiner \citep{Wang2004pmotifminer} & - & - & \checkmark \\ 
	 Ribeiro et al. \cite{Ribeiro2012Parallel} & \checkmark & both & \checkmark \\
	 Subenum \citep{Shahrivari2015Multicore} & \checkmark & \checkmark & -\\ 
      \hline
      \hline
    \end{tabular}
    \begin{tablenotes}
      \small
      \item[a] If the algorithm mines exact isomorphic subgraphs.
      \item[b] If the algorithm mines all the frequent subgraphs.
      \item[c] If the algorithm mines different types of subgraphs (in contrast to special types such as induced, closed, or maximal subgraphs).
    \end{tablenotes}
      
  \end{threeparttable}
  \end{center}
\end{table}

\section{Tools}
Some of the algorithms reviewed in this survey are implemented, and their corresponding code or tools have been made publicly available by the developers. The algorithms for which we could find a tool are listed in Table \ref{tab:tools} based on the classification in Table \ref{tab:classification}. However, we could not find any publicly available package or library integrating multiple algorithms with different functionalities. The only package we could find was ``subgraphMining'' \citep{samatova2013practical}, which is a library for R. This library provides an implementation of SUBDUE, gSpan, and SLEUTH. The gSpan is a popular algorithm for mining frequent subgraphs in network-transaction setting \cite{YanHan2002gSpan, YanHan2002gSpantech}. SLEUTH is proposed for mining frequent subtrees in a database of trees \citep{Zaki2005SLEUTH}.

\begin{table}[h]
\small{
  \begin{center}
    \caption{Publicly available tools for algorithms reviewed in this paper}
    \label{tab:tools}
    \resizebox{\textwidth}{!}{\begin{tabular}{p{3cm}|l|l} 
    \hline
    \hline
    \rowcolor{lightgray}
      \textbf{Name} & \textbf{Address} & \textbf{Platform}\\
      \hline
      \hline
      \hline 
\rowcolor{Gainsboro!100} 
\multicolumn{3}{l}{Static single network setting} \tabularnewline            
\hline 

\rowcolor{Gainsboro!50}
\multicolumn{3}{l}{Exact frequency algorithms} \tabularnewline      
SUBDUE \citep{HolderCookDjoko1994SUBDUE, CookHolder1994SUBDUE, CookHolderKetkar2006MiningGraphData} & \href{http://ailab.wsu.edu/subdue/}{http://ailab.wsu.edu/subdue/} & Python \& C\\
MAVisto \citep{schreiber2004towards, Schreiber2005MAVisto} & \href{http://mavisto.ipk-gatersleben.de/}{http://mavisto.ipk-gatersleben.de/}& Java \\
Kavosh \citep{Kashani2009Kavosh} & \href{http://lbb.ut.ac.ir/dynamic/uploads/soft/Kavosh.rar}{http://lbb.ut.ac.ir/dynamic/uploads/soft/Kavosh.rar} & C++  \\
\rowcolor{Gainsboro!50}
\multicolumn{3}{l}{Approximate frequency algorithms} \tabularnewline    
SEuS \citep{ghazizadeh2002seus} & \href{http://www.cs.umd.edu/projects/seus/}{http://www.cs.umd.edu/projects/seus/} & Java\\
mfinder \citep{Kashtanetal2004} & \href{https://www.weizmann.ac.il/mcb/UriAlon/download/network-motif-software/}{https://www.weizmann.ac.il/mcb/UriAlon/download/network-motif-software/} & C++\\
FANMOD \citep{WernickeRasche2006FANMOD} & \href{http://theinf1.informatik.uni-jena.de/~wernicke/motifs/}{http://theinf1.informatik.uni-jena.de/~wernicke/motifs/} & C++ \\
RAND-ESU \citep{wernicke2005faster} & \href{http://theinf1.informatik.uni-jena.de/~wernicke/motifs/}{http://theinf1.informatik.uni-jena.de/~wernicke/motifs/} & C++ \\
MODA \citep{OmidiFalkMasoudi2009MODA}  & \href{http://lbb.ut.ac.ir/dynamic/uploads/soft/MODA.rar}{http://lbb.ut.ac.ir/dynamic/uploads/soft/MODA.rar} & C\# .NET\\ 


\rowcolor{Gainsboro!50}
\multicolumn{3}{l}{Graphlet Mining} \tabularnewline           
ESCAPE \citep{Pinaretal2016ESCAPE} & \href{https://bitbucket.org/seshadhri/escape}{https://bitbucket.org/seshadhri/escape} & C++ \\
RAGE \citep{Marcus2012RAGE} & \href{http://www.eng.tau.ac.il/~shavitt/RAGE/Rage.htm}{http://www.eng.tau.ac.il/~shavitt/RAGE/Rage.htm} & Java  \\
Orca \citep{Hocevar2014graphlet} & \href{http://www.biolab.si/supp/orca/orca.html}{http://www.biolab.si/supp/orca/orca.html} & C++ \\
GRAFT \citep{Rahman2014Graft} & \href{https://github.com/DMGroup-IUPUI/GRAFT-Source}{https://github.com/DMGroup-IUPUI/GRAFT-Source} & C++ \\
Bressan et al. \cite{Bressanetal2017CountingGraphlets}  & \href{https://github.com/Steven--/graphlets}{https://github.com/Steven--/graphlets} & Java \\
\hline 
\rowcolor{Gainsboro!100} 
\multicolumn{3}{l}{Temporal single network setting} \tabularnewline            
\hline 

       		    
\rowcolor{Gainsboro!50}
\multicolumn{3}{l}{Dynamic topology} \tabularnewline
COMMIT \citep{Gurukaretal2015COMMIT}& \href{http://www.cse.iitd.ac.in/~sayan/software.html}{http://www.cse.iitd.ac.in/~sayan/software.html} & C++\\
Paranjape et al. \cite{Paranjapeetal2017Motifs}  & \href{http://snap.stanford.edu/temporal-motifs}{http://snap.stanford.edu/temporal-motifs} & C++ \\

\rowcolor{Gainsboro!50}
\multicolumn{3}{l}{Network data streams} \tabularnewline         
StreamFSM \citep{Rayetal2014StreamingGraphs}& \href{https://github.com/rayabhik83/StreamFSM}{https://github.com/rayabhik83/StreamFSM} & C++\\

\hline 
\rowcolor{Gainsboro!100}
\multicolumn{3}{l}{CPU-bounded and I/O bounded} \tabularnewline      
\hline 
     
PGD \citep{Ahmedetal2015Graphlet}& \href{https://github.com/nkahmed/pgd}{https://github.com/nkahmed/pgd} & C++ \\     
FASCIA \citep{SlotaMadduri2013Counting} & \href{http://fascia-psu.sourceforge.net/}{http://fascia-psu.sourceforge.net/} & C++ \\             
      \hline 
      \hline 
    \end{tabular}}
  \end{center}
  }
\end{table}

\section{Conclusion}
In this paper, we reviewed some of the most popular algorithms for mining frequent or significant motifs in a large network. The main challenges associated with this problem are related to the computational resources required to implement graph and subgraph isomorphisms in each iteration of the algorithm. The sampling-based approaches are proposed as a solution for coping with the complexities associated with graph and subgraph isomorphism problems in giant networks. However, the results in \citep{OmidiFalkMasoudi2009MODA} show that even sampling-based approaches might not be successful in the detection of larger motifs (more than ten vertices) and parallel computing might be a potential solution. Although several algorithms have been proposed for parallel, distributed, and disk-based mining of frequent subgraphs and some of them are reviewed in this paper, this area of research seems a promising research direction in the future. 

There is an increasing amount of interest in mining temporal networks. It is shown that the temporality of networks can increase the discriminability of motifs \citep{lietal2018temporalHIN}. In \citep{Kovanenetal2011Temporalmotifs}, it is discussed that temporal motifs are frequent enough to have significant impacts on the dynamics of the networks, and at the same time, it cannot be simply estimated by temporal correlations.

Another avenue of future research, in parallel to temporal networks mining, is mining frequent substructures in multi-layer networks \citep{de2013mathematical, kivela2014multilayer} or networks with multi-edges in both network-transaction and single network settings. There has been some research in this area, such as the MUGRAM algorithm \citep{Ingalalli2018multigraphs}, proposed for mining all the frequent subgraphs (or potentially sub-multi-networks) in a single multi-network and the algorithm proposed in \citep{ren2019finding} for motif discovery in multi-layer cellular interactions network. 

Furthermore, there are a few tools offering visualization capabilities for motif discovery problems. For example, MAVisto provides a visualization platform \citep{Schreiber2005MAVisto}, and the mfinder webpage offers a visualization tool (mDraw) complementing the mfinder. One of the reasons that developers have not invested in developing visualization tools is because most of motif discovery algorithms generate a large number of frequent patterns. Therefore, the development of visualization platforms to show the discovered motifs or their summary statistics seems a promising direction for future research.

Finally, it has to be noted that although the frequent subgraph mining and motif discovery implementations are applied to extend our insight into frequent common sub-functionalities of different systems, they are not free of critique. For example, in \citep{Artzy2004Commentmotifs}, it is discussed that how lack of representative null models can result in consideration of some subgraphs as frequent, and consequently, potentially wrong conclusions (also refer to \citep{Hallinan2005Motifs}). In contrast, the same subgraphs are considered infrequent using other null models (these critics are replied by Milo et al. \cite{Milo2004Response}), or, in \citep{Ingram2006motifs} in which it is discussed that the biological functions in gene networks cannot be simply predicted from their static structural properties or the static motifs discovered in these networks.

\section*{Acknowledgment}

This work was supported in part by the National Science Foundation under the Grant NSF-1741306, IIS-1650531, and DIBBs-1443019.  Any opinions, findings, and conclusions or recommendations expressed in this material are those of the author(s) and do not necessarily reflect the views of the National Science Foundation.

\pagebreak
\appendix

\section*{Appendix}

\renewcommand{\thesubsection}{\Alph{subsection}}

\subsection{Preliminary Concepts}\label{app_prel_concepts}
\begin{small}

\begin{longtable}{|p{2.5cm}|p{13cm}|}
\caption{A brief description of preliminary concepts}
\label{tab:Prelim_Concepts}\\
\hline
\hline
\textbf{Concept} & \textbf{Definition} \\
\hline
      \textbf{Network} & We define a network or graph as an ordered pair of $N=(V, E)$. The first term, $V$, called set of \textit{vertices} or \textit{nodes}, is composed of discrete elements representing the components of the system ($V=\lbrace v_1,v_2,\ldots, v_n\rbrace$). The second term, $E$, called set of \textit{edges}, \textit{links}, or \textit{connections}, represents the set of interactions between pair of vertices ($E=\lbrace e_1,e_2,\ldots, e_m\rbrace \subseteq V \times V$). Therefore, each edge connects two vertices ($e_k=\lbrace v_i,v_j\rbrace$). The vertices are connected by an edge are called \textit{adjacent} vertices. There might be multiple edges between each pair of vertices. In this case, these edges are called \textit{multi-edge} and the network is called a \textit{multi-graph}. The edges connecting a vertex to itself are called \textit{self-loops}. The networks without multi-edges and self-loops are called \textit{simple networks}. If the direction of edges in the network matters, then the edges are shown as ordered pairs ($e_k=(v_i,v_j)$), and the network is called \textit{directed network}. The $v_i$ is called the \textit{tail} and the $v_j$ is called the \textit{head} of edge. In the visualization of directed networks, the directed edges are shown as lines with arrowheads (the head of the arrow is on the head vertex). The vertices and edges might be labeled or have attributes.  In this case, two functions are defined to map the vertices and their edges to their corresponding labels \citep{InokuchiWashioMotoda2000}, and the network is considered a \textit{labeled} or \textit{weighted network}. It should be noted that the labels of vertices or edges are not necessarily distinct. For example, in a citation network with vertices representing authors with labels specifying the avenue of presentation, there might be many authors with identical labels (see figures below).
      
      \begin{minipage}{0.6\textwidth} 
\begin{figure}[H]
    
\centering 
\subfigure{ 
    \includegraphics[scale=0.3]{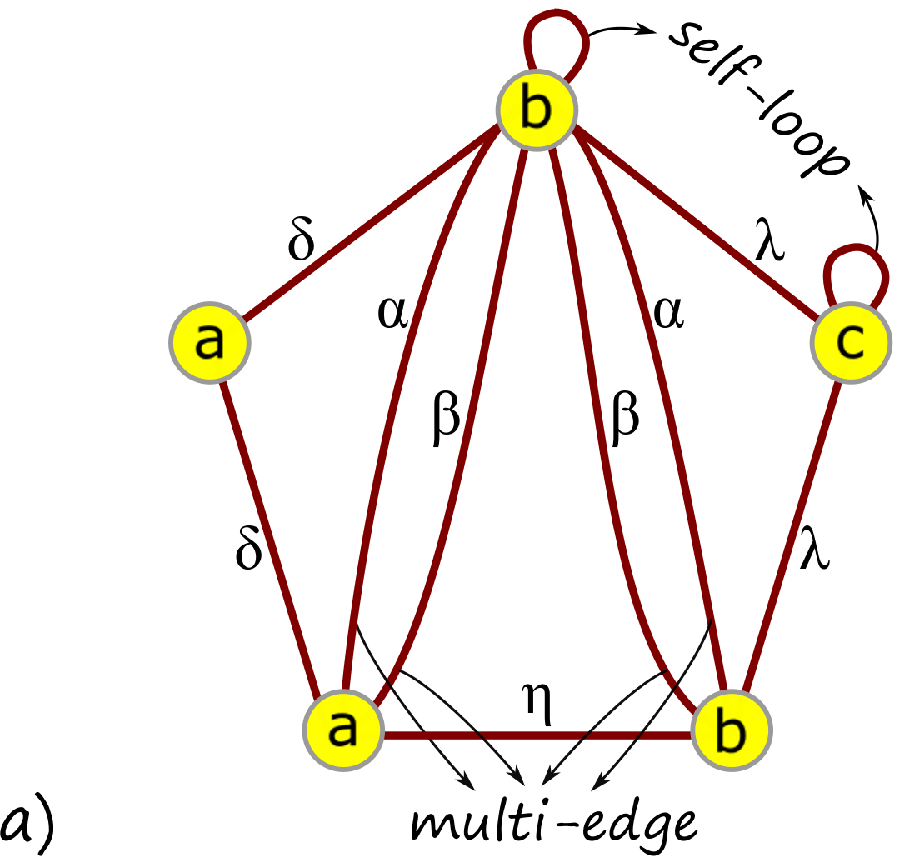} 
    \label{fig:subfig_undirected} 
} 
\subfigure{ 
   \includegraphics[scale=0.3]{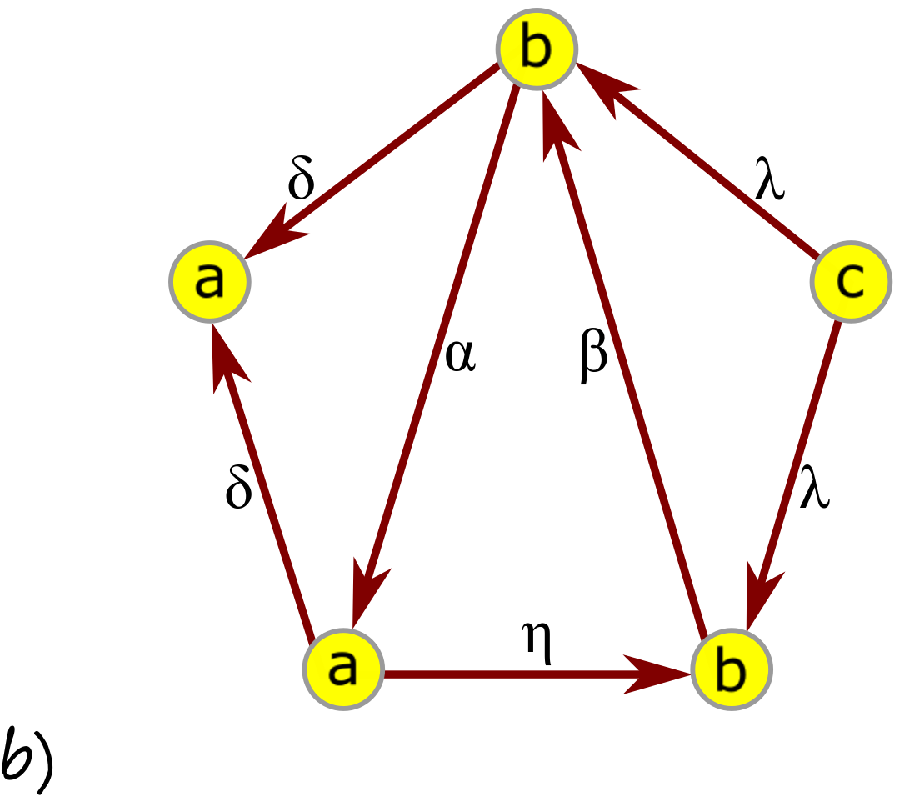} 
    \label{fig:subfig_directed} 
} 
\end{figure}
\end{minipage}

Examples of directed and undirected networks. \subref{fig:subfig_undirected}: An undirected labeled network with multi-edges, and self-loops. \subref{fig:subfig_directed}: A labeled directed network.
\\\hline

      \textbf{Degree} & In undirected networks, the number of adjacent vertices to each vertex is called the degree of the vertex. Because in directed networks, vertices might be head or tail of different edges, the numbers of edges that the vertex plays the role of head and tail for are called the indegree and the outdegree of the vertex, respectively.\\\hline

\textbf{Adjacency matrix} & An adjacency matrix is another form of representation of a network, defined as a $V\times V$ matrix. The elements of the matrix, $a_{ij}=1$ if there is an edge between $v_i$ and $v_j$, otherwise $a_{ij}=0$. The elements of the adjacency matrix can also be the labels or weights of the corresponding edges in the network. The adjacency matrix is symmetrical for undirected networks or may or may not be symmetrical for directed networks.\\\hline

\textbf{Walk, path, and cycle} & Starting from a vertex, a sequence of vertices might be traversed in the network in which every two consecutive edges have one vertex in common. This sequence of edges is called a \textit{walk}. If all the edges and internal vertices met in the sequence are unique, the walk is called a \textit{path}. A path with identical initial and final vertices is called a \textit{cycle}.\\\hline

\textbf{Connected networks, trees, and forests} & The networks with at least one path between every pair of vertices are called a \textit{connected network}. If the network is connected, but without any cycle, the network is called a \textit{tree}, and if there are no cycles in the network and the network is not connected, then the network is called a forest.\\\hline

\textbf{Graph isomorphism problem} & We call two networks $N_1=(V_1,E_1)$ and $N_2=(V_2,E_2)$ isomorphic if there is a bijective function $I$ which map the components of the networks onto each other. In other words, two networks $N_1$ and $N_2$ are isomorphic and shown as $(N_1 \simeq N_2)$ if there is a function $I:N_1 \rightarrow N_2$ and $\lbrace v_i,v_j\rbrace \in E_1 \Leftrightarrow \lbrace I(v_i),I(v_j)\rbrace \in E_2$. The function $I$ is called an \textit{isomorphism}. The collection of all the networks isomorphic to each other is called \textit{an isomorphism class}. The isomorphism of a network to itself is called an \textit{automorphism}. Also, the properties of the networks remained unchanged under the isomorphism are called \textit{graph isomorphism invariants}. Although the isomorphism problem can be solved in polynomial time for some special cases, such as tree isomorphism \cite{Johannes1994Isomorphism}, the general graph isomorphism problem is not known to be in the P or NP-complete \citep{Fortin96thegraph, Read1977isomorphism}. The evaluation of graph isomorphism is an important step in frequent subgraph mining and motif discovery \citep{KuramochiKarypis2001FSG}. In this step, it is checked that if two candidates generated are representing the same network. Therefore, to avoid redundancy of candidates, the check should be performed before finalizing the list of frequent subgraphs. One of the approaches adopted for graph isomorphism is the canonical labeling of networks and then comparing these labels for different networks. The canonical label of a network is a code that represents the network. This code is invariant to different representations of the network. Canonical labeling has the same complexity as graph isomorphism; however, using invariant properties under graph isomorphism, it is tried to reduce the complexity of the problem. For different implementation of canonical labeling refer to \citep{KuramochiKarypis2001FSG,  InokuchiWashioMotoda2000, matsuda2002BGBI, HuanWangPrins2003FFSM} and for a detailed example refer to \citep{KuramochiKarypis2004FSG, KuramochiKarypis2004FSGtechreport}.\\\hline

\textbf{Subgraph} & A subset of the components of a network is called a subgraph of the network. In other words, having $N_1=(V_1,E_1)$, any network $N_2=(V_2,E_2)$ for which $V_2 \subseteq V_1$ and $E_2 \subseteq E_1$ is considered a subgraph of $N_1$ and is shown as $N_2 \subseteq N_1$. For a subgraph $N_2$ of $N_1$, if $N_2 \neq N_1$, then $N_2$ is considered a proper subgraph of $N_1$.The network $N_2$ is called an induced subgraph of $N_1$ if $N_2 \subseteq N_1$, and $E_2$ is composed of all the edges in $N_1$ connecting pairs of vertices in $V_2$ (see figures below).

\begin{minipage}{0.6\textwidth}
\begin{figure}[H] 
    
\centering 
\subfigure{ 
    \includegraphics[scale=0.4]{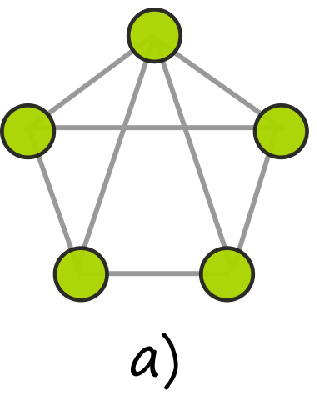} 
    \label{fig:graph} 
} 
\subfigure{ 
   \includegraphics[scale=0.4]{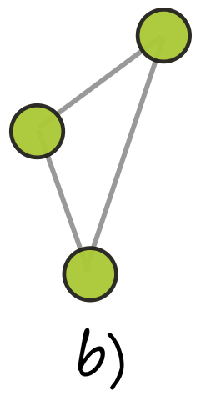} 
    \label{fig:graphinduced} 
} 
\subfigure{ 
   \includegraphics[scale=0.4]{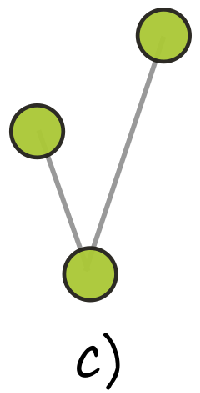} 
    \label{fig:graphgeneral} 
} 

\end{figure}
\end{minipage}

Network and two of its subgraphs. \subref{fig:graph}: A network composed of five vertices and eight edges. \subref{fig:graphinduced}: An induced subgraph of \subref{fig:graph} including all the edges in the original network connecting vertices in the subgraph. \subref{fig:graphgeneral}: A non-induced subgraph of \subref{fig:graph}. \\\hline
\textbf{Subgraph isomorphism problem} & The subgraph isomorphism problem is defined as finding if a network $N_1$ has a subgraph isomorphic to a second network $N_2$. Clearly, the network $N_1$ might have multiple subgraphs isomorphic to $N_2$. In this case, each of these subgraphs is called an \textit{instance} or \textit{embedding} of $N_2$ in $N_1$. This problem is in NP-complete \citep{GareyJohnson1980}. Similar to the graph isomorphism problem, there are subclasses of this problem that can be efficiently solved, such as subtree isomorphism \citep{Reyner1977,matula1978subtree}. However, there is no optimal solution to the general case. Ullmann \citep{Ullmann1976} proposes an algorithm used extensively for both graph and subgraph isomorphism problems. Also, multiple algorithms for subgraph isomorphism are re-implemented in \citep{Lee2012isomorphism}, and their performances are checked. They show that none of the evaluated algorithms is the best for all cases.\\\hline

\textbf{Temporal networks} & The changes possible in networks can be in the form of insertion and deletion of vertices and edges, and the relabeling of these components over time. These networks might be called network sequence, dynamic networks, evolving networks, and time-series networks in different literature. In \citep{Inokuchi2010FRISS}, a classification of these networks is provided in which in the sequence or time-series of networks, all the changes listed above are possible. In dynamic networks, the vertices in the networks are fixed. They cannot be inserted, deleted, or relabeled; however, the edges connecting vertices might change over time. In evolving networks, the vertices can be added but not removed. The same is true for edges; they cannot be removed after they are inserted in the networks. In this paper, we use the general term of temporal networks when we refer to networks that don't have a necessarily fixed list of vertices, edges, and labels over time.\\

\hline
\hline

\end{longtable}

\end{small}

\bibliographystyle{plainnat}

\begin{thebibliography}{167}
\providecommand{\natexlab}[1]{#1}
\providecommand{\url}[1]{\texttt{#1}}
\expandafter\ifx\csname urlstyle\endcsname\relax
  \providecommand{\doi}[1]{doi: #1}\else
  \providecommand{\doi}{doi: \begingroup \urlstyle{rm}\Url}\fi

\bibitem[Agrawal and Srikant(1994)]{agrawal1994fast}
Rakesh Agrawal and Ramakrishnan Srikant.
\newblock Fast algorithms for mining association rules in large databases.
\newblock In \emph{Proceedings of the 20th International Conference on Very
  Large Data Bases}, VLDB ’94, page 487–499, San Francisco, CA, USA, 1994.
  Morgan Kaufmann Publishers Inc.
\newblock ISBN 1558601538.

\bibitem[Ahmed et~al.(2015)Ahmed, Neville, Rossi, and
  Duffield]{Ahmedetal2015Graphlet}
Nesreen~K. Ahmed, Jennifer Neville, Ryan~A. Rossi, and Nick Duffield.
\newblock Efficient graphlet counting for large networks.
\newblock In \emph{2015 IEEE International Conference on Data Mining}, pages
  1--10, Nov 2015.
\newblock \doi{10.1109/ICDM.2015.141}.

\bibitem[Albert and Albert(2004)]{Albert2004motifs}
Istv{\'a}n Albert and R{\'e}ka Albert.
\newblock Conserved network motifs allow protein-protein interaction
  prediction.
\newblock \emph{Bioinformatics}, 20\penalty0 (18):\penalty0 3346--3352, 07
  2004.
\newblock \doi{10.1093/bioinformatics/bth402}.

\bibitem[Albert et~al.(1999)Albert, Jeong, and
  Barab{\'a}si]{AlbertJeongBarabasi1999}
Reka Albert, Hawoong Jeong, and Albert-L{\'a}szl{\'o} Barab{\'a}si.
\newblock Diameter of the world-wide web.
\newblock \emph{Nature}, page 130, 1999.
\newblock \doi{https://doi.org/10.1038/43601}.

\bibitem[Alon et~al.(1995)Alon, Yuster, and Zwick]{Alonetal1995Colorcoding}
Noga Alon, Raphael Yuster, and Uri Zwick.
\newblock Color-coding.
\newblock \emph{Journal of the ACM (JACM)}, 42\penalty0 (4):\penalty0
  844–856, July 1995.
\newblock ISSN 0004-5411.
\newblock \doi{10.1145/210332.210337}.
\newblock URL \url{https://doi.org/10.1145/210332.210337}.

\bibitem[Alon(2007)]{Alon2007motifs}
Uri Alon.
\newblock Network motifs: theory and experimental approaches.
\newblock \emph{Nature Reviews Genetics}, 8\penalty0 (6):\penalty0 450--461,
  2007.
\newblock \doi{10.1038/nrg2102}.

\bibitem[Artzy-Randrup et~al.(2004)Artzy-Randrup, Fleishman, Ben-Tal, and
  Stone]{Artzy2004Commentmotifs}
Yael Artzy-Randrup, Sarel~J. Fleishman, Nir Ben-Tal, and Lewi Stone.
\newblock Comment on "network motifs: Simple building blocks of complex
  networks" and "superfamilies of evolved and designed networks".
\newblock \emph{Science}, 305\penalty0 (5687):\penalty0 1107--1107, 2004.
\newblock ISSN 0036-8075.
\newblock \doi{10.1126/science.1099334}.
\newblock URL \url{https://science.sciencemag.org/content/305/5687/1107.3}.

\bibitem[Bajardi et~al.(2011)Bajardi, Barrat, Natale, Savini, and
  Colizza]{Bajardi2011Movements}
Paolo Bajardi, Alain Barrat, Fabrizio Natale, Lara Savini, and Vittoria
  Colizza.
\newblock Dynamical patterns of cattle trade movements.
\newblock \emph{PLOS ONE}, 6\penalty0 (5):\penalty0 1--19, 05 2011.
\newblock \doi{10.1371/journal.pone.0019869}.
\newblock URL \url{https://doi.org/10.1371/journal.pone.0019869}.

\bibitem[Barab{\'a}si and Albert(1999)]{BarabasiAlbert1999}
Albert-L{\'a}szl{\'o} Barab{\'a}si and R{\'e}ka Albert.
\newblock Emergence of scaling in random networks.
\newblock \emph{Science}, 286\penalty0 (5439):\penalty0 509--512, 1999.
\newblock ISSN 0036-8075.
\newblock \doi{10.1126/science.286.5439.509}.
\newblock URL \url{https://science.sciencemag.org/content/286/5439/509}.

\bibitem[Baskerville and Paczuski(2006)]{baskerville2006subgraph}
Kim Baskerville and Maya Paczuski.
\newblock Subgraph ensembles and motif discovery using an alternative heuristic
  for graph isomorphism.
\newblock \emph{Physical Review E}, 74:\penalty0 051903, Nov 2006.
\newblock \doi{10.1103/PhysRevE.74.051903}.
\newblock URL \url{https://link.aps.org/doi/10.1103/PhysRevE.74.051903}.

\bibitem[Bhuiyan et~al.(2012)Bhuiyan, Rahman, Rahman, and
  Al~Hasan]{Bhuiyan2012GUISE}
Mansurul~A. Bhuiyan, Mahmudur Rahman, Mahmuda Rahman, and Mohammad Al~Hasan.
\newblock Guise: Uniform sampling of graphlets for large graph analysis.
\newblock In \emph{Proceedings of the 2012 IEEE 12th International Conference
  on Data Mining}, ICDM ’12, page 91–100, USA, 2012. IEEE Computer Society.
\newblock ISBN 9780769549057.
\newblock \doi{10.1109/ICDM.2012.87}.
\newblock URL \url{https://doi.org/10.1109/ICDM.2012.87}.

\bibitem[Bollobas(1998)]{Bollobas1998}
Bela Bollobas.
\newblock \emph{Modern graph theory}, volume 184.
\newblock Springer, New York, 1998.
\newblock ISBN 9780387984919.

\bibitem[Bondy and Murty(1976)]{BondyMurty1976}
John~Adrian Bondy and Uppaluri Siva~Ramachandra Murty.
\newblock \emph{Graph theory with applications}.
\newblock American Elsevier Pub. Co, New York, 1976.
\newblock ISBN 9780444194510.

\bibitem[Borgelt and Berthold(2002)]{BorgeltBerthold2002}
Christian Borgelt and Michael~R Berthold.
\newblock Mining molecular fragments: finding relevant substructures of
  molecules.
\newblock In \emph{2002 IEEE International Conference on Data Mining, 2002.
  Proceedings.}, pages 51--58, Dec 2002.
\newblock \doi{10.1109/ICDM.2002.1183885}.

\bibitem[Borgwardt et~al.(2006)Borgwardt, Kriegel, and
  Wackersreuther]{BorgwardtKriegelWackersreuther2006}
Karsten~M Borgwardt, Hans-Peter Kriegel, and Peter Wackersreuther.
\newblock Pattern mining in frequent dynamic subgraphs.
\newblock In \emph{Sixth International Conference on Data Mining (ICDM'06)},
  pages 818--822, Dec 2006.
\newblock \doi{10.1109/ICDM.2006.124}.

\bibitem[Bressan et~al.(2017)Bressan, Chierichetti, Kumar, Leucci, and
  Panconesi]{Bressanetal2017CountingGraphlets}
Marco Bressan, Flavio Chierichetti, Ravi Kumar, Stefano Leucci, and Alessandro
  Panconesi.
\newblock Counting graphlets: Space vs time.
\newblock In \emph{Proceedings of the Tenth ACM International Conference on Web
  Search and Data Mining}, WSDM ’17, page 557–566, New York, NY, USA, 2017.
  Association for Computing Machinery.
\newblock ISBN 9781450346757.
\newblock \doi{10.1145/3018661.3018732}.
\newblock URL \url{https://doi.org/10.1145/3018661.3018732}.

\bibitem[Bringmann and Nijssen(2008)]{Bringmann2008Frequent}
Bj\"{o}rn Bringmann and Siegfried Nijssen.
\newblock \emph{What is Frequent in a Single Graph?}, page 858–863.
\newblock PAKDD’08. Springer-Verlag, Berlin, Heidelberg, 2008.
\newblock ISBN 3540681248.

\bibitem[Bruno et~al.(2010)Bruno, Palopoli, and Rombo]{Bruno2010Trends}
Francesco Bruno, Luigi Palopoli, and Simona~E. Rombo.
\newblock New trends in graph mining: Structural and node-colored network
  motifs.
\newblock \emph{International Journal of Knowledge Discovery in Bioinformatics
  (IJKDB)}, 1\penalty0 (1):\penalty0 81--99, 2010.

\bibitem[Chakravarthy et~al.(2004)Chakravarthy, Beera, and
  Balachandran]{Chakravarthyetal2004DBSubdue}
Sharma Chakravarthy, Ramji Beera, and Ramanathan Balachandran.
\newblock Db-subdue: Database approach to graph mining.
\newblock In Honghua Dai, Ramakrishnan Srikant, and Chengqi Zhang, editors,
  \emph{Advances in Knowledge Discovery and Data Mining}, pages 341--350,
  Berlin, Heidelberg, 2004. Springer.
\newblock ISBN 978-3-540-24775-3.

\bibitem[Chen et~al.(2006)Chen, Hsu, Lee, and Ng]{Chen2006NeMoFinder}
Jin Chen, Wynne Hsu, Mong~Li Lee, and See-Kiong Ng.
\newblock Nemofinder: Dissecting genome-wide protein-protein interactions with
  meso-scale network motifs.
\newblock In \emph{Proceedings of the 12th ACM SIGKDD International Conference
  on Knowledge Discovery and Data Mining}, KDD ’06, page 106–115, New York,
  NY, USA, 2006. Association for Computing Machinery.
\newblock ISBN 1595933395.
\newblock \doi{10.1145/1150402.1150418}.
\newblock URL \url{https://doi.org/10.1145/1150402.1150418}.

\bibitem[Chen et~al.(2016)Chen, Li, Wang, and Lui]{Chen2016graphlet}
Xiaowei Chen, Yongkun Li, Pinghui Wang, and John C.~S. Lui.
\newblock A general framework for estimating graphlet statistics via random
  walk.
\newblock \emph{Proceedings of the VLDB Endowment}, 10\penalty0 (3):\penalty0
  253–264, November 2016.
\newblock ISSN 2150-8097.
\newblock \doi{10.14778/3021924.3021940}.
\newblock URL \url{https://doi.org/10.14778/3021924.3021940}.

\bibitem[Cheng et~al.(2010)Cheng, Yan, and Han]{Cheng2010survey}
Hong Cheng, Xifeng Yan, and Jiawei Han.
\newblock Mining graph patterns.
\newblock In Charu~C. Aggarwal and Haixun Wang, editors, \emph{Managing and
  Mining Graph Data}, pages 365--392. Springer US, Boston, MA, 2010.
\newblock ISBN 978-1-4419-6045-0.
\newblock \doi{10.1007/978-1-4419-6045-0_12}.
\newblock URL \url{https://doi.org/10.1007/978-1-4419-6045-0_12}.

\bibitem[Chent et~al.(2007)Chent, Yan, Zhu, and Han]{Chenetal2007gApprox}
Chen Chent, Xifeng Yan, Feida Zhu, and Jiawei Han.
\newblock gapprox: Mining frequent approximate patterns from a massive network.
\newblock In \emph{Seventh IEEE International Conference on Data Mining (ICDM
  2007)}, pages 445--450, Oct 2007.
\newblock \doi{10.1109/ICDM.2007.36}.

\bibitem[Ciriello and Guerra(2008)]{Ciriello2008review}
Giovanni Ciriello and Concettina Guerra.
\newblock {A review on models and algorithms for motif discovery in
  protein–protein interaction networks}.
\newblock \emph{Briefings in Functional Genomics}, 7\penalty0 (2):\penalty0
  147--156, 04 2008.
\newblock ISSN 2041-2649.
\newblock \doi{10.1093/bfgp/eln015}.
\newblock URL \url{https://doi.org/10.1093/bfgp/eln015}.

\bibitem[Coatney and Parthasarathy(2005)]{Coatney2005MotifMiner}
Matt Coatney and Srinivasan Parthasarathy.
\newblock Motifminer: Efficient discovery of common substructures in
  biochemical molecules.
\newblock \emph{Knowledge and Information Systems}, 7\penalty0 (2):\penalty0
  202--223, Feb 2005.
\newblock ISSN 0219-3116.
\newblock \doi{10.1007/s10115-003-0119-4}.
\newblock URL \url{https://doi.org/10.1007/s10115-003-0119-4}.

\bibitem[Cook and Holder(1994)]{CookHolder1994SUBDUE}
Diane~J. Cook and Lawrence~B. Holder.
\newblock Substructure discovery using minimum description length and
  background knowledge.
\newblock \emph{Journal of Artificial Intelligence Research}, 1\penalty0
  (1):\penalty0 231–255, February 1994.
\newblock ISSN 1076-9757.

\bibitem[Cook et~al.(2001)Cook, Holder, Galal, and
  Maglothin]{Cooketal2001ParallelSUBDUE}
Diane~J. Cook, Lawrence~B. Holder, Gehad Galal, and Ron Maglothin.
\newblock Approaches to parallel graph-based knowledge discovery.
\newblock \emph{Journal of Parallel and Distributed Computing}, 61\penalty0
  (3):\penalty0 427--446, 2001.

\bibitem[Cook et~al.(2006)Cook, Holder, and
  Ketkar]{CookHolderKetkar2006MiningGraphData}
Diane~J. Cook, Lawrence~B. Holder, and Nikhil Ketkar.
\newblock \emph{Unsupervised and Supervised Pattern Learning in Graph Data},
  chapter~7, pages 159--181.
\newblock John Wiley \& Sons, Ltd, 2006.
\newblock ISBN 9780470073049.
\newblock \doi{10.1002/9780470073049.ch7}.
\newblock URL
  \url{https://onlinelibrary.wiley.com/doi/abs/10.1002/9780470073049.ch7}.

\bibitem[De~Domenico et~al.(2013)De~Domenico, Sol\'e-Ribalta, Cozzo, Kivel\"a,
  Moreno, Porter, G\'omez, and Arenas]{de2013mathematical}
Manlio De~Domenico, Albert Sol\'e-Ribalta, Emanuele Cozzo, Mikko Kivel\"a,
  Yamir Moreno, Mason~A. Porter, Sergio G\'omez, and Alex Arenas.
\newblock Mathematical formulation of multilayer networks.
\newblock \emph{Physical Review X}, 3:\penalty0 041022, Dec 2013.
\newblock \doi{10.1103/PhysRevX.3.041022}.
\newblock URL \url{https://link.aps.org/doi/10.1103/PhysRevX.3.041022}.

\bibitem[Diestel(2005)]{Diestel2005}
Reinhard Diestel.
\newblock \emph{Graph theory}, volume 173.
\newblock Springer, Berlin, third edition, 2005.
\newblock ISBN 9783540261827.

\bibitem[Elhesha and Kahveci(2016)]{Elhesha2016disjoint}
Rasha Elhesha and Tamer Kahveci.
\newblock Identification of large disjoint motifs in biological networks.
\newblock \emph{BMC bioinformatics}, 17\penalty0 (1):\penalty0 408, 2016.
\newblock \doi{https://doi.org/10.1186/s12859-016-1271-7}.

\bibitem[Elseidy et~al.(2014)Elseidy, Abdelhamid, Skiadopoulos, and
  Kalnis]{Elseidy2014GraMi}
Mohammed Elseidy, Ehab Abdelhamid, Spiros Skiadopoulos, and Panos Kalnis.
\newblock Grami: Frequent subgraph and pattern mining in a single large graph.
\newblock \emph{Proceedings of the VLDB Endowment}, 7\penalty0 (7):\penalty0
  517–528, March 2014.
\newblock ISSN 2150-8097.
\newblock \doi{10.14778/2732286.2732289}.
\newblock URL \url{https://doi.org/10.14778/2732286.2732289}.

\bibitem[Fiedler and Borgelt(2007)]{fiedler2007support}
Mathias Fiedler and Christian Borgelt.
\newblock Support computation for mining frequent subgraphs in a single graph.
\newblock In \emph{International Workshop on Mining and Learning with Graphs
  (MLG)}. Citeseer, 2007.

\bibitem[Fortin(1996)]{Fortin96thegraph}
Scott Fortin.
\newblock The graph isomorphism problem.
\newblock Technical Report TR 96-20, University of Alberta - Department of
  Computing Science, 1996.

\bibitem[Garey and Johnson(1979)]{GareyJohnson1980}
Michael~R. Garey and David~S. Johnson.
\newblock \emph{Computers and Intractability: A Guide to the Theory of
  NP-Completeness}.
\newblock W. H. Freeman \& Co., USA, 1979.
\newblock ISBN 0716710447.

\bibitem[Gawronski and Turcotte(2014)]{Gawronski2014RiboFSM}
Alex~R. Gawronski and Marcel Turcotte.
\newblock Ribofsm: frequent subgraph mining for the discovery of rna structures
  and interactions.
\newblock \emph{BMC bioinformatics}, 15 Suppl 13\penalty0 (S13):\penalty0
  S2--S2, 2014.
\newblock \doi{https://doi.org/10.1186/1471-2105-15-S13-S2}.

\bibitem[Ghazizadeh and Chawathe(2002)]{ghazizadeh2002seus}
Shayan Ghazizadeh and Sudarshan~S. Chawathe.
\newblock Seus: Structure extraction using summaries.
\newblock In Steffen Lange, Ken Satoh, and Carl~H. Smith, editors,
  \emph{Discovery Science}, pages 71--85, Berlin, Heidelberg, 2002. Springer.
\newblock ISBN 978-3-540-36182-4.

\bibitem[Grochow and Kellis(2007)]{GrochowKellis2007symmetrybreaking}
Joshua~A. Grochow and Manolis Kellis.
\newblock Network motif discovery using subgraph enumeration and
  symmetry-breaking.
\newblock In \emph{Proceedings of the 11th Annual International Conference on
  Research in Computational Molecular Biology}, RECOMB’07, page 92–106,
  Berlin, Heidelberg, 2007. Springer-Verlag.
\newblock ISBN 9783540716808.

\bibitem[Gross et~al.(2013)Gross, Yellen, and Zhang]{GrossYellenZhang2013}
Jonathan~L. Gross, Jay Yellen, and Ping Zhang.
\newblock \emph{Handbook of Graph Theory, 2nd Edition}.
\newblock Chapman and Hall/CRC, 2 edition, 2013.
\newblock ISBN 9781138199668.

\bibitem[Gurukar et~al.(2015)Gurukar, Ranu, and
  Ravindran]{Gurukaretal2015COMMIT}
Saket Gurukar, Sayan Ranu, and Balaraman Ravindran.
\newblock Commit: A scalable approach to mining communication motifs from
  dynamic networks.
\newblock In \emph{Proceedings of the 2015 ACM SIGMOD International Conference
  on Management of Data}, SIGMOD ’15, page 475–489, New York, NY, USA,
  2015. Association for Computing Machinery.
\newblock ISBN 9781450327589.
\newblock \doi{10.1145/2723372.2737791}.
\newblock URL \url{https://doi.org/10.1145/2723372.2737791}.

\bibitem[Güvenoglu and Bostanoglu(2018)]{Guvenoglu2018qualitative}
Büsra Güvenoglu and Belgin~E. Bostanoglu.
\newblock A qualitative survey on frequent subgraph mining.
\newblock \emph{Open Computer Science}, 8\penalty0 (1):\penalty0 194--209,
  2018.
\newblock \doi{https://doi.org/10.1515/comp-2018-0018}.

\bibitem[Han and Sethu(2016)]{Han2016Waddling}
Guyue Han and Harish Sethu.
\newblock Waddling random walk: Fast and accurate mining of motif statistics in
  large graphs.
\newblock In \emph{{IEEE} 16th International Conference on Data Mining, {ICDM}
  2016, December 12-15, 2016, Barcelona, Spain}, pages 181--190, 2016.
\newblock \doi{10.1109/ICDM.2016.0029}.
\newblock URL \url{https://doi.org/10.1109/ICDM.2016.0029}.

\bibitem[Han et~al.(2007)Han, Cheng, Xin, and Yan]{Han2007status}
Jiawei Han, Hong Cheng, Dong Xin, and Xifeng Yan.
\newblock Frequent pattern mining: current status and future directions.
\newblock \emph{Data Mining and Knowledge Discovery}, 15\penalty0 (1):\penalty0
  55--86, Aug 2007.
\newblock ISSN 1573-756X.
\newblock \doi{10.1007/s10618-006-0059-1}.
\newblock URL \url{https://doi.org/10.1007/s10618-006-0059-1}.

\bibitem[Hočevar and Demšar(2014)]{Hocevar2014graphlet}
Tomaž Hočevar and Janez Demšar.
\newblock {A combinatorial approach to graphlet counting}.
\newblock \emph{Bioinformatics}, 30\penalty0 (4):\penalty0 559--565, 12 2014.
\newblock ISSN 1367-4803.
\newblock \doi{10.1093/bioinformatics/btt717}.
\newblock URL \url{https://doi.org/10.1093/bioinformatics/btt717}.

\bibitem[Holder et~al.(1994)Holder, Cook, and Djoko]{HolderCookDjoko1994SUBDUE}
Lawrence~B. Holder, Diane~J. Cook, and Surnjani Djoko.
\newblock Substructure discovery in the subdue system.
\newblock In \emph{Proceedings of the 3rd International Conference on Knowledge
  Discovery and Data Mining}, AAAIWS’94, page 169–180. AAAI Press, 1994.

\bibitem[Holme(2015)]{Holme2015colloquium}
Petter Holme.
\newblock Modern temporal network theory: a colloquium.
\newblock \emph{The European Physical Journal B}, 88\penalty0 (9):\penalty0
  234, Sep 2015.
\newblock ISSN 1434-6036.
\newblock \doi{10.1140/epjb/e2015-60657-4}.
\newblock URL \url{https://doi.org/10.1140/epjb/e2015-60657-4}.

\bibitem[Holme and Saramäki(2012)]{Holme2012Temporal}
Petter Holme and Jari Saramäki.
\newblock Temporal networks.
\newblock \emph{Physics Reports}, 519\penalty0 (3):\penalty0 97 -- 125, 2012.
\newblock ISSN 0370-1573.
\newblock \doi{https://doi.org/10.1016/j.physrep.2012.03.001}.
\newblock URL
  \url{http://www.sciencedirect.com/science/article/pii/S0370157312000841}.
\newblock Temporal Networks.

\bibitem[Huan et~al.(2003)Huan, Wang, and Prins]{HuanWangPrins2003FFSM}
Jun Huan, Wei Wang, and Jan Prins.
\newblock Efficient mining of frequent subgraphs in the presence of
  isomorphism.
\newblock In \emph{Third IEEE International Conference on Data Mining}, pages
  549--552, Nov 2003.
\newblock \doi{10.1109/ICDM.2003.1250974}.

\bibitem[Huan et~al.(2004{\natexlab{a}})Huan, Wang, Prins, and
  Yang]{Huanetal2004SPIN}
Jun Huan, Wei Wang, Jan Prins, and Jiong Yang.
\newblock Spin: Mining maximal frequent subgraphs from graph databases.
\newblock In \emph{Proceedings of the Tenth ACM SIGKDD International Conference
  on Knowledge Discovery and Data Mining}, KDD ’04, page 581–586, New York,
  NY, USA, 2004{\natexlab{a}}. Association for Computing Machinery.
\newblock ISBN 1581138881.
\newblock \doi{10.1145/1014052.1014123}.
\newblock URL \url{https://doi.org/10.1145/1014052.1014123}.

\bibitem[Huan et~al.(2004{\natexlab{b}})Huan, Wang, Prins, and
  Yang]{Huanetal2004SPINtechreport}
Jun Huan, Wei Wang, Jan Prins, and Jiong Yang.
\newblock Spin: mining maximal frequent subgraphs from graph databases.
\newblock Technical Report TR04-018, Department of Computer Science, University
  of North Carolina at Chapel Hill, 2004{\natexlab{b}}.

\bibitem[Huang et~al.(2014)Huang, Cheng, and Wu]{Huang2014Traversals}
Silu Huang, James Cheng, and Huanhuan Wu.
\newblock Temporal graph traversals: Definitions, algorithms, and applications.
\newblock \emph{arXiv}, http://arxiv.org/abs/1401.1919, 2014.
\newblock URL \url{http://arxiv.org/abs/1401.1919}.

\bibitem[Ingalalli et~al.(2018)Ingalalli, Ienco, and
  Poncelet]{Ingalalli2018multigraphs}
Vijay Ingalalli, Dino Ienco, and Pascal Poncelet.
\newblock Mining frequent subgraphs in multigraphs.
\newblock \emph{Information Sciences}, 451-452:\penalty0 50--66, 2018.

\bibitem[Ingram et~al.(2006)Ingram, Stumpf, and Stark]{Ingram2006motifs}
Piers~J Ingram, Michael~PH Stumpf, and Jaroslav Stark.
\newblock Network motifs: structure does not determine function.
\newblock \emph{BMC Genomics}, 7\penalty0 (1):\penalty0 108--108, 2006.
\newblock \doi{https://doi.org/10.1186/1471-2164-7-108}.

\bibitem[Inokuchi and Washio(2010)]{Inokuchi2010FRISS}
Akihiro Inokuchi and Takashi Washio.
\newblock Mining frequent graph sequence patterns induced by vertices.
\newblock In \emph{Proceedings of the 2010 SIAM International Conference on
  Data Mining}, pages 466--477, 2010.
\newblock \doi{10.1137/1.9781611972801.41}.
\newblock URL \url{https://epubs.siam.org/doi/abs/10.1137/1.9781611972801.41}.

\bibitem[Inokuchi et~al.(2000)Inokuchi, Washio, and
  Motoda]{InokuchiWashioMotoda2000}
Akihiro Inokuchi, Takashi Washio, and Hiroshi Motoda.
\newblock An apriori-based algorithm for mining frequent substructures from
  graph data.
\newblock In Djamel~A. Zighed, Jan Komorowski, and Jan {\.{Z}}ytkow, editors,
  \emph{Principles of Data Mining and Knowledge Discovery}, pages 13--23,
  Berlin, Heidelberg, 2000. Springer.
\newblock ISBN 978-3-540-45372-7.

\bibitem[Inokuchi et~al.(2003)Inokuchi, Washio, and
  Motoda]{InokuchiWashioMotoda2003Complete}
Akihiro Inokuchi, Takashi Washio, and Hiroshi Motoda.
\newblock Complete mining of frequent patterns from graphs: Mining graph data.
\newblock \emph{Machine Learning}, 50\penalty0 (3):\penalty0 321--354, Mar
  2003.
\newblock ISSN 1573-0565.
\newblock \doi{10.1023/A:1021726221443}.
\newblock URL \url{https://doi.org/10.1023/A:1021726221443}.

\bibitem[Jiang et~al.(2012)Jiang, Coenen, and Zito]{Jiang2013survey}
Chuntao Jiang, Frans Coenen, and Michele Zito.
\newblock A survey of frequent subgraph mining algorithms.
\newblock \emph{The Knowledge Engineering Review}, 28\penalty0 (1):\penalty0
  75--105, 2012.
\newblock \doi{10.1017/S0269888912000331}.

\bibitem[Jin et~al.(2007{\natexlab{a}})Jin, McCallen, and
  Almaas]{JinMcCallenAlmaas2007TrendMotif}
Ruoming Jin, Scott McCallen, and Eivind Almaas.
\newblock Trend motif: A graph mining approach for analysis of dynamic complex
  networks.
\newblock In \emph{Seventh IEEE International Conference on Data Mining (ICDM
  2007)}, pages 541--546, Oct 2007{\natexlab{a}}.
\newblock \doi{10.1109/ICDM.2007.92}.

\bibitem[Jin et~al.(2007{\natexlab{b}})Jin, McCallen, and
  Almaas]{JinMcCallenAlmaas2007TrendMotiftechreport}
Ruoming Jin, Scott McCallen, and Eivind Almaas.
\newblock Trend motif: A graph mining approach for analysis of dynamic complex
  networks.
\newblock Technical Report TR-KSU-CS-2007-05, Kent State University,
  2007{\natexlab{b}}.

\bibitem[Kaluza et~al.(2010)Kaluza, K{\"o}lzsch, Gastner, and
  Blasius]{kaluza2010complex}
Pablo Kaluza, Andrea K{\"o}lzsch, Michael~T Gastner, and Bernd Blasius.
\newblock The complex network of global cargo ship movements.
\newblock \emph{Journal of the Royal Society Interface}, 7\penalty0
  (48):\penalty0 1093--1103, 2010.
\newblock \doi{https://doi.org/10.1098/rsif.2009.0495}.

\bibitem[Kashani et~al.(2009)Kashani, Ahrabian, Elahi, Nowzari-Dalini, Ansari,
  Asadi, Mohammadi, Schreiber, and Masoudi-Nejad]{Kashani2009Kavosh}
ZRM Kashani, H.~Ahrabian, E.~Elahi, A.~Nowzari-Dalini, ES~Ansari, S.~Asadi,
  S.~Mohammadi, F.~Schreiber, and A.~Masoudi-Nejad.
\newblock Kavosh: a new algorithm for finding network motifs.
\newblock \emph{BMC Bioinformatics}, 10\penalty0 (1):\penalty0 318--318, 2009.
\newblock \doi{https://doi.org/10.1186/1471-2105-10-318}.

\bibitem[Kashtan et~al.(2004{\natexlab{a}})Kashtan, Itzkovitz, Milo, and
  Alon]{Kashtan2004generalizations}
Nadav Kashtan, Shalev Itzkovitz, Ron Milo, and Uri Alon.
\newblock Topological generalizations of network motifs.
\newblock \emph{Physical Review E}, 70:\penalty0 031909, Sep
  2004{\natexlab{a}}.
\newblock \doi{10.1103/PhysRevE.70.031909}.
\newblock URL \url{https://link.aps.org/doi/10.1103/PhysRevE.70.031909}.

\bibitem[Kashtan et~al.(2004{\natexlab{b}})Kashtan, Itzkovitz, Milo, and
  Alon]{Kashtanetal2004}
Nadav Kashtan, Shalev Itzkovitz, Ron Milo, and Uri Alon.
\newblock Efficient sampling algorithm for estimating subgraph concentrations
  and detecting network motifs.
\newblock \emph{Bioinformatics}, 20\penalty0 (11):\penalty0 1746–1758, July
  2004{\natexlab{b}}.
\newblock ISSN 1367-4803.
\newblock \doi{10.1093/bioinformatics/bth163}.
\newblock URL \url{https://doi.org/10.1093/bioinformatics/bth163}.

\bibitem[Kelley et~al.(2003)Kelley, Sharan, Karp, Sittler, Root, Stockwell, and
  Ideker]{Kelleyetal2003Pathways}
Brian~P. Kelley, Roded Sharan, Richard~M. Karp, Taylor Sittler, David~E. Root,
  Brent~R. Stockwell, and Trey Ideker.
\newblock Conserved pathways within bacteria and yeast as revealed by global
  protein network alignment.
\newblock \emph{Proceedings of the National Academy of Sciences}, 100\penalty0
  (20):\penalty0 11394--11399, 2003.
\newblock ISSN 0027-8424.
\newblock \doi{10.1073/pnas.1534710100}.
\newblock URL \url{https://www.pnas.org/content/100/20/11394}.

\bibitem[Kempe et~al.(2002)Kempe, Kleinberg, and
  Kumar]{Kempeetal2002Connectivity}
David Kempe, Jon Kleinberg, and Amit Kumar.
\newblock Connectivity and inference problems for temporal networks.
\newblock \emph{Journal of Computer and System Sciences}, 64\penalty0
  (4):\penalty0 820 -- 842, 2002.
\newblock ISSN 0022-0000.
\newblock \doi{https://doi.org/10.1006/jcss.2002.1829}.
\newblock URL
  \url{http://www.sciencedirect.com/science/article/pii/S0022000002918295}.

\bibitem[Kivelä et~al.(2014)Kivelä, Arenas, Barthelemy, Gleeson, Moreno, and
  Porter]{kivela2014multilayer}
Mikko Kivelä, Alex Arenas, Marc Barthelemy, James~P. Gleeson, Yamir Moreno,
  and Mason~A. Porter.
\newblock {Multilayer networks}.
\newblock \emph{Journal of Complex Networks}, 2\penalty0 (3):\penalty0
  203--271, 07 2014.
\newblock ISSN 2051-1310.
\newblock \doi{10.1093/comnet/cnu016}.
\newblock URL \url{https://doi.org/10.1093/comnet/cnu016}.

\bibitem[K\"{o}bler et~al.(1994)K\"{o}bler, Sch\"{o}ning, and
  Tor\'{a}n]{Johannes1994Isomorphism}
Johannes K\"{o}bler, Uwe Sch\"{o}ning, and Jacobo Tor\'{a}n.
\newblock \emph{The Graph Isomorphism Problem: Its Structural Complexity}.
\newblock Birkhauser Verlag, CHE, 1994.
\newblock ISBN 0817636803.

\bibitem[Kostakos(2009)]{Kostakos2009Temporal}
Vassilis Kostakos.
\newblock Temporal graphs.
\newblock \emph{Physica A: Statistical Mechanics and its Applications},
  388\penalty0 (6):\penalty0 1007--1023, 2009.

\bibitem[Kovanen et~al.(2011)Kovanen, Karsai, Kaski, Kert{\'{e}}sz, and
  Saramäki]{Kovanenetal2011Temporalmotifs}
Lauri Kovanen, M{\'{a}}rton Karsai, Kimmo Kaski, J{\'{a}}nos Kert{\'{e}}sz, and
  Jari Saramäki.
\newblock Temporal motifs in time-dependent networks.
\newblock \emph{Journal of Statistical Mechanics: Theory and Experiment},
  2011\penalty0 (11):\penalty0 P11005, nov 2011.
\newblock \doi{10.1088/1742-5468/2011/11/p11005}.
\newblock URL \url{https://doi.org/10.1088%2F1742-5468%2F2011%2F11%2Fp11005}.

\bibitem[Krishna et~al.(2011)Krishna, Suri, and
  Athithan]{Krishna2011comparative}
Varun Krishna, N.~N. R.~Ranga Suri, and G.~Athithan.
\newblock A comparative survey of algorithms for frequent subgraph discovery.
\newblock \emph{Current Science}, 100\penalty0 (2):\penalty0 190--198, 2011.

\bibitem[Kuramochi and Karypis(2001)]{KuramochiKarypis2001FSG}
Michihiro Kuramochi and George Karypis.
\newblock Frequent subgraph discovery.
\newblock In \emph{Proceedings 2001 IEEE International Conference on Data
  Mining}, pages 313--320, Nov 2001.
\newblock \doi{10.1109/ICDM.2001.989534}.

\bibitem[Kuramochi and Karypis(2004{\natexlab{a}})]{Kuramochi2004Sparse}
Michihiro Kuramochi and George Karypis.
\newblock Finding frequent patterns in a large sparse graph.
\newblock \emph{Society for Industrial and Applied Mathematics. Proceedings of
  the SIAM International Conference on Data Mining}, page 345,
  2004{\natexlab{a}}.

\bibitem[Kuramochi and Karypis(2004{\natexlab{b}})]{KuramochiKarypis2004FSG}
Michihiro Kuramochi and George Karypis.
\newblock An efficient algorithm for discovering frequent subgraphs.
\newblock \emph{IEEE Transactions on Knowledge and Data Engineering},
  16\penalty0 (9):\penalty0 1038--1051, Sep. 2004{\natexlab{b}}.
\newblock ISSN 2326-3865.
\newblock \doi{10.1109/TKDE.2004.33}.

\bibitem[Kuramochi and
  Karypis(2004{\natexlab{c}})]{KuramochiKarypis2004FSGtechreport}
Michihiro Kuramochi and George Karypis.
\newblock An efficient algorithm for discovering frequent subgraphs.
\newblock Technical Report TR 02-026, Department of Computer Science and
  Engineering, University of Minnesota, 2004{\natexlab{c}}.

\bibitem[Kuramochi and Karypis(2004{\natexlab{d}})]{KuramochiKarypis2004GREW}
Michihiro Kuramochi and George Karypis.
\newblock Grew - a scalable frequent subgraph discovery algorithm.
\newblock In \emph{Fourth IEEE International Conference on Data Mining
  (ICDM'04)}, pages 439--442, Nov 2004{\natexlab{d}}.
\newblock \doi{10.1109/ICDM.2004.10024}.

\bibitem[Kuramochi and Karypis(2005)]{Kuramochi2005Sparse}
Michihiro Kuramochi and George Karypis.
\newblock Finding frequent patterns in a large sparse graph*.
\newblock \emph{Data Mining and Knowledge Discovery}, 11\penalty0 (3):\penalty0
  243--271, Nov 2005.
\newblock ISSN 1573-756X.
\newblock \doi{10.1007/s10618-005-0003-9}.
\newblock URL \url{https://doi.org/10.1007/s10618-005-0003-9}.

\bibitem[Kuramochi and Karypis(2006)]{KuramochiKarypis2006MiningGraphData}
Michihiro Kuramochi and George Karypis.
\newblock \emph{Finding Topological Frequent Patterns from Graph Datasets},
  chapter~6, pages 117--158.
\newblock John Wiley \& Sons, Ltd, 2006.
\newblock ISBN 9780470073049.
\newblock \doi{10.1002/9780470073049.ch6}.
\newblock URL
  \url{https://onlinelibrary.wiley.com/doi/abs/10.1002/9780470073049.ch6}.

\bibitem[{Lacroix} et~al.(2006){Lacroix}, {Fernandes}, and
  {Sagot}]{lacroix2006motif}
V.~{Lacroix}, C.~G. {Fernandes}, and M.~{Sagot}.
\newblock Motif search in graphs: Application to metabolic networks.
\newblock \emph{IEEE/ACM Transactions on Computational Biology and
  Bioinformatics}, 3\penalty0 (4):\penalty0 360--368, Oct 2006.
\newblock ISSN 2374-0043.
\newblock \doi{10.1109/TCBB.2006.55}.

\bibitem[Lee et~al.(2012)Lee, Han, Kasperovics, and Lee]{Lee2012isomorphism}
Jinsoo Lee, Wook-Shin Han, Romans Kasperovics, and Jeong-Hoon Lee.
\newblock An in-depth comparison of subgraph isomorphism algorithms in graph
  databases.
\newblock \emph{Proc. VLDB Endow.}, 6\penalty0 (2):\penalty0 133–144,
  December 2012.
\newblock ISSN 2150-8097.
\newblock \doi{10.14778/2535568.2448946}.
\newblock URL \url{https://doi.org/10.14778/2535568.2448946}.

\bibitem[Li et~al.(2017)Li, Cornelius, Liu, Wang, and
  Barab{\'a}si]{Lietal2017temporalnetworks}
A.~Li, S.~P. Cornelius, Y.-Y. Liu, L.~Wang, and A.-L. Barab{\'a}si.
\newblock The fundamental advantages of temporal networks.
\newblock 358\penalty0 (6366):\penalty0 1042--1046, 2017.
\newblock ISSN 0036-8075.
\newblock \doi{10.1126/science.aai7488}.

\bibitem[Li et~al.(2014)Li, Palchykov, Jiang, Kaski, Kert{\'e}sz, Miccich{\`e},
  Tumminello, Zhou, and Mantegna]{li2014statistically}
Ming-Xia Li, Vasyl Palchykov, Zhi-Qiang Jiang, Kimmo Kaski, J{\'a}nos
  Kert{\'e}sz, Salvatore Miccich{\`e}, Michele Tumminello, Wei-Xing Zhou, and
  Rosario~N Mantegna.
\newblock Statistically validated mobile communication networks: the evolution
  of motifs in european and chinese data.
\newblock \emph{New Journal of Physics}, 16\penalty0 (8):\penalty0 083038,
  2014.

\bibitem[Li et~al.(2018)Li, Lou, Shi, and Han]{lietal2018temporalHIN}
Yuchen Li, Zhengzhi Lou, Yu~Shi, and Jiawei Han.
\newblock Temporal motifs in heterogeneous information networks.
\newblock In \emph{MLG Workshop at Knowledge Discovery and Data Mining}, 2018.

\bibitem[Liu et~al.(2012)Liu, Cheung, and Liu]{Liuetal2015stochasticmotifs}
Kai Liu, William~K. Cheung, and Jiming Liu.
\newblock Detecting multiple stochastic network motifs in network data.
\newblock In \emph{Proceedings of the 16th Pacific-Asia Conference on Advances
  in Knowledge Discovery and Data Mining - Volume Part II}, PAKDD’12, page
  205–217, Berlin, Heidelberg, 2012. Springer-Verlag.
\newblock ISBN 9783642302190.
\newblock \doi{10.1007/978-3-642-30220-6_18}.
\newblock URL \url{https://doi.org/10.1007/978-3-642-30220-6_18}.

\bibitem[Mangan and Alon(2003)]{Mangan2003Forward}
Shmoolik Mangan and Uri Alon.
\newblock Structure and function of the feed-forward loop network motif.
\newblock \emph{Proceedings of the National Academy of Sciences}, 100\penalty0
  (21):\penalty0 11980--11985, 2003.
\newblock ISSN 0027-8424.
\newblock \doi{10.1073/pnas.2133841100}.
\newblock URL \url{https://www.pnas.org/content/100/21/11980}.

\bibitem[Marcus and Shavitt(2012)]{Marcus2012RAGE}
Dror Marcus and Yuval Shavitt.
\newblock Rage – a rapid graphlet enumerator for large networks.
\newblock \emph{Computer Networks}, 56\penalty0 (2):\penalty0 810 -- 819, 2012.
\newblock ISSN 1389-1286.
\newblock \doi{https://doi.org/10.1016/j.comnet.2011.08.019}.
\newblock URL
  \url{http://www.sciencedirect.com/science/article/pii/S1389128611003902}.

\bibitem[M{\"a}rtens et~al.(2017)M{\"a}rtens, Meier, Hillebrand, Tewarie, and
  Van~Mieghem]{martens2017brain}
Marcus M{\"a}rtens, Jil Meier, Arjan Hillebrand, Prejaas Tewarie, and Piet
  Van~Mieghem.
\newblock Brain network clustering with information flow motifs.
\newblock \emph{Applied Network Science}, 2\penalty0 (1):\penalty0 25, 2017.
\newblock \doi{https://doi.org/10.1007/s41109-017-0046-z}.

\bibitem[Masoudi-Nejad et~al.(2012)Masoudi-Nejad, Schreiber, and
  Kashani]{Masoudi2012review}
Ali Masoudi-Nejad, Falk Schreiber, and Zahra Razaghi~Moghadam Kashani.
\newblock Building blocks of biological networks: a review on major network
  motif discovery algorithms.
\newblock \emph{IET Systems Biology}, 6\penalty0 (5):\penalty0 164--174, Oct
  2012.
\newblock ISSN 1751-8857.
\newblock \doi{10.1049/iet-syb.2011.0011}.

\bibitem[Matias et~al.(2006)Matias, Schbath, Birmel{\'e}, Daudin, and
  Robin]{matias2006network}
Catherine Matias, Sophie Schbath, Etienne Birmel{\'e}, Jean-Jacques Daudin, and
  St{\'e}phane Robin.
\newblock Network motifs: mean and variance for the count.
\newblock \emph{REVSTAT--Statistical Journal}, 4\penalty0 (1):\penalty0 31--51,
  2006.

\bibitem[Matsuda et~al.(2000)Matsuda, Horiuchi, Motoda, and
  Washio]{matsuda2000GBI1}
Takashi Matsuda, Tadashi Horiuchi, Hiroshi Motoda, and Takashi Washio.
\newblock Extension of graph-based induction for general graph structured data.
\newblock In Takao Terano, Huan Liu, and Arbee L.~P. Chen, editors,
  \emph{Knowledge Discovery and Data Mining. Current Issues and New
  Applications}, pages 420--431, Berlin, Heidelberg, 2000. Springer.
\newblock ISBN 978-3-540-45571-4.

\bibitem[Matsuda et~al.(2002)Matsuda, Motoda, Yoshida, and
  Washio]{matsuda2002BGBI}
Takashi Matsuda, Hiroshi Motoda, Tetsuya Yoshida, and Takashi Washio.
\newblock Mining patterns from structured data by beam-wise graph-based
  induction.
\newblock In Steffen Lange, Ken Satoh, and Carl~H. Smith, editors,
  \emph{Discovery Science}, pages 422--429, Berlin, Heidelberg, 2002. Springer.
\newblock ISBN 978-3-540-36182-4.

\bibitem[Matula(1978)]{matula1978subtree}
David~W. Matula.
\newblock Subtree isomorphism in o(n5/2).
\newblock In B.~Alspach, P.~Hell, and D.J. Miller, editors, \emph{Algorithmic
  Aspects of Combinatorics}, volume~2 of \emph{Annals of Discrete Mathematics},
  pages 91 -- 106. Elsevier, 1978.
\newblock \doi{https://doi.org/10.1016/S0167-5060(08)70324-8}.
\newblock URL
  \url{http://www.sciencedirect.com/science/article/pii/S0167506008703248}.

\bibitem[McKay(1981)]{mckay1981}
Brendan~D. McKay.
\newblock \emph{Practical graph isomorphism}, volume~30.
\newblock Congressus Numerantium, 1981.

\bibitem[McKay and Piperno(2014)]{McKayPiperno2014}
Brendan~D. McKay and Adolfo Piperno.
\newblock Practical graph isomorphism, ii.
\newblock \emph{Journal of Symbolic Computation}, 60:\penalty0 94 -- 112, 2014.
\newblock ISSN 0747-7171.
\newblock \doi{https://doi.org/10.1016/j.jsc.2013.09.003}.
\newblock URL
  \url{http://www.sciencedirect.com/science/article/pii/S0747717113001193}.

\bibitem[Milo et~al.(2002)Milo, Shen-Orr, Itzkovitz, Kashtan, Chklovskii, and
  Alon]{MiloS2002motifs}
Ron Milo, Shai Shen-Orr, Shalev Itzkovitz, Nadav Kashtan, Dmitri Chklovskii,
  and Uri Alon.
\newblock Network motifs: Simple building blocks of complex networks.
\newblock \emph{Science}, 298\penalty0 (5594):\penalty0 824--827, 2002.
\newblock ISSN 0036-8075.
\newblock \doi{10.1126/science.298.5594.824}.
\newblock URL \url{https://science.sciencemag.org/content/298/5594/824}.

\bibitem[Milo et~al.(2003)Milo, Kashtan, Itzkovitz, Newman, and
  Alon]{Milo2003random}
Ron Milo, Nadav Kashtan, Shalev Itzkovitz, Mark~EJ Newman, and Uri Alon.
\newblock On the uniform generation of random graphs with prescribed degree
  sequences, 2003.

\bibitem[Milo et~al.(2004{\natexlab{a}})Milo, Itzkovitz, Kashtan, Levitt, and
  Alon]{Milo2004Response}
Ron Milo, Shalev Itzkovitz, Nadav Kashtan, Reuven Levitt, and Uri Alon.
\newblock Response to comment on "network motifs: Simple building blocks of
  complex networks" and "superfamilies of evolved and designed networks".
\newblock \emph{Science}, 305\penalty0 (5687):\penalty0 1107--1107,
  2004{\natexlab{a}}.
\newblock ISSN 0036-8075.
\newblock \doi{10.1126/science.1100519}.
\newblock URL \url{https://science.sciencemag.org/content/305/5687/1107.4}.

\bibitem[Milo et~al.(2004{\natexlab{b}})Milo, Itzkovitz, Kashtan, Levitt,
  Shen-Orr, Ayzenshtat, Sheffer, and Alon]{Milo2004Superfamilies}
Ron Milo, Shalev Itzkovitz, Nadav Kashtan, Reuven Levitt, Shai Shen-Orr, Inbal
  Ayzenshtat, Michal Sheffer, and Uri Alon.
\newblock Superfamilies of evolved and designed networks.
\newblock \emph{Science}, 303\penalty0 (5663):\penalty0 1538--1542,
  2004{\natexlab{b}}.
\newblock ISSN 0036-8075.
\newblock \doi{10.1126/science.1089167}.
\newblock URL \url{https://science.sciencemag.org/content/303/5663/1538}.

\bibitem[Mrzic et~al.(2018)Mrzic, Meysman, Bittremieux, Moris, Cule, Goethals,
  and Laukens]{Mrzic2018bioinformatics}
Aida Mrzic, Pieter Meysman, Wout Bittremieux, Pieter Moris, Boris Cule, Bart
  Goethals, and Kris Laukens.
\newblock Grasping frequent subgraph mining for bioinformatics applications.
\newblock \emph{BioData Mining}, 11\penalty0 (1):\penalty0 20--24, 2018.
\newblock \doi{https://doi.org/10.1186/s13040-018-0181-9}.

\bibitem[Mukherjee et~al.(2018)Mukherjee, Hasan, Boucher, and
  Kahveci]{Mukherjeeetal2018Countingmotifs}
Kingshuk Mukherjee, Md~M. Hasan, Christina Boucher, and Tamer Kahveci.
\newblock Counting motifs in dynamic networks.
\newblock \emph{BMC Systems Biology}, 12\penalty0 (Suppl 1):\penalty0 6--12,
  2018.
\newblock \doi{https://doi.org/10.1186/s12918-018-0533-6}.

\bibitem[Ng and Li(2009)]{Li2009Biological}
See-Kiong Ng and Xiao-Li Li.
\newblock \emph{Biological Data Mining in Protein Interaction Networks}.
\newblock Information Science Reference - Imprint of: IGI Publishing, Hershey,
  PA, 2009.
\newblock ISBN 1605663980.

\bibitem[Nguyen et~al.(2006)Nguyen, Ohara, Mogi, Motoda, and
  Washio]{Oharaetal2006MiningGraphData}
Phu~Chien Nguyen, Kouzou Ohara, Akira Mogi, Hiroshi Motoda, and Takashi Washio.
\newblock \emph{Constructing Decision Trees for Graph-Structured Data by
  Chunkingless Graph-Based Induction}, page 390–399.
\newblock PAKDD’06. Springer-Verlag, Berlin, Heidelberg, 2006.
\newblock ISBN 3540332065.
\newblock \doi{10.1007/11731139_45}.
\newblock URL \url{https://doi.org/10.1007/11731139_45}.

\bibitem[Nicosia et~al.(2013)Nicosia, Tang, Mascolo, Musolesi, Russo, and
  Latora]{Nicosia2013Temporal}
Vincenzo Nicosia, John Tang, Cecilia Mascolo, Mirco Musolesi, Giovanni Russo,
  and Vito Latora.
\newblock Graph metrics for temporal networks.
\newblock In \emph{Temporal networks}, pages 15--40. Springer, 2013.
\newblock \doi{10.1007/978-3-642-36461-7_2}.

\bibitem[Ohnishi et~al.(2010)Ohnishi, Takayasu, and Takayasu]{Ohnishi2010}
Takaaki Ohnishi, Hideki Takayasu, and Misako Takayasu.
\newblock Network motifs in an inter-firm network.
\newblock \emph{Journal of Economic Interaction and Coordination}, 5\penalty0
  (2):\penalty0 171--180, Dec 2010.
\newblock ISSN 1860-7128.
\newblock \doi{10.1007/s11403-010-0066-6}.
\newblock URL \url{https://doi.org/10.1007/s11403-010-0066-6}.

\bibitem[Omidi et~al.(2009)Omidi, Schreiber, and
  Masoudi-Nejad]{OmidiFalkMasoudi2009MODA}
Saeed Omidi, Falk Schreiber, and Ali Masoudi-Nejad.
\newblock Moda: An efficient algorithm for network motif discovery in
  biological networks.
\newblock \emph{Genes \& Genetic Systems}, 84\penalty0 (5):\penalty0 385--395,
  2009.
\newblock \doi{10.1266/ggs.84.385}.

\bibitem[Onnela et~al.(2005)Onnela, Saram\"aki, Kert\'esz, and
  Kaski]{Onnela2005Intensity}
Jukka-Pekka Onnela, Jari Saram\"aki, J\'anos Kert\'esz, and Kimmo Kaski.
\newblock Intensity and coherence of motifs in weighted complex networks.
\newblock \emph{Physical Review E}, 71:\penalty0 065103, Jun 2005.
\newblock \doi{10.1103/PhysRevE.71.065103}.
\newblock URL \url{https://link.aps.org/doi/10.1103/PhysRevE.71.065103}.

\bibitem[Ottino(2004)]{ottino2004engineering}
Julio~M Ottino.
\newblock Engineering complex systems.
\newblock \emph{Nature}, 427\penalty0 (6973):\penalty0 399, 2004.
\newblock \doi{https://doi.org/10.1038/427399a}.

\bibitem[Pan and Saram\"aki(2011)]{Pan2011temporal}
Raj~Kumar Pan and Jari Saram\"aki.
\newblock Path lengths, correlations, and centrality in temporal networks.
\newblock \emph{Physical Review. E}, 84:\penalty0 016105, Jul 2011.
\newblock \doi{10.1103/PhysRevE.84.016105}.
\newblock URL \url{https://link.aps.org/doi/10.1103/PhysRevE.84.016105}.

\bibitem[Paranjape et~al.(2017)Paranjape, Benson, and
  Leskovec]{Paranjapeetal2017Motifs}
Ashwin Paranjape, Austin~R. Benson, and Jure Leskovec.
\newblock Motifs in temporal networks.
\newblock In \emph{Proceedings of the Tenth ACM International Conference on Web
  Search and Data Mining}, WSDM ’17, page 601–610, New York, NY, USA, 2017.
  Association for Computing Machinery.
\newblock ISBN 9781450346757.
\newblock \doi{10.1145/3018661.3018731}.
\newblock URL \url{https://doi.org/10.1145/3018661.3018731}.

\bibitem[Parida(2007)]{Parida2007topologicalmotifs}
Laxmi Parida.
\newblock Discovering topological motifs using a compact notation.
\newblock \emph{Journal of Computational Biology}, 14\penalty0 (3):\penalty0
  300--323, 2007.
\newblock \doi{10.1089/cmb.2006.0142}.
\newblock URL \url{https://doi.org/10.1089/cmb.2006.0142}.
\newblock PMID: 17563313.

\bibitem[Parthasarathy and Coatney(2002)]{Parthasarathy2002}
Srinivasan Parthasarathy and Matt Coatney.
\newblock Efficient discovery of common substructures in macromolecules.
\newblock In \emph{2002 IEEE International Conference on Data Mining}, pages
  362--369, Dec 2002.
\newblock \doi{10.1109/ICDM.2002.1183924}.

\bibitem[Parthasarathy et~al.(2010)Parthasarathy, Tatikonda, and
  Ucar]{parthasarathy2010survey}
Srinivasan Parthasarathy, Shirish Tatikonda, and Duygu Ucar.
\newblock A survey of graph mining techniques for biological datasets.
\newblock In \emph{Managing and mining graph data}, pages 547--580. Springer,
  2010.
\newblock \doi{10.1007/978-1-4419-6045-0_18}.
\newblock URL \url{https://doi.org/10.1007/978-1-4419-6045-0_18}.

\bibitem[Paulau et~al.(2015)Paulau, Feenders, and Blasius]{paulau2015motif}
Pavel~V Paulau, Christoph Feenders, and Bernd Blasius.
\newblock Motif analysis in directed ordered networks and applications to food
  webs.
\newblock \emph{Scientific Reports}, 5:\penalty0 11926, 2015.
\newblock \doi{https://doi.org/10.1038/srep11926}.

\bibitem[Picard et~al.(2008)Picard, Daudin, Koskas, Schbath, and
  Robin]{Picardetal2008exceptionality}
Franck Picard, J-J Daudin, Michel Koskas, Sophie Schbath, and Stephane Robin.
\newblock Assessing the exceptionality of network motifs.
\newblock \emph{Journal of Computational Biology}, 15\penalty0 (1):\penalty0
  1--20, 2008.
\newblock \doi{10.1089/cmb.2007.0137}.
\newblock URL \url{https://doi.org/10.1089/cmb.2007.0137}.
\newblock PMID: 18257674.

\bibitem[Pinar et~al.(2017)Pinar, Seshadhri, and Vishal]{Pinaretal2016ESCAPE}
Ali Pinar, C~Seshadhri, and Vaidyanathan Vishal.
\newblock Escape: Efficiently counting all 5-vertex subgraphs.
\newblock In \emph{Proceedings of the 26th International Conference on World
  Wide Web}, WWW ’17, page 1431–1440, Republic and Canton of Geneva, CHE,
  2017. International World Wide Web Conferences Steering Committee.
\newblock ISBN 9781450349130.
\newblock \doi{10.1145/3038912.3052597}.
\newblock URL \url{https://doi.org/10.1145/3038912.3052597}.

\bibitem[Pržulj(2007)]{Przulj2007graphletdistribution}
Nataša Pržulj.
\newblock {Biological network comparison using graphlet degree distribution}.
\newblock \emph{Bioinformatics}, 23\penalty0 (2):\penalty0 e177--e183, 01 2007.
\newblock ISSN 1367-4803.
\newblock \doi{10.1093/bioinformatics/btl301}.
\newblock URL \url{https://doi.org/10.1093/bioinformatics/btl301}.

\bibitem[Quevillon et~al.(2015)Quevillon, Hanks, Bansal, and
  Hughes]{quevillon2015social}
Lauren~E Quevillon, Ephraim~M Hanks, Shweta Bansal, and David~P Hughes.
\newblock Social, spatial, and temporal organization in a complex insect
  society.
\newblock \emph{Scientific Reports}, 5:\penalty0 13393, 2015.
\newblock \doi{https://doi.org/10.1038/srep13393}.

\bibitem[Rahman et~al.(2014)Rahman, Bhuiyan, and Al~Hasan]{Rahman2014Graft}
Mahmudur Rahman, Mansurul~A. Bhuiyan, and Mohammad Al~Hasan.
\newblock Graft: An efficient graphlet counting method for large graph
  analysis.
\newblock \emph{IEEE Transactions on Knowledge and Data Engineering},
  26\penalty0 (10):\penalty0 2466--2478, Oct 2014.
\newblock ISSN 2326-3865.
\newblock \doi{10.1109/TKDE.2013.2297929}.

\bibitem[Raj and Prabhakar(2015)]{Ramraj2015Survey}
Ram Raj and R~Prabhakar.
\newblock Frequent subgraph mining algorithms – a survey.
\newblock \emph{Procedia Computer Science}, 47:\penalty0 197 -- 204, 2015.
\newblock ISSN 1877-0509.
\newblock \doi{https://doi.org/10.1016/j.procs.2015.03.198}.
\newblock URL
  \url{http://www.sciencedirect.com/science/article/pii/S1877050915004664}.
\newblock Graph Algorithms, High Performance Implementations and Its
  Applications ( ICGHIA 2014 ).

\bibitem[Ravandi et~al.(2019)Ravandi, Mili, and Springer]{Ravandi2019driver}
Babak Ravandi, Fatma Mili, and John~A Springer.
\newblock {Identifying and using driver nodes in temporal networks}.
\newblock \emph{Journal of Complex Networks}, 7\penalty0 (5):\penalty0
  720--748, 03 2019.
\newblock ISSN 2051-1329.
\newblock \doi{10.1093/comnet/cnz004}.

\bibitem[Ray et~al.(2014)Ray, Holder, and
  Choudhury]{Rayetal2014StreamingGraphs}
Abhik Ray, Lawrence~B. Holder, and Sutanay Choudhury.
\newblock Frequent subgraph discovery in large attributed streaming graphs.
\newblock In \emph{Proceedings of the 3rd International Conference on Big Data,
  Streams and Heterogeneous Source Mining: Algorithms, Systems, Programming
  Models and Applications - Volume 36}, BIGMINE'14, pages 166--181. JMLRorg,
  2014.

\bibitem[Read and Corneil(1977)]{Read1977isomorphism}
Ronald~C. Read and Derek~G. Corneil.
\newblock The graph isomorphism disease.
\newblock \emph{Journal of Graph Theory}, 1\penalty0 (4):\penalty0 339--363,
  1977.
\newblock \doi{10.1002/jgt.3190010410}.
\newblock URL
  \url{https://onlinelibrary.wiley.com/doi/abs/10.1002/jgt.3190010410}.

\bibitem[Redmond and Cunningham(2013)]{Redmond2013Isomorphism}
Ursula Redmond and P\'{a}draig Cunningham.
\newblock Temporal subgraph isomorphism.
\newblock In \emph{Proceedings of the 2013 IEEE/ACM International Conference on
  Advances in Social Networks Analysis and Mining}, ASONAM ’13, page
  1451–1452, New York, NY, USA, 2013. Association for Computing Machinery.
\newblock ISBN 9781450322409.
\newblock \doi{10.1145/2492517.2492586}.
\newblock URL \url{https://doi.org/10.1145/2492517.2492586}.

\bibitem[Rehman et~al.(2014)Rehman, Asghar, Zhuang, and
  Fong]{Rehman2014Performance}
Saif~U. Rehman, Sohail Asghar, Yan Zhuang, and Simon Fong.
\newblock Performance evaluation of frequent subgraph discovery techniques.
\newblock \emph{Mathematical Problems in Engineering}, 2014:\penalty0 1--6,
  2014.
\newblock \doi{https://doi.org/10.1155/2014/869198}.

\bibitem[Ren et~al.(2019)Ren, Sarkar, Ay, Dobra, and Kahveci]{ren2019finding}
Yuanfang Ren, Aisharjya Sarkar, Ahmet Ay, Alin Dobra, and Tamer Kahveci.
\newblock Finding conserved patterns in multilayer networks.
\newblock In \emph{Proceedings of the 10th ACM International Conference on
  Bioinformatics, Computational Biology and Health Informatics}, BCB ’19,
  page 97–102, New York, NY, USA, 2019. Association for Computing Machinery.
\newblock ISBN 9781450366663.
\newblock \doi{10.1145/3307339.3342184}.
\newblock URL \url{https://doi.org/10.1145/3307339.3342184}.

\bibitem[Reyner(1977)]{Reyner1977}
Steven~W. Reyner.
\newblock An analysis of a good algorithm for the subtree problem.
\newblock \emph{SIAM Journal on Computing}, 6\penalty0 (4):\penalty0 730--732,
  1977.
\newblock \doi{https://doi.org/10.1137/0206053}.

\bibitem[Ribeiro et~al.(2009)Ribeiro, Silva, and Kaiser]{Ribeiro2009Strategies}
Pedro Ribeiro, Fernando Silva, and Marcus Kaiser.
\newblock Strategies for network motifs discovery.
\newblock In \emph{2009 Fifth IEEE International Conference on e-Science},
  pages 80--87, Dec 2009.
\newblock \doi{10.1109/e-Science.2009.20}.

\bibitem[Ribeiro et~al.(2012)Ribeiro, Silva, and Lopes]{Ribeiro2012Parallel}
Pedro Ribeiro, Fernando Silva, and Luís Lopes.
\newblock Parallel discovery of network motifs.
\newblock \emph{Journal of Parallel and Distributed Computing}, 72\penalty0
  (2):\penalty0 144 -- 154, 2012.
\newblock ISSN 0743-7315.
\newblock \doi{https://doi.org/10.1016/j.jpdc.2011.08.007}.
\newblock URL
  \url{http://www.sciencedirect.com/science/article/pii/S0743731511001729}.

\bibitem[Rissanen(1989)]{Rissanen1989}
Jorma Rissanen.
\newblock \emph{Stochastic Complexity in Statistical Inquiry Theory}.
\newblock World Scientific Publishing Co., Inc., USA, 1989.
\newblock ISBN 9971508591.

\bibitem[S. and Jackway(2005)]{Hallinan2005Motifs}
Hallinan~Jennifer S. and P~T. Jackway.
\newblock Network motifs, feedback loops and the dynamics of genetic regulatory
  networks.
\newblock In \emph{2005 IEEE Symposium on Computational Intelligence in
  Bioinformatics and Computational Biology}, pages 1--7, Nov 2005.
\newblock \doi{10.1109/CIBCB.2005.1594903}.

\bibitem[Samatova et~al.(2013)Samatova, Hendrix, Jenkins, Padmanabhan, and
  Chakraborty]{samatova2013practical}
Nagiza~F Samatova, William Hendrix, John Jenkins, Kanchana Padmanabhan, and
  Arpan Chakraborty.
\newblock \emph{Practical graph mining with R}.
\newblock CRC Press, 2013.

\bibitem[Schneider et~al.(2013)Schneider, Belik, Couronné, Smoreda, and
  González]{Schneider2013human}
Christian~M. Schneider, Vitaly Belik, Thomas Couronné, Zbigniew Smoreda, and
  Marta~C. González.
\newblock Unravelling daily human mobility motifs.
\newblock \emph{Journal of The Royal Society Interface}, 10\penalty0
  (84):\penalty0 20130246, 2013.
\newblock \doi{10.1098/rsif.2013.0246}.
\newblock URL
  \url{https://royalsocietypublishing.org/doi/abs/10.1098/rsif.2013.0246}.

\bibitem[Schreiber and Schwöbbermeyer(2004)]{schreiber2004towards}
Falk Schreiber and Henning Schwöbbermeyer.
\newblock Towards motif detection in networks: Frequency concepts and flexible
  search.
\newblock In \emph{Proceedings of the International Workshop on Network Tools
  and Applications in Biology (NETTAB04}, pages 91--102, 2004.

\bibitem[Schreiber and Schwöbbermeyer(2005)]{Schreiber2005MAVisto}
Falk Schreiber and Henning Schwöbbermeyer.
\newblock {MAVisto: a tool for the exploration of network motifs}.
\newblock \emph{Bioinformatics}, 21\penalty0 (17):\penalty0 3572--3574, 07
  2005.
\newblock ISSN 1367-4803.
\newblock \doi{10.1093/bioinformatics/bti556}.
\newblock URL \url{https://doi.org/10.1093/bioinformatics/bti556}.

\bibitem[Schreiber and Schw{\"o}bbermeyer(2005)]{Schreiber2005Frequency}
Falk Schreiber and Henning Schw{\"o}bbermeyer.
\newblock Frequency concepts and pattern detection for the analysis of motifs
  in networks.
\newblock In Corrado Priami, Emanuela Merelli, Pablo Gonzalez, and Andrea
  Omicini, editors, \emph{Transactions on Computational Systems Biology III},
  pages 89--104, Berlin, Heidelberg, 2005. Springer.
\newblock ISBN 978-3-540-31446-2.

\bibitem[Shahrivari and Jalili(2015)]{Shahrivari2015Multicore}
Saeed Shahrivari and Saeed Jalili.
\newblock Fast parallel all-subgraph enumeration using multicore machines.
\newblock \emph{Scientific Programming}, 2015, January 2015.
\newblock ISSN 1058-9244.
\newblock \doi{10.1155/2015/901321}.
\newblock URL \url{https://doi.org/10.1155/2015/901321}.

\bibitem[Sharan et~al.(2005)Sharan, Suthram, Kelley, Kuhn, McCuine, Uetz,
  Sittler, Karp, and Ideker]{Sharanetal2005Conserved}
Roded Sharan, Silpa Suthram, Ryan~M. Kelley, Tanja Kuhn, Scott McCuine, Peter
  Uetz, Taylor Sittler, Richard~M. Karp, and Trey Ideker.
\newblock Conserved patterns of protein interaction in multiple species.
\newblock \emph{Proceedings of the National Academy of Sciences}, 102\penalty0
  (6):\penalty0 1974--1979, 2005.
\newblock ISSN 0027-8424.
\newblock \doi{10.1073/pnas.0409522102}.
\newblock URL \url{https://www.pnas.org/content/102/6/1974}.

\bibitem[Shen-Orr et~al.(2002)Shen-Orr, Alon, Mangan, and
  Milo]{Shen2002transcriptional}
Shai~S. Shen-Orr, Uri Alon, Shmoolik Mangan, and Ron Milo.
\newblock Network motifs in the transcriptional regulation network of
  escherichia coli.
\newblock \emph{Nature Genetics}, 31\penalty0 (1):\penalty0 64--68, 2002.
\newblock \doi{https://doi.org/10.1038/ng881}.

\bibitem[Slota and Madduri(2013)]{SlotaMadduri2013Counting}
George~M. Slota and Kamesh Madduri.
\newblock Fast approximate subgraph counting and enumeration.
\newblock In \emph{2013 42nd International Conference on Parallel Processing},
  pages 210--219, Oct 2013.
\newblock \doi{10.1109/ICPP.2013.30}.

\bibitem[Sporns and Kötter(2004)]{Sporns2004brain}
Olaf Sporns and Rolf Kötter.
\newblock Motifs in brain networks.
\newblock \emph{PLOS Biology}, 2\penalty0 (11), 10 2004.
\newblock \doi{10.1371/journal.pbio.0020369}.
\newblock URL \url{https://doi.org/10.1371/journal.pbio.0020369}.

\bibitem[Sporns et~al.(2007)Sporns, Honey, and
  K{\"o}tter]{sporns2007identification}
Olaf Sporns, Christopher~J Honey, and Rolf K{\"o}tter.
\newblock Identification and classification of hubs in brain networks.
\newblock \emph{PLoS ONE}, 2\penalty0 (10):\penalty0 e1049, 2007.
\newblock \doi{https://doi.org/10.1371/journal.pone.0001049}.

\bibitem[Tang et~al.(2010{\natexlab{a}})Tang, Musolesi, Mascolo, and
  Latora]{Tangetal2010TemporalDistance}
John Tang, Mirco Musolesi, Cecilia Mascolo, and Vito Latora.
\newblock Characterising temporal distance and reachability in mobile and
  online social networks, January 2010{\natexlab{a}}.
\newblock ISSN 0146-4833.
\newblock URL \url{https://doi.org/10.1145/1672308.1672329}.

\bibitem[Tang et~al.(2010{\natexlab{b}})Tang, Musolesi, Mascolo, Latora, and
  Nicosia]{Tangetal2010centralitymetrics}
John Tang, Mirco Musolesi, Cecilia Mascolo, Vito Latora, and Vincenzo Nicosia.
\newblock Analysing information flows and key mediators through temporal
  centrality metrics.
\newblock In \emph{Proceedings of the 3rd Workshop on Social Network Systems},
  SNS ’10, New York, NY, USA, 2010{\natexlab{b}}. Association for Computing
  Machinery.
\newblock ISBN 9781450300803.
\newblock \doi{10.1145/1852658.1852661}.
\newblock URL \url{https://doi.org/10.1145/1852658.1852661}.

\bibitem[Ullmann(1976)]{Ullmann1976}
Julian~R Ullmann.
\newblock An algorithm for subgraph isomorphism.
\newblock \emph{Journal of the ACM (JACM)}, 23\penalty0 (1):\penalty0 31–42,
  January 1976.
\newblock ISSN 0004-5411.
\newblock \doi{10.1145/321921.321925}.
\newblock URL \url{https://doi.org/10.1145/321921.321925}.

\bibitem[Vanetik et~al.(2006)Vanetik, Shimony, and
  Gudes]{VanetikShimonyGudes2006}
N.~Vanetik, S.~E. Shimony, and E.~Gudes.
\newblock Support measures for graph data.
\newblock \emph{Data Mining and Knowledge Discovery}, 13\penalty0 (2):\penalty0
  243–260, September 2006.
\newblock ISSN 1384-5810.
\newblock \doi{10.1007/s10618-006-0044-8}.
\newblock URL \url{https://doi.org/10.1007/s10618-006-0044-8}.

\bibitem[Vanetik et~al.(2002)Vanetik, Gudes, and
  Shimony]{VanetikGudesShimony2002}
Natalia Vanetik, Ehud Gudes, and Solomon~Eyal Shimony.
\newblock Computing frequent graph patterns from semistructured data.
\newblock In \emph{2002 IEEE International Conference on Data Mining}, pages
  458--465, Dec 2002.
\newblock \doi{10.1109/ICDM.2002.1183988}.

\bibitem[V{\'a}zquez et~al.(2004)V{\'a}zquez, Dobrin, Sergi, Eckmann, Oltvai,
  and Barab{\'a}si]{Topological2004Complex}
A.~V{\'a}zquez, R.~Dobrin, D.~Sergi, J.-P. Eckmann, Z.~N. Oltvai, and A.-L.
  Barab{\'a}si.
\newblock The topological relationship between the large-scale attributes and
  local interaction patterns of complex networks.
\newblock \emph{Proceedings of the National Academy of Sciences}, 101\penalty0
  (52):\penalty0 17940--17945, 2004.
\newblock ISSN 0027-8424.
\newblock \doi{10.1073/pnas.0406024101}.
\newblock URL \url{https://www.pnas.org/content/101/52/17940}.

\bibitem[Wan and Mamitsuka(2009)]{Wan2009Motifs}
Raymond Wan and Hiroshi Mamitsuka.
\newblock Discovering network motifs in protein interaction networks.
\newblock In \emph{Biological Data Mining in Protein Interaction Networks},
  pages 117--143. IGI Global, 2009.
\newblock \doi{DOI: 10.4018/978-1-60566-398-2.ch008}.

\bibitem[Wang and Parthasarathy(2004)]{Wang2004pmotifminer}
Chao Wang and Srinivasan Parthasarathy.
\newblock Parallel algorithms for mining frequent structural motifs in
  scientific data.
\newblock In \emph{Proceedings of the 18th Annual International Conference on
  Supercomputing}, ICS ’04, page 31–40, New York, NY, USA, 2004.
  Association for Computing Machinery.
\newblock ISBN 1581138393.
\newblock \doi{10.1145/1006209.1006215}.
\newblock URL \url{https://doi.org/10.1145/1006209.1006215}.

\bibitem[Wang et~al.(2014)Wang, Lui, Ribeiro, Towsley, Zhao, and
  Guan]{Wang2014Statistics}
Pinghui Wang, John C.~S. Lui, Bruno Ribeiro, Don Towsley, Junzhou Zhao, and
  Xiaohong Guan.
\newblock Efficiently estimating motif statistics of large networks.
\newblock \emph{ACM Transactions on Knowledge Discovery from Data}, 9\penalty0
  (2), September 2014.
\newblock ISSN 1556-4681.
\newblock \doi{10.1145/2629564}.
\newblock URL \url{https://doi.org/10.1145/2629564}.

\bibitem[Washio and Motoda(2003)]{Washio2003graph}
Takashi Washio and Hiroshi Motoda.
\newblock State of the art of graph-based data mining.
\newblock \emph{ACM SIGKDD Explorations Newsletter}, 5\penalty0 (1):\penalty0
  59–68, July 2003.
\newblock ISSN 1931-0145.
\newblock \doi{10.1145/959242.959249}.
\newblock URL \url{https://doi.org/10.1145/959242.959249}.

\bibitem[Waters and Fewell(2012)]{Waters2012Insect}
James~S. Waters and Jennifer~H. Fewell.
\newblock Information processing in social insect networks.
\newblock \emph{PLOS ONE}, 7\penalty0 (7):\penalty0 1--7, 07 2012.
\newblock \doi{10.1371/journal.pone.0040337}.
\newblock URL \url{https://doi.org/10.1371/journal.pone.0040337}.

\bibitem[Watts and Strogatz(1998)]{WattsStrogatz1998}
Duncan~J. Watts and Steven~H. Strogatz.
\newblock Collective dynamics of `small-world' networks.
\newblock \emph{Nature}, 393\penalty0 (6684):\penalty0 440--442, 1998.
\newblock \doi{https://doi.org/10.1038/30918}.

\bibitem[Wernicke(2005)]{wernicke2005faster}
Sebastian Wernicke.
\newblock A faster algorithm for detecting network motifs.
\newblock In Rita Casadio and Gene Myers, editors, \emph{Algorithms in
  Bioinformatics}, pages 165--177, Berlin, Heidelberg, 2005. Springer.
\newblock ISBN 978-3-540-31812-5.

\bibitem[Wernicke(2006)]{FANMOD_Wernicke2006}
Sebastian Wernicke.
\newblock Efficient detection of network motifs.
\newblock \emph{IEEE/ACM Transactions on Computational Biology and
  Bioinformatics}, 3\penalty0 (4):\penalty0 347–359, October 2006.
\newblock ISSN 1545-5963.
\newblock \doi{10.1109/TCBB.2006.51}.
\newblock URL \url{https://doi.org/10.1109/TCBB.2006.51}.

\bibitem[Wernicke and Rasche(2006)]{WernickeRasche2006FANMOD}
Sebastian Wernicke and Florian Rasche.
\newblock {FANMOD: a tool for fast network motif detection}.
\newblock \emph{Bioinformatics}, 22\penalty0 (9):\penalty0 1152--1153, 02 2006.
\newblock ISSN 1367-4803.
\newblock \doi{10.1093/bioinformatics/btl038}.
\newblock URL \url{https://doi.org/10.1093/bioinformatics/btl038}.

\bibitem[Willy(2009)]{Willy2009Discovering}
Hugo Willy.
\newblock Discovering interaction motifs from protein interaction networks.
\newblock In \emph{Biological Data Mining in Protein Interaction Networks},
  pages 99--116. IGI Global, 2009.
\newblock \doi{10.4018/978-1-60566-398-2.ch007}.

\bibitem[W{\"o}rlein et~al.(2005)W{\"o}rlein, Meinl, Fischer, and
  Philippsen]{Worlein2005Quantitative}
Marc W{\"o}rlein, Thorsten Meinl, Ingrid Fischer, and Michael Philippsen.
\newblock A quantitative comparison of the subgraph miners mofa, gspan, ffsm,
  and gaston.
\newblock In Al{\'i}pio~M{\'a}rio Jorge, Lu{\'i}s Torgo, Pavel Brazdil, Rui
  Camacho, and Jo{\~a}o Gama, editors, \emph{Knowledge Discovery in Databases:
  PKDD 2005}, pages 392--403, Berlin, Heidelberg, 2005. Springer.

\bibitem[Wuchty et~al.(2003)Wuchty, Oltvai, and
  Barab{\'a}si]{wuchty2003evolutionary}
Stephan Wuchty, Zolt{\'a}n~N Oltvai, and Albert-L{\'a}szl{\'o} Barab{\'a}si.
\newblock Evolutionary conservation of motif constituents in the yeast protein
  interaction network.
\newblock \emph{Nature Genetics}, 35\penalty0 (2):\penalty0 176--179, 2003.
\newblock \doi{https://doi.org/10.1038/ng1242}.

\bibitem[Yan and Han(2002{\natexlab{a}})]{YanHan2002gSpan}
Xifeng Yan and Jiawei Han.
\newblock gspan: graph-based substructure pattern mining.
\newblock In \emph{2002 IEEE International Conference on Data Mining, 2002.
  Proceedings.}, pages 721--724, Dec 2002{\natexlab{a}}.
\newblock \doi{10.1109/ICDM.2002.1184038}.

\bibitem[Yan and Han(2002{\natexlab{b}})]{YanHan2002gSpantech}
Xifeng Yan and Jiawei Han.
\newblock gspan: graph-based substructure pattern mining.
\newblock Technical Report UIUCDCS-R-2002-2296, University of Illinois at
  Urbana-Champaign, 2002{\natexlab{b}}.

\bibitem[Yan and Han(2006)]{yan2006discovery}
Xifeng Yan and Jiawei Han.
\newblock \emph{Discovery of Frequent Substructures}, chapter~5, pages 97--115.
\newblock John Wiley \& Sons, Ltd, 2006.
\newblock ISBN 9780470073049.
\newblock \doi{10.1002/9780470073049.ch5}.
\newblock URL
  \url{https://onlinelibrary.wiley.com/doi/abs/10.1002/9780470073049.ch5}.

\bibitem[Yoshida and Motoda(1995)]{YoshidaMotoda1995CLIP}
Ken'ichi Yoshida and Hiroshi Motoda.
\newblock Clip: concept learning from inference patterns.
\newblock \emph{Artificial Intelligence}, 75\penalty0 (1):\penalty0 63 -- 92,
  1995.
\newblock ISSN 0004-3702.
\newblock \doi{https://doi.org/10.1016/0004-3702(94)00066-A}.
\newblock URL
  \url{http://www.sciencedirect.com/science/article/pii/000437029400066A}.
\newblock AI Research in Japan.

\bibitem[Yoshida et~al.(1994)Yoshida, Motoda, and Indurkhya]{yoshida1994graph}
Kenichi Yoshida, Hiroshi Motoda, and Nitin Indurkhya.
\newblock Graph-based induction as a unified learning framework.
\newblock \emph{Applied Intelligence}, 4\penalty0 (3):\penalty0 297--316, 1994.

\bibitem[You et~al.(2006)You, Holder, and Cook]{Youetal2006MetabolicPathways}
Chang~Hun You, Lawrence~B Holder, and Diane~J Cook.
\newblock Application of graph-based data mining to metabolic pathways.
\newblock In \emph{Sixth IEEE International Conference on Data Mining -
  Workshops (ICDMW'06)}, pages 169--173, Dec 2006.
\newblock \doi{10.1109/ICDMW.2006.31}.

\bibitem[Zaki(2004)]{Zaki2005SLEUTH}
Mohammed~J. Zaki.
\newblock Efficiently mining frequent embedded unordered trees.
\newblock \emph{Fundamenta Informaticae}, 66\penalty0 (1–2):\penalty0
  33–52, November 2004.
\newblock ISSN 0169-2968.

\bibitem[Zhao et~al.(2010)Zhao, Tian, He, Oliver, Jin, and
  Lee]{Zhaoetal2010Communicationmotifs}
Qiankun Zhao, Yuan Tian, Qi~He, Nuria Oliver, Ruoming Jin, and Wang-Chien Lee.
\newblock Communication motifs: A tool to characterize social communications.
\newblock In \emph{Proceedings of the 19th ACM International Conference on
  Information and Knowledge Management}, CIKM ’10, page 1645–1648, New
  York, NY, USA, 2010. Association for Computing Machinery.
\newblock ISBN 9781450300995.
\newblock \doi{10.1145/1871437.1871694}.
\newblock URL \url{https://doi.org/10.1145/1871437.1871694}.

\bibitem[Zhu et~al.(2011)Zhu, Qu, Lo, Yan, Han, and Yu]{zhu2011mining}
Feida Zhu, Qiang Qu, David Lo, Xifeng Yan, Jiawei Han, and Philip~S Yu.
\newblock Mining top-k large structural patterns in a massive network.
\newblock \emph{Proceedings of the VLDB Endowment}, 4\penalty0 (11):\penalty0
  807--818, 2011.

\end{thebibliography}

\end{document}